\documentclass[final,3p,authoryear,letter]{elsarticle}%
\usepackage{amsfonts}
\usepackage{empheq}
\usepackage[utf8]{inputenc}
\usepackage[hidelinks]{hyperref}
\usepackage{amsmath,amssymb}
\usepackage{float}
\usepackage{yfonts}
\usepackage{changepage}
\usepackage{caption}
\usepackage{subcaption}
\usepackage{tikz}
\usepackage{eufrak}
\usepackage{pgfplots}
\usepackage{mathrsfs}
\usepackage{multirow}
\usepackage{tikz}
\usepackage{amsmath}
\usepackage{amssymb}
\usepackage{graphicx}%
\usetikzlibrary{external}
\providecommand{\U}[1]{\protect\rule{.1in}{.1in}}
\pgfplotsset{compat=1.14}

\include{References}
\begin{document}

\begin{frontmatter}
\title{On the network orientational affinity assumption in polymers   and the  micro-macro connection through the chain stretch}
\author[Add1]{V\'ictor Jes\'us Amores}
\author[Add1]{Khanh Nguyen}
\author[Add1,Add2]{Francisco Javier Mont\'ans \corref{cor1}}
\address[Add1]{E.T.S. de Ingenier\'ia Aeron\'autica y del Espacio. Universidad Polit\'ecnica de Madrid, Pza. Cardenal Cisneros 3, 28040-Madrid, Spain.}
\address[Add2]{Dept. of Mechanical and Aerospace Engineering. Herbert Wertheim College of Engineering, University of Florida, Gainesville, FL 32611, USA}
\cortext[cor1]{Corresponding author, e-mail: fco.montans@upm.es}

\begin{abstract}
We question the network affinity   assumption in modeling chain orientations under polymer deformations, and  the use of the stretch measure projected from the right Cauchy-Green deformation tensor (or  non-affine micro-stretches derived from that measure) as a basic state variable for the micro-macro transition. These ingredients are standard, taken from the statistical theory of polymers, and used in most micromechanical polymer network and soft tissue models.

This  assumption imposes a constraint in the network which results in an anisotropic distribution of the orientation of the chains and, hence, in an additional configurational entropy that should be included. This additional entropy would result in an additional stress tensor. But an arguably more natural alternative, in line with the typical assumption for the chain behavior itself and with the disregard of these forces, is to consider that
the network may fluctuate unconstrained to adapt to macroscopic deformations. This way, the isotropic statistical distribution of the orientation of the chains is maintained unconstrained during deformation and no additional stress is imposed. We show that this free-fluctuating network
assumption is equivalent to consider the stretch projected from the stretch tensor (instead of the right Cauchy-Green deformation tensor) as the state variable for the deformation of the network chains.

Employing our recent data-driven macro-micro-macro approach, we show very important differences in predictions using both assumed behaviors, and demonstrate that with the free-fluctuating network
assumption, we can obtain accurate predictions for all tests in polymers using just one test curve to calibrate the model. With the same macro-micro-macro approach employing the network affinity assumption, we are capable of capturing accurately only the test used for calibration of the model, but not the overall polymer behavior.
Further numerical examples are developed to give supporting evidence for the  unconstrained orientation assumption.
\end{abstract}
\end{frontmatter}

\section{Introduction\label{SEC:1}}

\label{Introduction}
Polymers have an intrinsic entropic nature (see e.g. Sec. IV in \cite{Anthony}, Fig. 2.8 in \cite{Treloar_book} and \cite{Gusev}), meaning that they are formed of
interconnected networks of large chains which vibrate freely and, statistically, produce
forces, nonvanishing in average, when such movement is constrained by a given continuum deformation \citep{Flory,Ward_book,Argon,Mark_Eman_book}. This entropic nature results in the characteristic soft behavior, to which the internal energy plays only a minor role in the intramolecular interactions \citep[Ch.13]{Flory,Treloar_book}. The practical and accurate modelling of polymers has been a very complex task which involved many researchers during the last century, resulting in a number of models ``approximately equal to the number of researchers who work (or worked) in the field'' \cite{Volokh}.  In essence, the two main approaches to model the stress-strain behavior of polymers are the phenomenological approach and the micromechanical approach.

Even though there are earlier works, the phenomenological   \cite{Mooney} model is probably the first model to give a relevant correlation to several types of tests at moderate stretches. Furthermore, the simplest form including only the first invariant $I_1$ of the Cauchy-Green deformation tensor $\boldsymbol{C}$, is coincident with the Gaussian micromechanical \emph{affine} approach, and it is well-known as the Neo-Hookean model. However, Mooney, and later many other authors \citep[among  others]{Rivlin,RivlinSaunders,Hart-Smith,Yeoh,Yeah-Fleming,Horgan-Sacco,Kroon,Itskov}, already noted the need to include the second invariant $I_2$ of $\boldsymbol{C}$ in the stored energy function for capturing well several deformation modes, an issue often referred to as \emph{``small deviations from the statistical theory''} (Claim C0) \citep[Sec. 10.8]{Yeah-Fleming,Treloar_book}. This \emph{deviation}, which is related to a term in $I_2$, was made apparent by Mooney in his ``Mooney plots'', in which experiments in polymers consistently exhibited a slope not predicted by the classical statistical theory. This was further analyzed by  \cite{RivlinSaunders} and \cite{Obata}. However, ``in contrast to the original success of the statistical theory, the failure to secure any very significant understanding of the relatively rather small deviations from the theory, despite repeated attempts over a period of 30 years [now 75], is disappointing'' \citep[p.229]{Treloar_book}. Nonetheless, Mooney-Rivlin approaches, including the $I_2$ term, are commonly used in finite element codes because of their practical fitting capability. A phenomenological  alternative to the Mooney-Rivlin invariants (or other similar alternative invariants, \cite{Criscione}, \cite{Shariff_inv}) are models based on principal stretches, specially those using the \cite{VL} decomposition. Two well-known models of this type are those from \cite{Ogden_model} and \cite{Shariff}. Remarkably, these models are known to perform very well in polymers, specially when they are characterized from several test types \citep{OSS,Verron,Itskov}. Indeed, Ogden's model is also very common in finite element codes and a frequent choice by engineering modelers.

The micromechanical theory starts probably with the work of \cite{Staudinger}, who recognized the polymer long chain structure and coined the word \emph{macromolecule}. It was   followed by the understanding of polymers through the kinetic/entropic theory of reversible deformations (Sec. 6 in \cite{Meyer}) and by the treatment of polymers as free joint chains in \cite{Kuhn1,Kuhn2,Kuhn-Grun}; the latter introducing the non-Gaussian distribution. The statistical distribution of chain lengths and the stress-strain relation, were given by  \cite{Wall}. The non-Gaussian stress-strain relation based on the Langevin function was given  by \cite{James-Guth}, who also introduced the three-chain model.  \cite{James-Guth} further addressed in an appendix a problem relevant to the micro-macro transition in polymer networks:\emph{ the affinity in displacements of junction points} of the network. They made a mathematical study assuming a Gaussian approach and that the joints at the boundary are fixed, whereas the rest of the junction points are free to fluctuate with the network around an equilibrium position. The conclusion of this early study is discussed and summarized in Sec. 4.7 of \cite{Treloar_book}: junction points  (their statistical equilibrium position) move in average in a pattern affine with (as  attached to)  the continuum, and the loads over the junction points are just those from the chains. Hence, at least in the Gaussian (moderate stretches) domain, this finding supports the computation of the chain stretch $\lambda_{ch}$ from the continuum deformation gradient, just projecting $\boldsymbol{C}$ on the original end-to-end direction of the  chain. Since then, this is the approach followed by most works (\cite{Treloar_book,Ward_book,Holzapfel,Volokh}). Furthermore, \cite{Flory-Rehner} developed a non-affine four-chain model. The chains connected the vertices of a representative tetrahedron cell with a central point. The junction points at the vertices followed affine displacements, but the central point moved freely as to maximize the entropy of the system. Their conclusion was again that the non-affine movement of the central point had little contribution to the entropy of the Gaussian model. But Treloar used an evolution of this model to the non-Gaussian domain (large stretches) and compared the average chain tension for the affine and non-affine cases. He observed that the upturn of the tension-stretch curve moved considerably to large stretches for the non-affine case when compared to the affine one, delaying chain locking effects and producing  large discrepancies in the chain tensions in that part of the domain; see Fig. 6.10 in \cite{Treloar_book} and therein references. Importantly, this work showed that, whereas the affinity assumption seemed not to affect substantially predictions at low-to-moderate stretches, it may be overconstraining the models at moderate-to-large stretches.

\cite{Flory76} (see also therein references) questioned the distinction between ``free'' junction points and \emph{macroscopic} constraints fixing \emph{microscopic} junction points (in essence, macroscopic entities which fix the microscopic ones that themselves constitute the macroscopic entities). In consequence he  further developed the phantom network model, which however just resulted in a modification of the constant leading the first invariant, see Eq. (40) in \cite{Flory76}. A more detailed treatment of the phantom network theory was introduced by \cite{Flory-Erman} adding a term accounting for the constraints, no longer proportional to the first invariant of $\boldsymbol{C}$, but still based on the Gaussian framework. Indeed this later improvement has been introduced in the 8-chain model to better capture the low stretches behavior by including $I_2$-like effects \citep[Fig. 6]{Arruda-Boyce-review}, yet at the cost of introducing two additional fitting parameters. But the original \cite{Arruda-Boyce-8chain} 8-chain model is one of the most successful models precisely because it has only two parameters which can be obtained from a single test. The model may be considered as affine because it consists of 8 chains in the diagonals of a cube with sides oriented towards the principal stretches, with the central joint point conceptually allowed to move freely, but subject to the symmetry of the cell (all chains get the same stretch). Cube vertices move  affinely with the continuum and, as previous micromechanical affine models, the strain energy only depends on the first invariant. However, deformations in the 8 chains just represent average deformations of the chains in all directions, so it may also be considered conceptually a non-affine model \citep{Arruda-Boyce-review,Microsphere} with a ``single non-affine network stretch $\lambda=\sqrt{I_1/3}$''  as \emph{micro-macro} transition \citep{Microsphere}. Indeed, this ``non-affinity'' of the \cite{Arruda-Boyce-8chain} model is highlighted as a possible reason behind the success of this model when compared to the well-known but \emph{``somewhat surprising lack of success of the full network model''} (Claim C1) \citep[p.509]{Arruda-Boyce-review}. The (affine) full network model (``micro-sphere'' or ``complete assembly of chains'') was developed by Treloar and co-workers, among others; see \cite{Treloar_book} and therein references. It is  an attractive idea for current computational power so it has been extended to soft tissues (e.g. \cite{Alastrue,Saez,Menzel,EllenKuhl}). But the performance of the affine version motivated the \cite{Microsphere} remark that \emph{``it is a well-known fact that the} [chain stretch] \emph{affinity assumption yields a model response that is not in agreement with experimental observations, in particular in the range of large deformations''} (Claim C2); see also \cite{Wu-Giessen}. This claim is in line with the commented previous findings by Treloar.

To overcome the deficiencies of the affinity assumption, more complex models have been introduced which aimed to: (1) correct the affine deformation of the surface (weight) associated to the chain during deformations (motivated on slip links or forest chains and with a practical effect that may be equivalent to include the second invariant in phenomenological models,  cf. Eq. (22) in \cite{Itskov}, Eq. (14) in \cite{Kroon}); (2) correct the affine chain stretch obtained from the continuum right Cauchy-Green deformation tensor (a similar effect as to include fluctuating junction points, inhomogeneous subnetworks or fluctuating micromechanical fields,  cf. Eqs. (14) and (17) of \cite{Itskov}, Eq. (11) in \cite{Kroon}). The first issue is addressed by tube models \citep{Edwards67,Edwards-Vilgis,Kaliske-Heinrich,Qin}. Their probability density function and energy in terms of tube diameter can be found in  \cite{Doi-Edwards}.  The second issue has been addressed with non-affine stretches \citep{Rubinstein,Microsphere,Lang}. The non-affine microsphere model \citep{Microsphere} has all these ingredients (non-affine tube and non-affine chain stretch) to capture all deformation modes at both moderate and large stretches. Other models based on the same or similar ideas have been recently proposed \citep{Kroon,Lopez-Pamies,Itskov,Verron-Gros,Xiang}, which add to classical alternatives as the slip-link (entanglements) model \citep{Ball}. Other mixed models include similar improvements as, e.g. including $I_2$ dependency in the  \cite{Gent} model by \cite{Horgan-Sacco,Horgan-Sch}.

In summary, all the improvements on the basic Mooney and Gaussian models have been developed with the purpose of better capturing the experimentally observed mechanical behavior. Conclusions respect to micromechanics assumptions have been derived from the improved predictive power under different deformation modes when such assumptions are considered versus when they are not. In this regard, there are several papers in which the predictive power of different models is assessed; the works of \cite{Steinmann} and \cite{Verron} (to which the predictions from \cite{Bergstrom} for Treloar's tests and \cite{Itskov} for \cite{Kawabata}  tests can be added) are excellent summaries, and of special interest to understand the state of the art in this respect. \cite{Steinmann} show predictive capabilities of 13 models respect to Treloar's data, with special attention when only one test is employed in the characterization and thereafter the model is employed to predict other tests.  Except for the 8-chain model and the \cite{Carroll} model (who puts emphasis again in the relevance of the second invariant to capture different test modes), predictions for other tests not employed to fit the material parameters, are deficient, including those of the affine microsphere or full network model (see C1 above). Motivated on the observation that \emph{``it is now well-established that a unique experiment is not sufficient to characterize a rubber-like material even assuming that it is elastic''} (Claim C3),  \cite{Verron} focused on the overall predictive power of 20 models when experimental data from several tests are employed simultaneously in the characterization. \cite{Verron}  used a complex fitting procedure, using  simultaneously gradient and  genetic algorithms and different test weightings when needed for achieving best fits. Furthermore, both the \cite{Treloar_exp} and \cite{Kawabata} data were employed. The \cite{Kawabata} experiments, on a similar material than that used by \cite{Treloar_exp},  are considered ``a very tough'' benchmark \citep{Itskov} because only three models (four including \cite{Itskov}) are known to capture both experimental sets with the same material parameters. In particular, the \cite{Kawabata} tests comprise all possible deformation states for an incompressible material within the tested experimental ranges.

Whereas conceptually one would expect some practical modelling benefit from the detailed micromechanical approach (i.e. less tests as in the \cite{Arruda-Boyce-8chain} model), until \cite{MMM}  no material model has been capable of fitting  the  \cite{Kawabata} experiments, or similar ones, using only one test curve.
Remarkably,
the phenomenological \cite{Ogden_model} (Fig. 3 in \cite{Verron}) and the \cite{Shariff} (Fig. 13 in \cite{Itskov}) models perform as well as  complex non-affine micromechanical models (Figs. 4, 11 and 12 in \cite{Itskov}). Also remarkably, the constrained junctions model (based on the non-affine ``free joints'' phantom network), behaved very well in its Gaussian domain (cf. Fig 5 in \cite{Verron})

Despite (and in view of) all the previous comments, we see a paradoxical duality in the micromechanical models which, as commented, has also concerned many authors in the past, motivating models like the phantom network model or the slip-link model. On one hand, chains are considered fully within the statistical entropic approach, with links free to fluctuate (maybe ``tube-constrained'' by forest chains), but junctions are assumed fixed to the continuum (the affinity assumption), somehow mixing an entropic view for chains with a more conventional view for the network structure. The practical implications relate to the use of affine chain representative weights and the use of an affine stretch from the Cauchy-Green deformation tensor. These lead to the mentioned unsatisfactory results, and to the several theories of increasing complexity to correct both aspects of the affinity assumption, both regarding a modified assigned surface (tube diameter) and a modified affine stretch of the type $\Lambda^{q/2}$, where $\Lambda$ is the squared stretch from the projection of $\boldsymbol{C}$ in the chain direction. There is experimental evidence in collagenous tissues and polymer hydrogels  that rotations may be non-affine (e.g. \cite{Wen,Gilbert,Billiar}   and therein references). However,  we are not aware of any attempt to treat the network itself in the same entropic way as the chains; i.e. to consider that chains may orientate randomly in the network (the orientations as statistical values), and that a change in their original angular distribution should be related to a nonvanishing work-conjugate generalized force of entropic origin.

Noteworthy, in the statistical theory, the chain length itself results to be not relevant in the chain behavior: only the chain stretch and the possible number of microstates (which in the unconstrained theory is related to the orientational distribution of links) play a role in the chain tension. If the network itself is allowed to fluctuate, and the chains within that network are allowed to adapt their direction to maximize entropy, there is an additional configurational entropy associated to the relative orientation of the chains. For example all chains oriented towards the same direction have much less entropy than an isotropic statistical  distribution of chains (the distribution accepted in the undeformed configuration). The free reorientation of chains seems to us natural because, in fact, the chain itself has no defined ``orientation'': the end-to-end vector $\boldsymbol{r}$ is just the links average direction ($r$ is the constraint for the links angular distribution). Then, we see the ``chain direction'' not as a fixed end-to-end vector of a particular chain, but as a dummy variable to account for the chains (or chain segments) statistically oriented towards that direction. This is conceptually similar to moving the plunger in a gas container: the  velocity vector distribution of the particles remains isotropic at equilibrium. The chain behavior would just depend on the available configurations which do not only depend on the average number of links between joints, but also on other constraints affecting the chain extension (e.g. forest chains). The change in orientational density of chains (direction representative weight)
would be tied to a respective change of the network configurational entropy and to a related generalized force (stress). We assume that this generalized force may be neglected in most cases (it vanishes or is small statistically, at least whereas the chains entropy does not drop sufficiently), resulting in a non-affine model respect to the chains orientation. This non-affinity is just related to the preservation of the isotropy (maximum entropy) in that orientational density, but chain stretches remain otherwise affine under this assumption.

 Of course this theory may (should) be seen as a questionable theory; so in this paper we \emph{do} question it. As \cite{Heinrich-Kaliske-critica} noted, ``it is not unusual  for people to reconsider issues that almost everyone else in a field
considers well established [...]. This is a useful endeavor [...] but [...] the results  must demonstrate improvements over the models to be replaced''. The proposed theory questions the widely adopted affine chain orientations distribution, and along it the widely accepted claims (C0), (C1), (C2) and (C3) above, which are ubiquitous  in the field. These are based on previous experience by many authors, and that we also felt unquestionable until the findings herein reported. Then, the first step must be to demonstrate that the assumption results in improvements over those well established theories.
In \cite{MMM} we developed a macro-micro-macro (MMM) approach to model polymers. Based on these assumptions, the approach performs a reverse-engineering of the average chain behavior from a \emph{single} macroscopic test curve (\emph{any} type of test which has a sufficient stretch range) (contradicting C3). This function embeds all effects in the longitudinal chain behavior (mean number of configurations due to links, entanglements and other constraints), but uses an affine stretch if chain orientations are considered to remain isotropic (contradicting C2). We have shown in \cite{MMM} that the chain function, reverse-engineered from a single test curve, reproduces to excellent accuracy (better than the above-mentioned models) all  the true biaxial cases from \cite{Kawabata} via angular integration , including low and large stretch ranges, and both tension axes. The method uses the WYPiWYG approach \citep{WYPIWYG,LatorreWYIPWYG,Romero,Pena,CrespoIJES} to determine the chain function just solving a linear system of equations (i.e. no classical material parameters or nonlinear optimization procedures: this is a quasi-automatic procedure). \ Moreover, the reverse-engineered chain function is essentially the same using any test curve on the material to calibrate the model; see discussion in \cite{MMM}, where we also report a comparison of predictions with the chain function obtained  from Treloar's tests. We also show therein that the model performs equally well in two types of silicones tested with a similar series of true biaxial tests by \cite{Kawamura}. Hence, in practice, this very simple approach seems to outperform the most sophisticated models.

Then, the question we address in this paper is whether the predictive power of the MMM approach is just due to numerically reverse-engineer the chain function (so the performance could be attributed  only to a better, phenomenological fit) or to the previous non-affine orientational assumption (more important, but which practical implementation is even simpler than that of the affine assumption because, e.g. it preserves the optimality of the integration rules in the sphere). Indeed, the MMM model is similar to a full network model, known to fail to provide good results employing an affine angular distribution of chains, but performing in an excellent manner (contradicting C1) if an unrestricted chains orientation distribution is assumed.

The rest of the paper is organized as follows. First, in Sec. {\ref{SEC2} } we recall basic facts as the Gibbs form of the entropy understood as a function of probability distributions, and the observation that if the system is unconstrained, entropy is maximized for the isotropic distribution. We then address the additivity of the entropy of a system made of subsystems (i.e. distributions of chains made themselves of distributions of links). We recall the orientational distribution of links in a chain to emphasize that the deviation from the statistical isotropic distribution (the chain constraint) is responsible for the tension in the chain. In consequence, we show that a non--isotropic distribution of chains in the network would result in an additional entropy (if we preserve the statistical view) with an associated additional stress. These are neglected in the affine deformations approach. To address the practical consequence of the different approaches, in Sec. \ref{SEC3} we show that whereas the homogenization of the affine deformation uses the micro-macro chain stretch connection from $\boldsymbol{C}$, the unconstrained chain orientational approach is accounted for just by employing a micro-macro chain stretch connection from $\boldsymbol{U}=\sqrt{\boldsymbol{C}}$. Since one may think that differences between both approaches should be modest (i.e. just using a chain stretch obtained from $\boldsymbol{C}$ or from $\boldsymbol{U}$ which are equal in principal directions), in Sec. \ref{toy example} we address those differences using the same Neo-Hookean model but based on the different micro-stretches. Here we show that the non-affine Neo-Hookean-like model predicts the previously unexplained slope in Mooney plots related to the $I_2$ invariant (contradicting claim C0). Finally, in view of these differences, we asses in Sec. \ref{SEC:5} the practical impact of considering both the affine chains orientation assumption and the unconstrained isotropic one to capture multiaxial behavior of real polymers. Several additional simulations are performed to give supporting evidence on the assumption. Whereas such evidence may be considered as nonconclusive because it is based on a specific full network approach, it may be one of the reasons why, despite of the above-mentioned efforts, no model has been capable of predicting the behavior of polymers using a single test curve to determine the material parameters; furthermore, they necessitated of non-afffine micro-stretch and micro-area corrections to predict that behavior even when using several tests simultaneously to obtain the material parameters.

We recommend to the reader interested mostly in the practical implications of this work to start with Secs. \ref{toy example} and \ref{SEC:5} (which can be read independently), then proceed to Sec.  \ref{SEC3}, and if interested in the theoretical motivation to address Sec. {\ref{SEC2}}.

\section{Entropy of chains and networks\label{SEC2}}

\subsection{Boltzmann entropy\label{Sec:Boltzmann entropy}}

The
entropy of the system may be understood employing Boltzmann's theory following
the principles of statistical mechanics. According to Boltzmann, the entropy
$S$ of a system may be computed from the quantity $W$ of possible equivalent
system states from microstates corresponding to that system macrostate%

\begin{equation}
S=k_{B}\ln W \label{Boltmannentropy}%
\end{equation}
where $k_{B}$ is Boltzmann's constant. The entropy may be considered as a measure of internal equilibrium in the system, or as a measure of homogeneity within the system. In general, if we have a system made of
$N$ components, and $i=1,...,n$ different microstates (slots), each microstate with
$N_{i}$ components such that $N=\sum_{i=1}^{n}N_i$, from the theory of combinatory, the number of possible
equivalent states, is\footnote{The actual original expression, that preserved additivity of the entropy in mixtures, does not contain $N!$,  see \cite{MullerWeiss_book}, Sec. 16.2, p. 171, but we keep the best known and intuitive expression for the discussion.\label{fn1}}%

\begin{equation}
W=\binom{N}{N_{1}}\binom{N-N_{1}}{N_{2}}...\binom{N-\sum_{i=1}^{n-1}N_{1}%
}{N_{n}}=\dfrac{N!}{%
{\textstyle\prod\limits_{i=1}^{n}}
N_{i}!} \label{Wdef}%
\end{equation}
where the number of possible different states is $n^{N}$. Hence, from Eq.\eqref{Boltmannentropy}, the entropy is%
\begin{equation}
S=k_{B}\ln N!-k_{B}\sum_{i=1}^{n}\ln N_{i}!
\label{Eq3}
\end{equation}

For $N_{i}$ sufficiently large, we
can think of a continuous distribution so we can apply the Stirling
approximation $\ln N!\approx N\ln N-N$. Using this approximation in Eq.
(\ref{Eq3}) and considering the case of interest below such that
$N=\sum_{i=1}^{n}N_{i}$, we have the Gibbs form of the entropy
\begin{equation}
S=k_{B}\left[  N\ln N-\sum_{i=1}^{n}N_{i}\ln N_{i}\right]  =-k_{B}N\sum
_{i=1}^{n}\rho_i\ln\rho_i \label{Seq1}%
\end{equation}
where $\rho_{i}=N_{i}/N$ is the \emph{relative} density in the microstate (discrete distribution).
Then taking $C=k_{B}N$, we can also write $S=-C\sum_{i=1}^{n}\rho_{i}\ln
\rho_{i}$, and if the number of microstates $n$ is large, we can take a distribution %
\begin{equation}
S=-C\int_{\Theta}\rho\left(  \theta\right)  \ln\rho\left(  \theta\right)
d\theta\geq0 \label{Seq2}%
\end{equation}
where $\Theta$ is the continuous domain of microstates and $\rho(\theta)$ is the continuous distribution. This last formula is
connected to Shannon's entropy of information (in that case typically  with a logarithm in base 2).

\subsection{Unconstrained maximum entropy
distribution\label{Sec: maximum entropy distribution}}

Assume that $\theta\in\Theta\equiv\left(  \theta_{1},\theta_{2}\right)  $. The
principle of maximum entropy states that at equilibrium
\begin{equation}
\rho\left(  \theta\right)  \text{ such that }S(\rho(\theta))=-C\int_{\theta_{1}}^{\theta
_{2}}\rho\left(  \theta\right)  \ln\rho\left(  \theta\right)  d\theta
\rightarrow\max
\end{equation}
meaning that the most probable $\rho\left(  \theta\right)  $ is such that a
variation $\delta\rho\left(  \theta\right)  $ over it implies that there is no
variation in the entropy, i.e. $\delta S\left(  \delta\rho\left(
\theta\right)  \right)  |_{\rho\left(  \theta\right)  }=0$ for all $\delta
\rho\left(  \theta\right)  $. Computing the first variation of the term
$\ln\left(  \rho\text{$\left(  \theta\right)  $}\right)  $$\rho\left(
\theta\right)  $ we obtain $\delta\left\{  \ln\text{$\left(  \rho
\text{$\left(  \theta\right)  $}\right)  $$\rho\left(  \theta\right)  $%
}\right\}  =\left[  1+\ln\left(  \rho\left(  \theta\right)  \right)  \right]
\delta\rho\left(  \theta\right)  $ and the variation of the entropy functional is%

\begin{equation}
\delta S\left(  \delta\left(  \theta\right)  \right)  =-C\int_{\theta_{1}%
}^{\theta_{2}}\left[  1+\ln\left(  \rho\left(  \theta\right)  \right)
\right]  \delta\rho\left(  \theta\right)  d\theta=0\label{Eq7}
\end{equation}
From the definition of $\rho\left(  \theta\right)  $ (as the continuous
version of $N_{i}/N$, a probability density), we must have
\begin{equation}
\int_{\theta_{1}}^{\theta_{2}}\rho\left(  \theta\right)  d\theta
=1\;\Rightarrow\;\delta\int_{\theta_{1}}^{\theta_{2}}\rho\left(
\theta\right)  d\theta=\int_{\theta_{1}}^{\theta_{2}}\delta\rho\left(
\theta\right)  d\theta=0 \label{Eq8}
\end{equation}
However, the variation $\delta\rho\left(  \theta\right)  $ is arbitrary (e.g.
may vanish in an arbitrary part of the domain), so in general Eqs. \eqref{Eq7} and \eqref{Eq8} must be fulfilled simultaneously in any arbitrary domain $\Theta^{\ast}$. Hence
$1+\ln\left(  \rho\left(  \theta\right)  \right)  =\alpha\text{, constant for
all }\theta$,
or equivalently  $\rho\left(  \theta\right)  $ is the constant distribution, a well-known result,
whose value must be $\rho\left(  \theta\right)  ={1}/{(\theta_{2}-\theta_{1})}$ to comply with Eq. \eqref{Eq8}.

In summary, the principle of maximum entropy gives a uniform
(\textquotedblleft isotropic\textquotedblright) distribution as the most probable distribution. Note that a similar conclusion is obtained from Eq. (\ref{Seq1}) if we
consider $N_{i}$ a continuous variable respect to which a derivative may be
made%
\begin{equation}
\frac{\partial S}{\partial N_{i}}=-k_{B}\left(  \ln\frac{N_{i}}{N}+1\right)
=0\;\Rightarrow\;\rho_{i}=\frac{N_{i}}{N}=\tilde{C}\exp\left(  -1\right)
=\gamma\end{equation}
so all $N_{i}$ and all $\rho_{i}$ are equal and $\tilde{C,}\gamma$ are constants
obtained from the requirement that $\sum_{i}\rho_{i}=1$ (which may also
be introduced as a constraint). Note that if all $N_i$ are in the same slot, the entropy is minimal (zero), and the maximum entropy, obtained for the isotropic distribution, is proportional to $N$.
\subsection{Configurational entropy at two scales\label{SEC:2-level entropy}}
In polymers we deal with long molecules modeled as chains, and we compute the entropy of the chains according to their possible random link configurations. But chains themselves form a network, and they may also orientate randomly in that network. If two subsystems $(1)$ and $(2)$, each with its own entropy are combined, but no configurational entropy of the combination is accounted for, the entropies are just summed up because each subsystem just adds slots (microstates) for their own components, totaling $n_1+n_2$ slots. Because we assumed subsystems do not mix, the total combinations of the system are $N^{(1)}!N^{(2)}!$, i.e.

\begin{equation}
S=k_B\ln\frac{N^{(1)}!N^{(2)}!}{\prod\limits_{i=1}^{n_1}N_i^{(1)}!\prod\limits_{i=1}^{n_2}N_i^{(2)}!}
=k_B\ln\frac{N^{(1)}!}{\prod\limits_{i=1}^{n_1}N_i^{(1)}!}
+\ln\frac{N^{(2)}!}{\prod\limits_{i=1}^{n_2}N_i^{(2)}!}=S^{(1)}+S^{(2)}
\end{equation}
Hence, in the literature the entropy of ensembles of chains is typically the sum of the entropies of the individual chains. However, when the subsystems themselves may also combine as systems, we are in practice adding a third ``system'', which in this example consists of two elements, namely $(1)$ and $(2)$, so say $N^{(3)}=2$. These subsystems may be placed in $n_3$ slots, so we have
that the entropy of this new $(3)$ system is $
S^{(3)}=k_B\ln({N^{(3)}!}/{\prod_{i=1}^{n_3}N_i^{(3)}!})$.

In general, the total entropy of the ensemble consisting of $Z$ subsystems, and one system labelled ``$Z+1$''  combining (but do no mixing) those subsystems is
\begin{equation}
S=k_B\ln\frac{\prod\limits_{z=1}^{Z+1}{N^{(z)}!}}{\prod\limits_{z=1}^{Z+1}\prod\limits_{i=1}^{n_z}{N_i^{(z)}!}}
=\sum_{z=1}^{Z+1}k_B\ln\frac{N^{(z)}!}{\prod\limits_{i=1}^{n_z}N_i^{(z)}!}=S^{(Z+1)}+\sum_{z=1}^{Z}S^{(z)}
\end{equation}
where $S^{(Z+1)}$ is the additional configurational entropy of the combination of immiscible subsystems. This entropy may be considered if chains orient statistically in the network. Of course, if $Z$ is large, $S^{(Z+1)}$ may be neglected as long as $N^{(Z+1)}$ is in the same order as $N^{(z)}$, unless the entropies of the subsystems drop sufficiently so that of the system becomes dominant.
\subsection{The entropy of a single chain}

This statistical approach is typically employed to study the behavior of rubber-like
materials considering that a rubber molecule can be understood as a chain
compound of $N$ independent random links, and the material as a cross-linked network of such chains. The changes in the internal energy
$U$  are neglected (cf. Fig. 8a in \cite{Anthony}, Fig. 2.8 in \cite{Treloar_book} or Fig. 6.3 of \cite{Argon}) when compared to the possible changes in the
entropic energy $-TS$, where $T$ is the absolute temperature; i.e. the power in isothermal cases (as considered along this work) is $
{\dot\Psi}=-T\dot{S}$. Then, the forces in the system $\mathfrak{f}$ are due to changes in its
entropy $\mathfrak{f}\left(  \bullet\right)  =-T\partial S/\partial\left(
\bullet\right)  $, where $\left(  \bullet\right)  $ are the set of selected
kinematic variables describing the change of entropy and $\mathfrak{f}\left(
\bullet\right)  $ are their work-conjugate generalized forces.
 The treatment of the chain entropy and related tension is well known, but it is presented here in terms of orientation of links to emphasize the trinomial ``kinematic constraint $\Leftrightarrow$ biased-orientational-distribution $\Leftrightarrow$ generalized force'',  making a parallelism with that of the network itself as a system of chains.

The possible microstates for a link are its possible orientations in space. To
make the reasoning simpler a discretization in the space of possible
directions can be made. That way, the polar $\theta$, and the azimuthal $\varphi$
axis, which are continuous, are discretized with the values $\theta=\theta
_{1},\theta_{2},\ldots,\theta_{p}$ and $\varphi=\varphi_{1},\varphi_{2}%
,\ldots,\varphi_{q}$. With this approach, the space is represented by a grid
with $p\times q$ possible combinations of the values $\left(  \theta
_{i},\varphi_{j}\right)  $. Let $N_{\theta_{1}\varphi_{2}}$ be the number
of chains that share the spatial orientation defined by the polar angle
$\theta_{1}$ and the azimuthal angle $\varphi_{2}$.  Boltzmann's entropy of the chain is%
\begin{equation}
S_{ch}=k_{B}\ln W\text{ \ with \ }W=\dfrac{N!}{\prod\limits_{i=1}^{p}%
\prod\limits_{j=1}^{q}N_{\theta_{i}\varphi_{j}}!}%
\end{equation}
Following the steps in Sec. \ref{Sec:Boltzmann entropy}, see Eq. (\ref{Seq1}),
and defining the density $\rho\left(  \theta_{i},\varphi_{j}\right)
=N_{\theta_{i}\varphi_{j}}/N$, we get%
\begin{equation}
S_{ch}=k_{B}\ln W=-k_{B}\sum_{i,j=1}^{p,q}N_{\theta_{i}\varphi_{i}}\ln\left(
\frac{N_{\theta_{i}\varphi_{i}}}{N}\right)  =-k_{B}N\sum_{i,j=1}^{p,q}%
\rho\left(  \theta_{i},\varphi_{j}\right)  \ln\rho\left(  \theta_{i}%
,\varphi_{j}\right)
\end{equation}
It is worth to note that if we split the chain into two chains, see Fig.
\ref{chain_into_2.eps}, with equivalent distributions ($N^{\left(  1\right)
}=\alpha N$ and $N^{\left(  2\right)  }=\left(  1-\alpha\right)  N$ means
$N_{\theta_{i}\varphi_{i}}^{\left(  1\right)  }=\alpha N_{\theta_{i}%
\varphi_{i}}$ and $N_{\theta_{i}\varphi_{i}}^{\left(  2\right)  }=\left(
1-\alpha\right)  N_{\theta_{i}\varphi_{i}}$), we have that the entropy of the
system made from both chains is the same as that of the single chain; i.e. only the
orientational distribution densities $\rho\left(  \theta_{i},\varphi_{j}\right)$ are relevant.
  For continuous distributions---see Eq. (\ref{Seq2})%
\begin{equation}
S_{ch}=-k_{B}N\int_{\Omega}\rho_{links}\left(  \theta,\varphi\right)  \ln
\rho_{links}\left(  \theta,\varphi\right)  d\Omega
\end{equation}
where $d\Omega=\sin\theta d\theta d\varphi$ is the solid angle of the unit
sphere (i.e. really the distribution must be taken per unit surface, not per angle aperture, so all slots are equal). The
\emph{unconstrained} distribution that complies with the maximum entropy is
the constant distribution, see Sec. \ref{Sec: maximum entropy distribution},
i.e. $\rho_{links}\left(  \theta,\varphi\right)  =1/\Omega=1/(4\pi)$, which means that a given link has the same probability to be oriented at any
given angle.

We now introduce a kinematic variable describing the macrostate and the constrained  entropy
of the chain. This variable is the end-to-end vector distance $\boldsymbol{r}%
=r\boldsymbol{\hat{r}}$, with modulus $r$ and direction $\boldsymbol{\hat{r}}$. Then,
we require that the projection of all links in the direction of $\boldsymbol{r}$
sum $r$. To maximize the entropy under this
constraint, we generate the Lagrangean%
\begin{equation}
L=-k_{B}\sum_{i,j=1}^{p,q}N_{\theta_{i}\varphi_{j}}\ln\left(
\dfrac{N_{\theta_{i}\varphi_{j}}}{N}\right)+f_{T}\left(  r-\sum_{i,j=1}^{p,q}N_{\theta_{i}\varphi_{j}}l\cos\theta
_{i}\right)
\end{equation}
where $l$ is the 1D length of the link, $\theta$ is now the angle between
the link and $\boldsymbol{\hat{r}}$ and  $f_T$ is the Lagrange multiplier (the  force associated to the constraint). Then, from ${\partial L}/{\partial N_{\theta_{i}\varphi_{j}}}=0 $, normalizing  $\sum\rho_{links}\left(
\theta_{i},\varphi_{j}\right)  =1$,  we get
--- see Fig \ref{single_chain_distribution.eps}:
\begin{figure}
[ptb]
\begin{center}
\includegraphics[
width = 0.55\textwidth
]%
{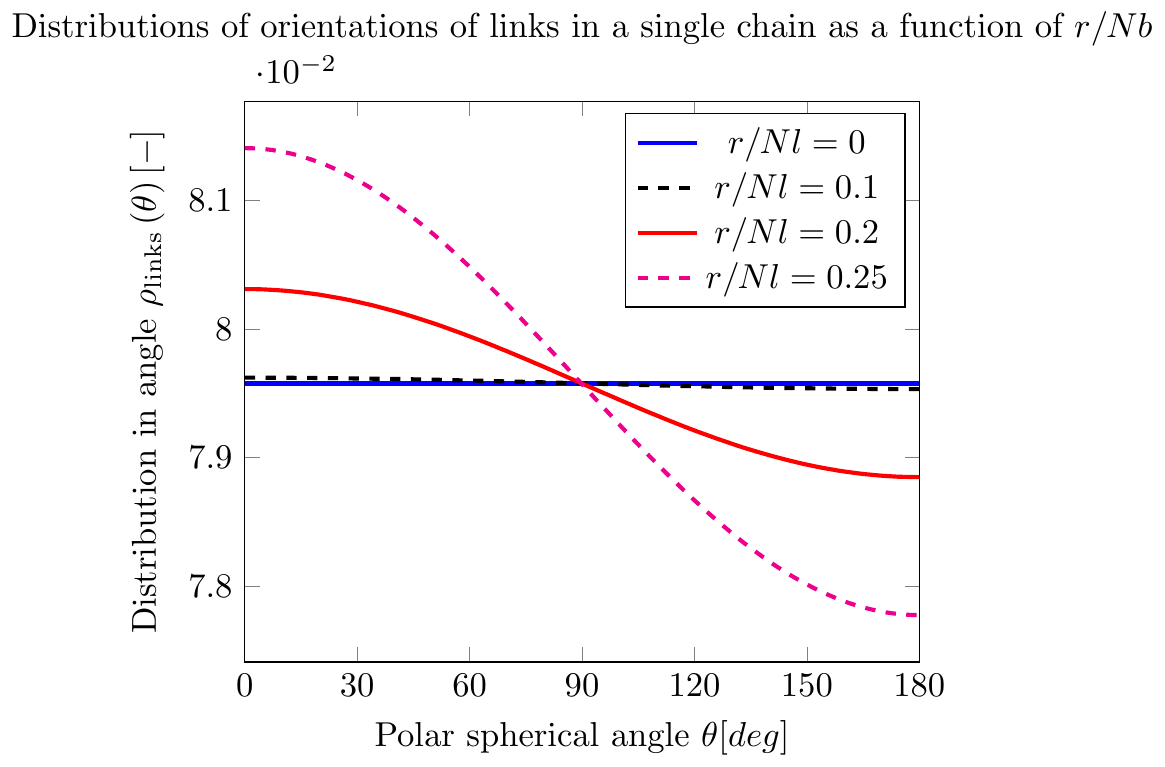}%
\caption{Distribution of the orientation of the links in a single chain as a
function of the fractional extension $r/Nl$. The distribution is isotropic in the angle  $\varphi$ around the chain axis. In average, the links are oriented in direction $\boldsymbol{r}$.}%
\label{single_chain_distribution.eps}%
\end{center}
\end{figure}
\begin{equation}
\rho_{links}\left(  \theta_{i},\varphi_{j}\right)  =\dfrac{N_{\theta_{i}\varphi_{j}}}{N}=\frac{\exp\left(
-\beta\cos\theta_{i}\right)  }{\sum\limits_{i,j=1}^{p,q}\exp\left(  -\beta
\cos\theta_{i}\right)  }%
\end{equation}
with $\beta=f_T l/k_{B}$. Considering $N_{\theta_{i}\varphi_{j}}=\bar{C}%
N\exp\left(  -\beta\cos\theta_{i}\right)  $, with  $\bar{C}=1/\sum\nolimits_{i,j=1}^{p,q}\exp\left(  -\beta
\cos\theta_{i}\right)  $, the remaining equation is%
\begin{equation}
r=\sum_{i,j=1}^{p,q}N_{\theta_{i}\varphi_{j}}l\cos\theta_{i}=N\bar{C}%
\sum_{i,j=1}^{p,q}\exp\left(  -\beta\cos\theta_{i}\right)  l\cos\theta_{i}%
\end{equation}
This equation can be written %
\begin{equation}
\frac{r}{Nl}=-\frac{d}{d\beta}\left[  \ln\sum_{i,j=1}^{p,q}\exp\left(
-\beta\cos\theta_{i}\right)  \right]\label{EQ:rNl}
\end{equation}
as it is immediate to check performing the derivative. Assuming $N$ large, we
can take a continuous distribution in the solid angle (to account for equal slots)%
\begin{equation}
\sum_{i,j=1}^{p,q}\exp\left(  -\beta\cos\theta_{i}\right)  \Rightarrow\int
_{0}^{2\pi}\int_{0}^{\pi}\exp\left(  -\beta\cos\theta\right)  \sin\theta
d\theta d\varphi=4\pi\frac{\sinh\beta}{\beta}%
\end{equation}
and using Eq. \eqref{EQ:rNl}%
\begin{equation}
\frac{r}{Nl}=-\frac{d}{d\beta}\ln\left(  4\pi\frac{\sinh\beta}{\beta}\right)
=\frac{1}{\beta}-\coth\beta=-\mathcal{L}\left(  \beta\right)
\end{equation}
where $\mathcal{L}\left(  \bullet\right)  $ is the well-known Langevin
function, so the tension $f$ in the chain $f=-TdS/dr=-Tf_{T}$ takes the well-known expression%
\begin{equation}
\beta=-\mathcal{L}^{-1}\left(  \frac{r}{Nl}\right)  \Rightarrow f:=-f_{T}%
T=-\frac{\beta k_{B}}{l}T=\frac{k_{B}T}{l}\mathcal{L}^{-1}\left(  \frac{r}%
{Nl}\right)  %
\end{equation}
A more accurate description  follows the Rayleigh distribution; see \cite{Itskov} and Sec. 6.5 of \cite{Treloar_book}.  From a
reference configuration with $r=~^{0}r$, the tension is obtained with
$r=\lambda_{ch}~^{0}r$, where $\lambda_{ch}$ is the stretch in the chain
respect to the reference configuration. As it is well known, in the Gaussian statistical theory (valid for small fractional extensions typical in unloaded configurations), the unloaded mean chain length is $\,^0r=\sqrt{ N}l$; see e.g. Eq. (3.11) in \cite{Treloar_book}.

Consider again the partition of the chain in aligned pieces (statistically equivalent segments, meaning equal distribution of links), each with a respective constraint $r^{(k)}$ such that $r=\sum r^{(k)}$
 and proportional lengths $Nl=\sum N^{(k)}l$. Following the same steps, we get that each chain partition has a tension
$f^{\left(  k\right)  }=\left(  k_{B}T/l\right)  \mathcal{L}^{-1}(
\lambda^{(k)}_{ch}/\sqrt{N^{(k)}} )$. The total chain length itself $Nl$ (or $N^{(k)}l$) does not  play a relevant role: the link length $l$ only affects the leading constant just as $k_B$, $T$,  but $N^{(k)}$ sets the unstretched location in the inverse Langevin function through $1/\sqrt{N^{(k)}}$ (accounting for the relative number of available configurations, or to tell how far $\lambda_{ch}$ is from zero entropy). Densely cross-linked or entangled polymers will have a smaller equivalent $N$ (for the representative chain) than loose ones, meaning less possible random configurations for the chain and the whole network. Of course the orientation for each
force is  along the respective mean chain orientation, that of $\boldsymbol{r}^{\left(
k\right)  }$. Interestingly, if we consider the middle point moving in a direction perpendicular to $\boldsymbol{r}$, there would be no
increase in the whole chain $r$ (see Fig.\ref{chain_into_2.eps}), but each of the
parts would increase their tension. This paradox, is  typically solved requiring that chain ends can only be joints of several chains in
the network, so internal links in chains can move freely (and very loosely because $\sqrt{N}<<N$) in a non-affine
manner, but chain joints are forced to move in an affine manner,
\textquotedblleft tied\textquotedblright\ to the continuum. This results in a
complete different treatment for chains themselves respect to the network as
an entity: free fluctuating internal links in a fixed, nonfluctuating network configuration. However, this paradox may also be resolved, as seen below, when the
tensions from all the chains are integrated in space allowing the network to
fluctuate and considering a chain length mainly as a representation of the possible configurations of the average chain (or chain parts) in the network oriented in such direction. Moreover, the number of links $N$ would be just  related to the number of available configurations of the representative single chain, covering the additional possible configurations of the network as a whole, or the decrease of them due to interaction with forest chains.
\begin{figure}
[ptb]
\begin{center}
\includegraphics[width=0.25\textwidth]%
{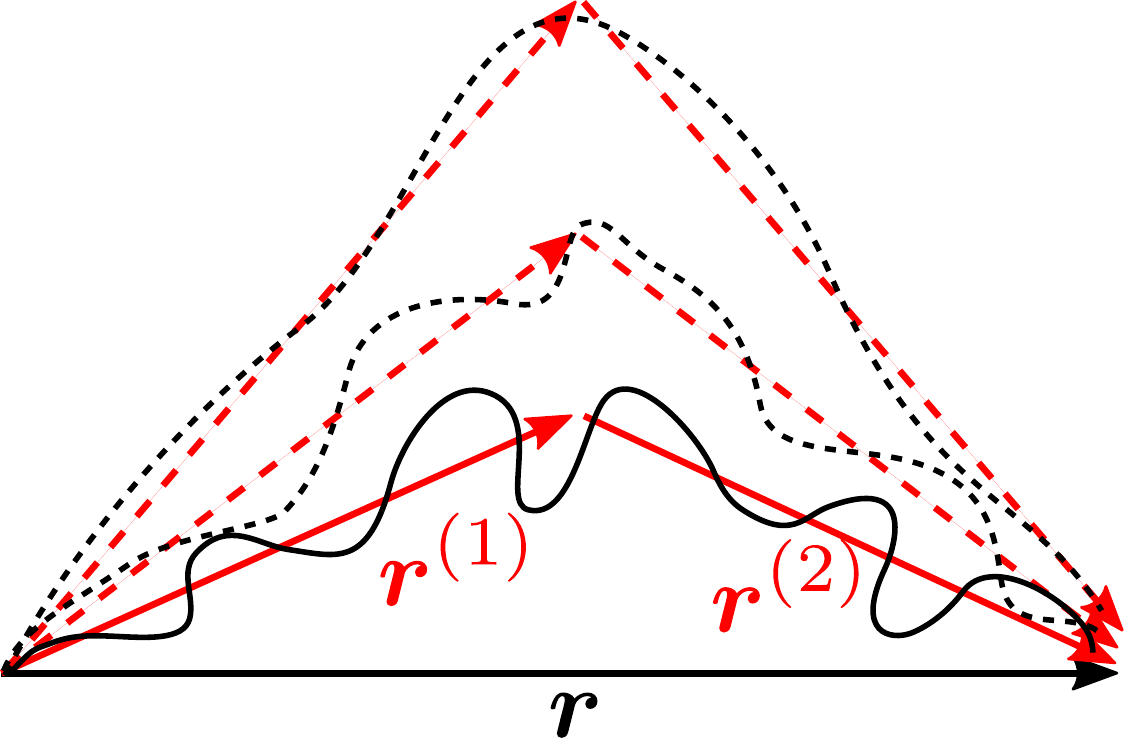}%
\caption{Considering a single chain as two different linked chains. Relevance
of the determined end-to-end vector $\boldsymbol{r}$. Note that each $\boldsymbol{r}$ has the mean orientation of the links of the respective chain fractions.}%
\label{chain_into_2.eps}%
\end{center}
\end{figure}

\subsection{The additional configurational entropy of the network \label{Sec:networkentropy}}%

\begin{figure}[htpb!]
\begin{minipage}{\textwidth}
\centering
\includegraphics[width=.41\textwidth
]%
{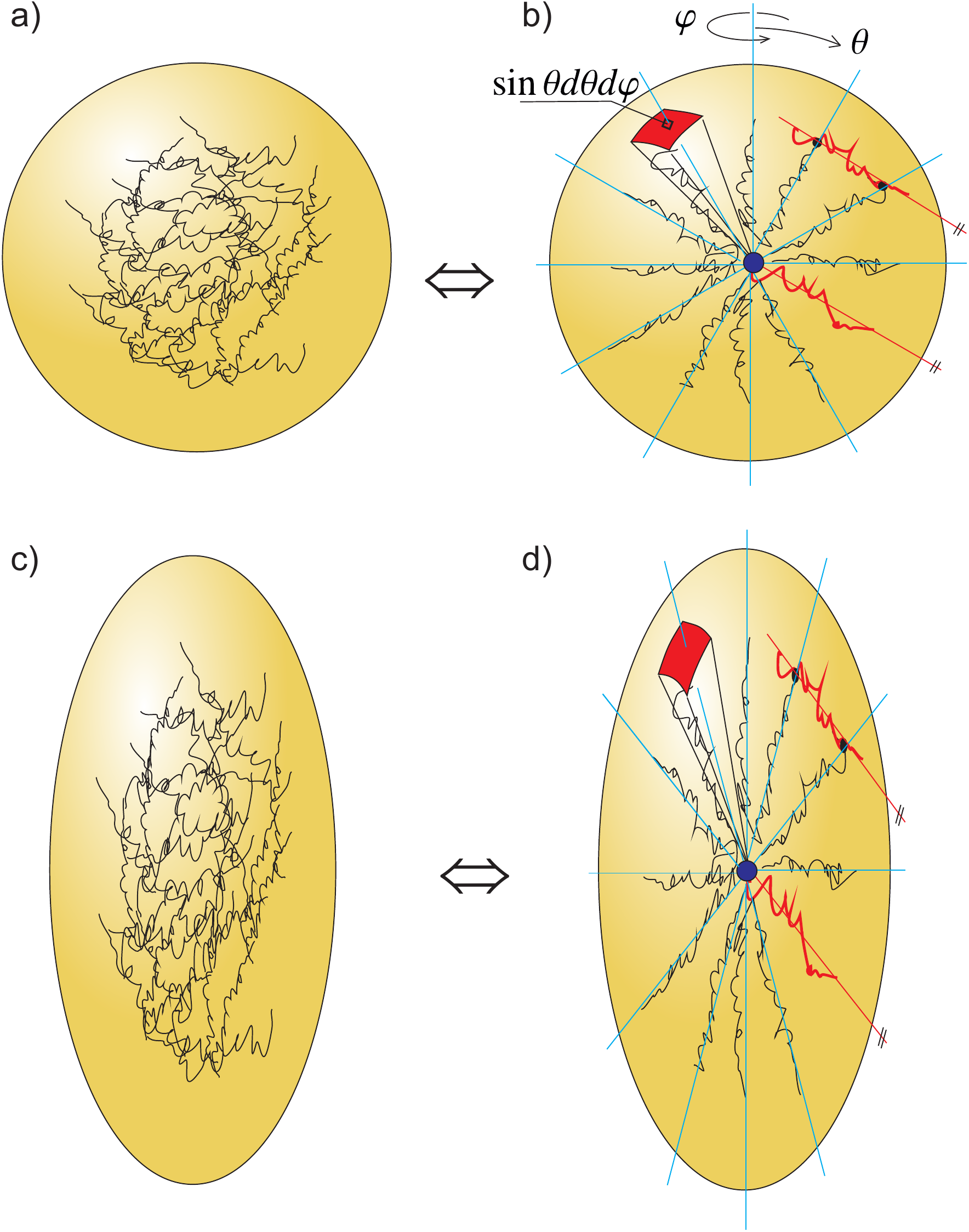}\quad
\includegraphics[width=.42\textwidth
]{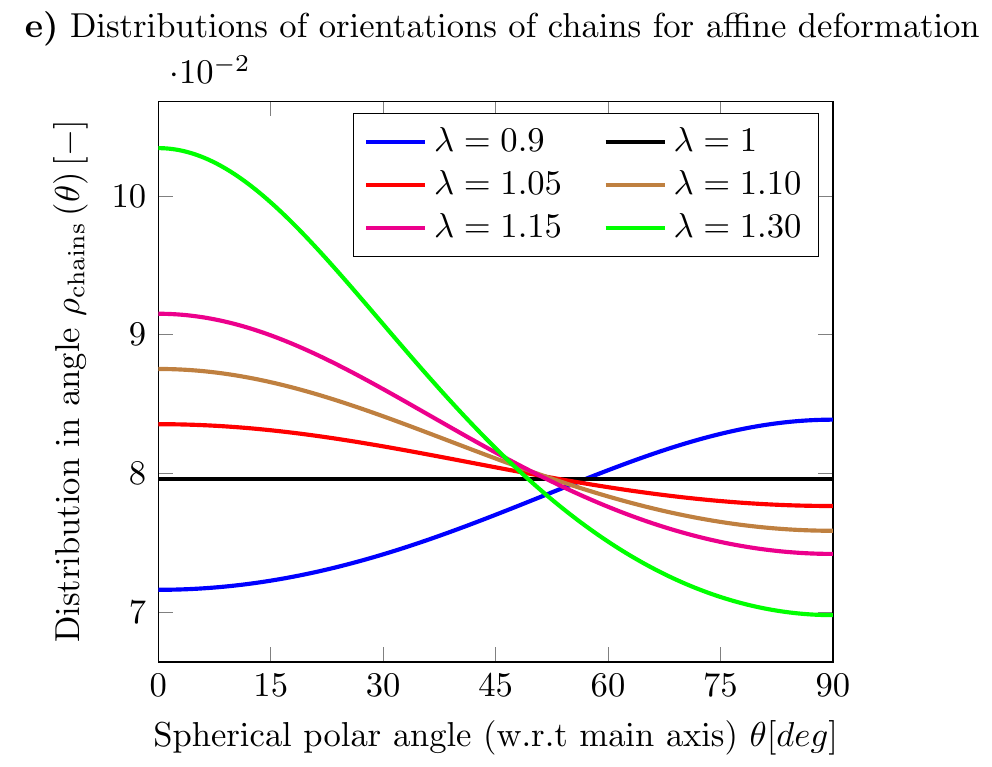}%

\caption{a) Network of chains around a point in the continuum. b) Equivalent
isotropic distribution of chains with one end in the continuum point and solid
angle $d\Omega=\sin\theta d\theta d\varphi$. c) Deformed continuum around a
continuum point, and deformation of the network assuming affine deformations.
d) Deformation of the chains and their new angular distribution assuming
affine deformations in the network. e) Distribution density of the chains $\rho_\text{chains}\left(
\theta\right)  $  for affine deformations, according to the orientation in the space. Note that for
$\lambda=1$ the uniform distribution is recovered. The affine deformation
assumption changes the spatial distribution of the chains according to the
affine deformation constraint.}%
\label{chains_in_spherev0}%
\end{minipage}
\end{figure}

The steps from the previous section are well known, and result in a tension in the chain from the non-isotropic orientation of the chain links produced by the kinematic constraint in the chain $r$. In Fig. \ref{chains_in_spherev0}a we show a sketch of a chain network around
a given point. Obviously the arrangement is random in the unloading
configuration so it is assumed to be isotropic, meaning that there is no
preferred orientation in the chains (or more specifically no preferred orientation in the end-to-end or
joint-to-joint vector $\boldsymbol{r}$), so they are equally distributed in the solid angle $d\Omega$; see Fig. \ref{chains_in_spherev0}b. Note
that there is no need to explicitly take into account chains attaching other
chains, because all of them are already considered in the isotropic
configuration (e.g. red chain in Fig. \ref{chains_in_spherev0}b).

During the deformation in a solid, the assumption of
spatially isotropic distribution of chains may be considered  not valid any more. Affine network deformations assume so, where at least some chain
ends of the network are  \textquotedblleft tied\textquotedblright\ to the ``solid''
(which is assumed to be the network itself) instead of maintaining the average isotropic distribution; see Figs.
\ref{chains_in_spherev0}c and \ref{chains_in_spherev0}d. This means that there
is a constraint in the spatial orientation of the chains in the network given by the macroscopic stretches. As we
did with $r$ for constraining the links orientation in the single chain, we
introduce the macromechanical kinetic variable describing a chains orientation constraint, say $\lambda$. For simplicity in the
discussion, we will assume only one degree of freedom, one constraint $\lambda$  typical from a tensile test along $z$ (in our assumption of an isotropic distribution, this force vanishes). This distribution changes the orientation of the chains
following the pattern of the affine deformations performed along $z$, i.e. a
chain of coordinates (now angles refer to the macroscopic tensile axis $z$)%
\begin{equation}
\boldsymbol{r}_{0}=r_{0}\boldsymbol{\hat{r}}_{0}=r_{0}\left[
\begin{array}
[c]{c}%
\cos\varphi\sin\theta\\
\sin\varphi\sin\theta\\
\cos\theta
\end{array}
\right]
\end{equation}
will change as $\boldsymbol{r}=\boldsymbol{U}_{conf}\boldsymbol{r}_{0}$,
and the differential of surface $d\boldsymbol{s}_{0}=ds_{0}\boldsymbol{\hat{r}}%
_{0}=d\Omega\boldsymbol{\hat{r}}_{0}$, according to Nanson's formula,  as $\boldsymbol{U}_{conf}^{-1}d\boldsymbol{s}_0$ where $\boldsymbol{U}_{conf}=diag(1/\sqrt{\lambda},1/\sqrt{\lambda},\lambda)$.
Under the affinity assumption $\boldsymbol{U}=\boldsymbol{U}_{conf}$. We normalize the density distributions in both the undeformed $\Omega$ and the deformed $\Theta$
configurations as%
\begin{equation}
1=\int_{\Omega}\rho_{\Omega}\left(  \theta,\varphi\right)  d\Omega
=\int_{\Theta}\rho_{\Theta}\left(  \theta,\varphi\right)  d\Theta
\end{equation}
If we transform slot-to-slot following the affine relation, since
$\rho_{\Omega}\left(  \theta,\varphi\right)  $ $=1/\Omega$
\begin{equation}
\rho_{\Theta}\left(  \theta,\varphi\right)  =\rho_{\Omega}\left(
\theta,\varphi\right)  \frac{d\Omega}{d\Theta}=\frac{1}{\Omega} \frac{d\Omega}{d\Theta}%
\end{equation}
which means that an increase of surface results in a decrease of chain density towards that orientation (they become more far apart). In general, the surface transformation, as a result of Nanson's formula, is%

\begin{equation}
\frac{d\Theta}{d\Omega}=\frac{1}{2}\sqrt{%
\begin{array}
[c]{c}%
\dfrac{2}{\lambda_{3}^{2}}+\dfrac{1}{\lambda_{2}^{2}}+\dfrac{1}{\lambda
_{1}^{2}}+\left(  \dfrac{2}{\lambda_{3}^{2}}-\dfrac{1}{\lambda_{2}^{2}}%
-\dfrac{1}{\lambda_{1}^{2}}\right)  \cos2\theta-\left(  \dfrac{1}{\lambda
_{1}^{2}}-\dfrac{1}{\lambda_{2}^{2}}\right)  \cos2\varphi+...\\
...+\left(  \dfrac{1}{2\lambda_{1}^{2}}-\dfrac{1}{2\lambda_{2}^{2}}\right)
\cos\left(  2\theta-2\varphi\right)  +\left(  \dfrac{1}{2\lambda_{1}^{2}%
}-\dfrac{1}{2\lambda_{2}^{2}}\right)  \cos\left(  2\theta+2\varphi\right)
\end{array}
}%
\end{equation}
For a prolate ellipsoid, typical of a
tensile test under consideration, with $\lambda_{1}=\lambda_{2}<\lambda_{3}$%
\begin{equation}
\frac{d\Theta}{d\Omega}=\frac{1}{\sqrt{2}}\sqrt{\dfrac{1}{\lambda_{3}^{2}%
}+\dfrac{1}{\lambda_{1}^{2}}+\left(  \dfrac{1}{\lambda_{3}^{2}}-\dfrac
{1}{\lambda_{1}^{2}}\right)  \cos2\theta}%
\end{equation}
so taking $\lambda_{3}=\lambda=1/\lambda_{1}^{2}=1/\lambda_{2}^{2}$%

\begin{equation}
\rho_{\mathrm{chains}}\left( \lambda, \theta\right) \equiv\rho_{\Theta}\left(  \lambda,\theta\right)  =\frac{\sqrt{2}\lambda}{4\pi
\sqrt{\left(  \lambda^{3}+1\right)  +\left(  1-\lambda^{3}\right)  \cos
2\theta}}
\end{equation}

$\allowbreak$ This distribution is plotted in Fig. \ref{chains_in_spherev0}e for different values of $\lambda$. Note that the affinity assumption produces a strongly anisotropic orientational distribution. The entropy due to the
reorientation of chains is%
\begin{equation}
S_{conf}=-C_{c}\int_{0}^{2\pi}\int_{0}^{\pi}\rho_{chains}\left(
\lambda,\theta\right)  \ln\rho_{chains}\left(  \lambda,\theta\right)
\sin\theta d\theta d\varphi
\end{equation}
For having an intuitive view, we define an equivalent configurational stress tensor as%
\begin{equation}
\boldsymbol{P}_{conf}=-T\frac{dS_{conf}}{d\boldsymbol{U}_{conf}}=-T\sum_{i=1}%
^{3}\frac{dS_{conf}}{d\lambda}\left(  \frac{d\lambda_{i}}{d\lambda}\right)
^{-1}\boldsymbol{\hat{n}}_{i}\otimes\boldsymbol{\hat{n}}_{i}%
\end{equation}
with $(d\lambda_1/d\lambda)^{-1}=(d\lambda_2/d\lambda)^{-1}=-2\lambda^{\frac{3}{2}}$ and
$(d\lambda_3/d\lambda)^{-1}=1$ and%
\begin{align}
\frac{dS_{conf}}{d\lambda}  & =-C_{c}\int_{0}^{2\pi}\int_{0}^{\pi}\frac
{d}{d\lambda}\left[  \rho_{chains}\left(  \lambda,\theta\right)  \ln
\rho_{chains}\left(  \lambda,\theta\right)  \right]  \sin\theta d\theta
d\varphi\nonumber\\
& =C_{c}\int_{0}^{2\pi}\int_{0}^{\pi}P_{chc\theta}\left(  \lambda
,\theta\right)  \sin\theta d\theta d\varphi
\end{align}
with%
\begin{equation}
P_{chc\theta}\left(  \lambda,\theta\right)  =-\frac{\sqrt{2}}{8\pi}%
\allowbreak\frac{3\left(  1+\cos2\theta\right)  -\alpha^{2}}{\alpha^{3}%
}\left(  1+\ln\frac{\lambda}{2\sqrt{2}\pi\alpha}\right)
\end{equation}
where $\alpha=\sqrt{\cos2\theta+\lambda^{3}-\lambda^{3}\cos2\theta+1}$. Note that $-TdS_{conf}/d\lambda$ is the generalized force due to the orientational constraint. To
 arrive to a compact expression (the accuracy is not relevant for the discussion), we
take the Taylor series in $\left(  \lambda-1\right)  $. The constant term in
the integral vanishes, and the other terms are%
\begin{equation}
\frac{dS_{conf}}{d\lambda}=0-\frac{C_{c}}{5}\left(  \lambda-1\right)  \left(
2\ln4\pi-1\right)  -\frac{C_{c}}{70}\left(  \lambda-1\right)  ^{2}\left(
44\ln4\pi-18\right)  +O\left(  \left(  \lambda-1\right)  ^{3}\right)
\end{equation}
Then, retaining the first term, dominant for small stretches, we can convert $-TdS_{conf}/d\lambda$ to the  orientational configuration stress tensor
(up to the pressure-like Lagrangian whose related constraint should be enforced in the complete
stress tensor), which is%
\begin{equation}
\boldsymbol{P}_{conf}=-T\frac{dS_{conf}}{d\boldsymbol{U}_{conf}}=T\hat{C}\left(
\lambda-1\right)  \left[
\begin{array}
[c]{ccc}%
-2\lambda^{\frac{3}{2}} &  & \\
& -2\lambda^{\frac{3}{2}} & \\
&  & 1
\end{array}
\right]  +O\left(  \left(  \lambda-1\right)  ^{2}\right)
\label{Pconf}\end{equation}
where $\hat{C}$ is a new constant.%

Equation (\ref{Pconf}) simply shows the observation that the constraint associated to an affine
redistribution of the orientation of the chains produces a stress tensor tied
to the \emph{additional configurational entropy of the network}. This is a result that
should be expected when comparing with the force obtained from the constraint
$r$ associated to the \emph{configurational entropy of the single chain}.
Remarkably, affine models do not consider (at least explicitly) the generalized force associated to the entropic term $-TdS_{conf}$ related to the
configurational entropy of the network, despite deviating from the isotropic
distribution and constraining the chains distribution to that of the
continuum. If this network configurational stress is not considered, then the
(unconstrained) isotropic distribution in the orientation of the chains should
be preserved. Note that the tensor, if considered per-chain,  means
that the equivalent force in a chain for an affine model in which the
configurational entropy is considered, is in general not aligned with
$\boldsymbol{r}$. Conceptually, this would be the force tied to the  biased statistical rotations of the chains.

\section{Macro-micro connection: the chain stretch\label{SEC3}}

The previous section describes the tension in the chains according to the
fractional extension $r/\left(  Nl\right)  $, where $r$ is the end-to-end
chain distance, the modulus of $\boldsymbol{r}$. In this section we perform the
connection of the stretch and tension in the chains to the deformation and
stress in the continuum.

\subsection{Homogenization of chain tensions}

Consider $
\boldsymbol{r}\left(  t\right)  =\lambda_{ch}\left(  t\right)
\boldsymbol{\hat{r}}\left(  t\right)
$
with modulus $\lambda_{ch}\left(  t\right)  $ and with direction $\boldsymbol{\hat{r}}\left(  t\right)
=\boldsymbol{r}\left(  t\right)  /\lambda_{ch}\left(  t\right)  $. The derivative is%
\begin{equation}
\boldsymbol{\dot{r}}   =\dot{\lambda}_{ch}\left(  t\right)  \boldsymbol{\hat{r}}\left(  t\right)
+\lambda_{ch}\left(  t\right)  \frac{d\boldsymbol{\hat{r}}\left(  t\right)  }{dt}%
\label{eqrpunto}\end{equation}
where we may consider the Gaussian solution $|\boldsymbol{r}\left(  0\right)|   =\sqrt{N}l$, so $
\lambda_{ch}\left(  t\right)  ={\left\vert \boldsymbol{r}\left(  t\right)
\right\vert }/{\sqrt{N}l}$, but in practice it is irrelevant to this discussion, so for generality we may just think of relative values, e.g. $r(0)=\lambda_{ch}(0)=1$. Defining $\boldsymbol{t}_{ch}$ as the tension vector in the chain,
the power spent in changing the  entropy of a single chain
is %

\begin{equation}
\mathcal{P}_{ch}=\boldsymbol{t}_{ch}\cdot\boldsymbol{\dot{r}}=\dot{\lambda}%
_{ch}\boldsymbol{t}_{ch}\cdot\boldsymbol{\hat{r}}+\lambda_{ch}\boldsymbol{t}_{ch}%
\cdot\frac{d\boldsymbol{\hat{r}}}{dt}\label{eqpch}%
\end{equation}
Define the averaging (homogenization) operator as%
\begin{equation}
\bar{\mathcal{A}}\equiv\left\langle \mathcal{A}\right\rangle =\int_{\Omega}\mathcal{A}\frac{d\Omega
}{\Omega}\label{homogeneization operator}%
\end{equation}
where $\mathcal{A}$ is any entity (scalar, vectorial or tensorial) and
$d\Omega/\Omega$ is the weight of each contribution, assumed proportional to its share of solid area. Then, in order to connect the
micro structure with the continuum, the following identity is proposed%
\begin{equation}
\mathcal{P}_{ext}=\tfrac{1}{2}\boldsymbol{S}:\boldsymbol{\dot C}=\mathcal{\bar{P}}\equiv\left\langle \mathcal{P}_{ch}\right\rangle
=\int_{\Omega}\mathcal{P}_{ch}\frac{d\Omega}{\Omega}=\int_{\Omega}\dot
{\lambda}_{ch}\boldsymbol{t}_{ch}\cdot\boldsymbol{\hat{r}}\frac{d\Omega}{\Omega}%
+\int_{\Omega}\lambda_{ch}\boldsymbol{t}_{ch}\cdot\frac{d\boldsymbol{\hat{r}}}%
{dt}\frac{d\Omega}{\Omega}\label{Powerint}%
\end{equation}

From the previous expression, two addends can be identified. The first one is the
power spent in the chains length change, $\lambda_{ch}$. The second term
represents the work spent in the rotations of the chains $d\boldsymbol{\hat{r}}/dt$. No assumption about the origin of $\boldsymbol{t}_{ch}$
has been made so far. If the
existence of a potential function for every chain is assumed, i.e., there
exists $\psi_{ch}\left(  \boldsymbol{r}\right)$ (e.g. such that $\dot{\psi}_{ch}=-T\dot{S}_{ch}$ as typically assumed), by frame invariance we must also assume that
the energy stored in a chain is independent of the observer, $\psi
_{ch}\left(  \boldsymbol{r}\right)  =\psi_{ch}\left(  \boldsymbol{Q}\boldsymbol{r}\right)
\,\forall\,\boldsymbol{Q}\in\mathrm{Orth}^{+}$. The vector modulus is the only
invariant with respect to the transformations in $\mathrm{Orth}^{+}$,
therefore, to satisfy the previous condition, $\psi_{ch}\left(  \boldsymbol{r}%
\right)  =\psi_{ch}\left(  \left\vert \boldsymbol{r}\right\vert \right)
=\psi_{ch}\left(  \lambda_{ch}\right)  $. If $\boldsymbol{t}_{ch}$ derives from such energy, it is the work
conjugate of $\boldsymbol{{r}}$, so $\boldsymbol{t}_{ch}:=d\psi_{ch}%
/d\boldsymbol{r}$:%

\begin{equation}
\boldsymbol{t}_{ch}:=\frac{d\psi_{ch}\left(  \left\vert \boldsymbol{r}\right\vert
\right)  }{d\boldsymbol{r}}=\frac{d\psi_{ch}\left(  \left\vert \boldsymbol{r}%
\right\vert \right)  }{d\left\vert \boldsymbol{r}\right\vert }\frac{d\left\vert
\boldsymbol{r}\right\vert }{d\boldsymbol{r}}=\frac{d\psi_{ch}\left(  \lambda
_{ch}\right)  }{d\lambda_{ch}}\frac{d\left\vert \boldsymbol{r}\right\vert
}{d\boldsymbol{r}}=\frac{d\psi_{ch}}{d\lambda_{ch}}\boldsymbol{\hat{r}}\label{tch}%
\end{equation}
where we used $d\left(  \sqrt{\boldsymbol{r}\cdot\boldsymbol{r}}\right)
/d\boldsymbol{r}=\boldsymbol{\hat{r}}$ and $d\boldsymbol{\hat{r}}/d\boldsymbol{r}%
=\boldsymbol{I}-\boldsymbol{\hat{r}}\otimes\boldsymbol{\hat{r}}=\boldsymbol{P}_{\boldsymbol{\hat
{r}}}$ is the projector on the plane perpendicular to $\boldsymbol{\hat{r}}$.
Therefore, Eqs. (\ref{Powerint}) and (\ref{tch}), along the condition
$\boldsymbol{\hat{r}}\perp d\boldsymbol{\hat{r}}/dt$, give%
\begin{equation}
\mathcal{P}_{ext}=\mathcal{\bar{P}}=\left\langle \mathcal{P}_{ch}\right\rangle
=\int_{\Omega}\mathcal{P}_{ch}\frac{d\Omega}{\Omega}=\int_{\Omega}\dot
{\lambda}_{ch}\frac{d\psi_{ch}}{d\lambda_{ch}}\frac{d\Omega}{\Omega
}=\frac{1}{\Omega}\int_{\Omega}\dot{\psi}_{ch}d\Omega\label{powersimply}%
\end{equation}
Noteworthy, the chains rotation term in the power equation is zero because the
individual rotation of every single chain considered in the model cannot
produce a variation in the entropy of that chain if only the chain entropy change is
considered (i.e. the network entropy configuration due to chains orientation
is neglected).

The micromechanical variable $\lambda_{ch}$ remains to be linked to the
continuum (micro-macro connection) and it can be done  in multiple ways. We consider two approaches. One way to
determine the continuum strain energy of the solid from the microstructural
quantities could be an averaging scheme. Considering only the configurational entropy of the chains themselves to build the overall one, we have the potential %
\begin{equation}
\Psi\left(  \boldsymbol{V}\right)=\Psi\left(  \boldsymbol{U}\right)  =\left\langle \psi_{ch}\right\rangle =\frac
{1}{\Omega}\int_{\Omega}\psi_{ch}\left(  \boldsymbol{r}\right)  d\Omega=\frac
{1}{\Omega}\int_{\Omega}\psi_{ch}\left(  \boldsymbol{\hat{r}},\boldsymbol{U}\right)
d\Omega  \label{Def PSI1}%
\end{equation}
where $\Omega$ is  the unit sphere surface, $\psi_{ch}\left(  \boldsymbol{r}%
\right)  $ is the energy density of the chains laid along direction
$\boldsymbol{\hat{r}}$ and $d\Omega$ is the surface differential of unit sphere
surface (isotropic distribution), with direction (perpendicular to)
$\boldsymbol{\hat{r}}$, so $d\Omega/\Omega$ is the corresponding weight assigned
to the chains in direction $\boldsymbol{\hat{r}}$. The tensor $\boldsymbol{U}$ is the
stretch of the continuum, and by $\psi_{ch}\left(  \boldsymbol{\hat{r}}%
_{},\boldsymbol{U}\right)  $ we are stating that in a way determined below, the
chain energy is related to the continuum stretches given by the stretch
tensor from the polar decomposition.
Note that we use frequently the same symbols for different functions with same
physical meaning. We do so to avoid proliferation of symbols and decorations; we leave
the arguments explicit if confusion is possible.

If we assign to the chain a constant surface $d\Omega$ (representing the weight or
reference density of chains with the given orientation $\boldsymbol{\hat{r}}$),
then the force is $d\boldsymbol{f}_{ch}=\boldsymbol{t}_{ch}\left(  \boldsymbol{r}\right)
d\Omega$. Then, we can define a stress tensor%
\begin{equation}
\boldsymbol{P}_{ch}=\frac{d\psi_{ch}}{d\lambda_{ch}}\boldsymbol{\hat{r}}%
\otimes\boldsymbol{\hat{r}}=:P_{ch}\boldsymbol{\hat{r}}\otimes\boldsymbol{\hat{r}}\text{
\ such that }d\boldsymbol{f}_{ch}=\boldsymbol{P}_{ch}\cdot\boldsymbol{\hat{r}}%
~d\Omega\label{Pch}%
\end{equation}
The force for the chains at orientation $\boldsymbol{\hat{r}}$ is%
\begin{equation}
d\boldsymbol{f}_{ch}=\boldsymbol{P}_{ch}\cdot~\boldsymbol{\hat{r}}~d\Omega={{\boldsymbol{t}_{ch}}}d\Omega=\frac{d\psi_{ch}}{d\lambda_{ch}%
}\boldsymbol{\hat{r}}d\Omega
\end{equation}
and $d\boldsymbol{f}_{ch}$ is to be integrated. The averaged power (power
density) is ---cf. Eq. \eqref{powersimply}%
\begin{align}
\left.  \mathcal{\bar{P}}=\frac{1}{\Omega}\int_{\Omega}{{d\boldsymbol{f}_{ch}\cdot\dot{\lambda}_{ch}\boldsymbol{\hat{r}}}d\Omega%
}\right.   &  =\frac{1}{\Omega}\int_{\Omega}\dot{\lambda}_{ch}\boldsymbol{\hat{r}%
}\cdot{{\boldsymbol{P}_{ch}\cdot\boldsymbol{\hat{r}}}d\Omega}\nonumber\\
&  =\frac{1}{\Omega}\int_{\Omega}\frac{d\psi_{ch}}{d\lambda_{ch}}\dot{\lambda
}_{ch}d\Omega=\frac{1}{\Omega}\int_{\Omega}\dot{\psi}_{ch}d\Omega
\label{Avg power}%
\end{align}
Thus, let us define the\ average stress tensor ---see Eq. \eqref{Def PSI1}%
\begin{equation}
\boldsymbol{\bar{P}}_{ch}=\left\langle \boldsymbol{P}_{ch}\right\rangle =\int_{\Omega
}\boldsymbol{P}_{ch}\frac{d\Omega}{\Omega}=\int_{\Omega}\frac{d\psi_{ch}}%
{d\lambda_{ch}}\boldsymbol{\hat{r}}\otimes\boldsymbol{\hat{r}}\frac{d\Omega}{\Omega}%
=:\frac{d\Psi}{d\boldsymbol{U}}\label{Pchmean}\end{equation}
During a tensile,
biaxial or pure shear test in an incompressible continuum, (to eliminate macroscopic rotations in the reasoning such that
$\boldsymbol{R}=\boldsymbol{I}$ and $\boldsymbol{F}=\boldsymbol{U}$), the Piola stress is%
\begin{equation}
\boldsymbol{P}=\frac{d\Psi}{d\boldsymbol{F}}+p\boldsymbol{F}^{-1}=\frac{d\Psi}%
{d\boldsymbol{U}}+p\boldsymbol{U}^{-1}\Rightarrow\;P_{i}=\frac{\partial\Psi}{\partial\lambda_{i}}+\frac
{1}{\lambda_{i}}p
\end{equation}
where $p$ is the
pressure-like Lagrange multiplier for the incompressibility condition. Then\begin{align}
\left.  P_{i}=\frac{1}{\lambda_{i}}p+\frac{\partial\Psi}{\partial\lambda_{i}%
}\right.   &  =\frac{1}{\lambda_{i}}p+\boldsymbol{\hat{n}}_{i}\cdot\boldsymbol{\bar{P}}_{ch}%
\cdot\boldsymbol{\hat{n}}_{i}\nonumber\\
&  =\frac{1}{\lambda_{i}}p+\int_{\Omega}\frac{d\psi_{ch}}{d\lambda_{ch}%
}\left(  \boldsymbol{\hat{r}}\cdot\boldsymbol{\hat{n}}_{i}\right)  ^{2}\frac{d\Omega
}{\Omega}\label{Pi-0}%
\end{align}
where $\boldsymbol{\hat{r}}$ is the orientation of chains, to which a constant
isotropic weight of $d\Omega/\Omega$ has been assigned.

\subsection{The isotropic micro-macro connection for the chain stretch}

Taking the following definition for the stretch
\begin{equation}
\lambda_{ch}\left(  \boldsymbol{\hat{r}}\right)  =\boldsymbol{U}:\boldsymbol{\hat{r}%
}\otimes\boldsymbol{\hat{r}}=\left(  \sum_{i=1}^{3}\lambda_{i}\boldsymbol{\hat{n}}%
_{i}\otimes\boldsymbol{\hat{n}}_{i}\right)  :\left(  \boldsymbol{\hat{r}}%
\otimes\boldsymbol{\hat{r}}\right)  =\sum_{i=1}^{3}\lambda_{i}\left(
\boldsymbol{\hat{r}}\cdot\boldsymbol{\hat{n}}_{i}\right)  ^{2}%
\end{equation}
it is immediate to check that we recover Eq. \eqref{Pi-0} by direct use of the chain rule%
\begin{equation}
P_{i}\equiv\frac{1}{\lambda_{i}}p+\frac{d\Psi}{d\lambda_{i}}=\frac{1}%
{\lambda_{i}}p+\frac{1}{\Omega}\int_{\Omega}\frac{d\psi_{ch}}{d\lambda_{ch}%
}\frac{d\lambda_{ch}}{d\lambda_{i}}d\Omega=\frac{1}{\lambda_{i}}p+\frac
{1}{\Omega}\int_{\Omega}\frac{d\psi_{ch}}{d\lambda_{ch}}\left(  \boldsymbol{\hat
{r}}\cdot\boldsymbol{\hat{n}}_{i}\right)  ^{2}d\Omega\tag{\ref{Pi-0}}
\end{equation}
which is consistent also with the definition in Eq. (\ref{Def PSI1}). In essence,
$\boldsymbol{\hat{r}}$ plays the role of a direction in the sphere (not that of a particular chain) and
$\lambda_{ch}$ of the stretch associated to that direction. No change in weights is assumed: physically it may be interpreted as that the isotropic distribution is maintained, because no force is considered to rotate the chains and entropy is only considered regarding the constraint $r$ in direction $\boldsymbol{\hat r}$ for each chain.

Note that once the
relation $P_{i}\left(  \lambda_{1},\lambda_{2}\right)  $ has been obtained, it
is of course valid not only for the mentioned tests, but for any boundary or deformation
condition found in any stress integration point in a finite element program.
Obviously, the stress power density is also equivalent. Since in these tests,
principal directions are rotationless and there is no volumetric power (by
incompressibility), we verify%
\begin{align}
\left.  \mathcal{P}\equiv\dot{\Psi}=%
{\textstyle\sum\limits_{i=1}^{3}}
P_{i}\dot{\lambda}_{i}\right.   &  =%
{\textstyle\sum\limits_{i=1}^{3}}
\frac{d\Psi}{d\lambda_{i}}\dot{\lambda}_{i}+%
{\textstyle\sum\limits_{i=1}^{3}}
p\frac{\dot{\lambda}_{i}}{\lambda_{i}}\nonumber\\
&  =%
{\textstyle\sum\limits_{i=1}^{3}}
\left(  \frac{1}{\Omega}\int_{\Omega}\frac{d\psi_{ch}}{d\lambda_{ch}}%
\frac{d\lambda_{ch}}{d\lambda_{i}}d\Omega\right)  ~\dot{\lambda}_{i}+p%
{\textstyle\sum\limits_{i=1}^{3}}
\frac{d}{dt}\left(  \ln\lambda_{i}\right)  \\
&  =\frac{1}{\Omega}\int_{\Omega}\frac{d\psi_{ch}}{d\lambda_{ch}}\left(
{\textstyle\sum\limits_{i=1}^{3}}
\frac{d\lambda_{ch}}{d\lambda_{i}}\dot{\lambda}_{i}\right)  ~d\Omega
+0\nonumber\\
&  =\frac{1}{\Omega}\int_{\Omega}\frac{d\psi_{ch}}{d\lambda_{ch}}\dot{\lambda
}_{ch}~d\Omega  \quad=\frac{1}{\Omega}\int_{\Omega}\dot{\psi}_{ch}~d\Omega=\mathcal{\bar{P}%
}\text{ \ (see Eqs. \eqref{powersimply} and \eqref{Avg power})\label{EQPwr}}%
\end{align}

\subsection{The affine micro-macro connection for the chain stretch}
\begin{figure}[tb!]
\centering
\includegraphics[width=.85\textwidth]{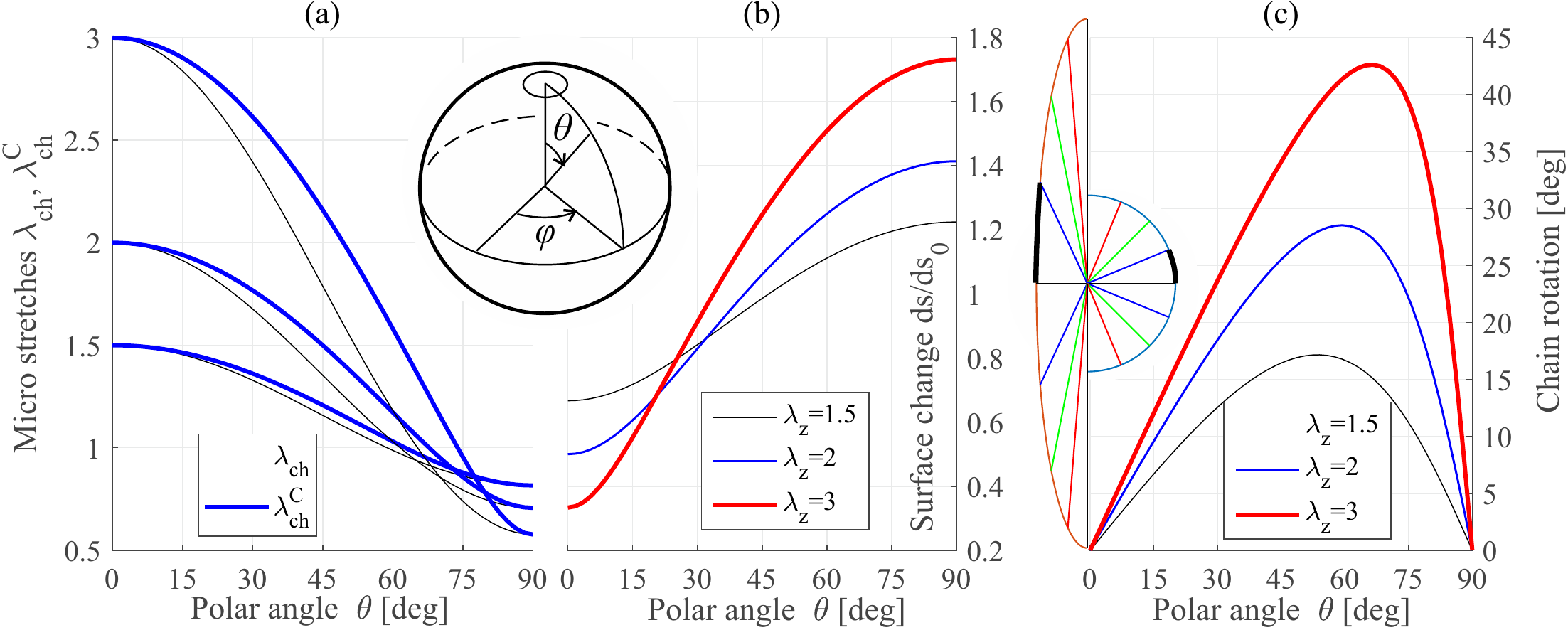}\quad
\caption{a) Difference between $\lambda_{ch}=\boldsymbol{\hat{r}}\cdot \boldsymbol{U} \cdot \boldsymbol{\hat{r}}$ and $\lambda_{ch}^C=\sqrt{\boldsymbol{\hat{r}}\cdot \boldsymbol{C} \cdot \boldsymbol{\hat{r}}}$ for different uniaxial stretches $\lambda_z=1.5$, $\lambda_z=2$, $\lambda_z=3$. b) Change of surface due to surface-affine deformations. c) Rotation of chains if they follow affine deformations.}%
\label{Affinity_composite2}%
\end{figure}
The usually employed alternative is to assume affinity in the right
Cauchy-Green deformation tensor obtained from the following micro-macro
relation. Because now  $\boldsymbol{\hat{r}}$ is attached to the continuum and a specific chain, in this case it is important to distinguish the original direction
$^{0}\boldsymbol{\hat{r}}$ corresponding to the continuum-undeformed
configuration, and the current deformed one $\boldsymbol{\hat{r}}$
\begin{equation}
\lambda_{ch}\left(  \boldsymbol{\hat{r}}\right)  ~\boldsymbol{\hat{r}}=\boldsymbol{RU}%
\cdot~^{0}\boldsymbol{\hat{r}}%
\end{equation}
so, multiplying by the transpose to cancel the continuum rotations, we
identify the affine squared stretches  from the Cauchy-Green
deformation tensor in the chain direction
\begin{equation}
\Lambda_{ch}=~^{0}\boldsymbol{\hat{r}}\cdot\boldsymbol{C}\cdot~^{0}\boldsymbol{\hat{r}%
}:=\left(  \lambda_{ch}^{C}\right)  ^{2}=\sum_{i=1}^{3}\Lambda_{i}\left(  \boldsymbol{\hat{n}}%
_{i}\cdot~^{0}\boldsymbol{\hat{r}}\right)  ^{2}%
\end{equation}
where $\Lambda_{i}$ are the eigenvalues of $\boldsymbol{C}$. These stretches assume deformations affine with the continuum deformation gradient. In general, for chains not aligned with the principal directions ---see Fig. \ref{Affinity_composite2}a

\begin{equation}
\lambda_{ch}^{C}\left(  ^{0}\boldsymbol{\hat{r}}\right)  =\sqrt{^{0}%
\boldsymbol{\hat{r}}\cdot\boldsymbol{C}\cdot~^{0}\boldsymbol{\hat{r}}}\neq~^{0}%
\boldsymbol{\hat{r}}\cdot\boldsymbol{U}\cdot~^{0}\boldsymbol{\hat{r}}=\lambda_{ch}\left(
^{0}\boldsymbol{\hat{r}}\right)  \label{Lambdasnotequal}%
\end{equation}
Note that even though
$\boldsymbol{R}=\boldsymbol{I}$ we have that necessarily a superposed rotation
$\boldsymbol{R}_{ch}\neq\boldsymbol{I}$ for chains not aligned in principal
directions, because  $\boldsymbol{U}\cdot~^{0}\boldsymbol{\hat{r}}_{}$ does not have
the direction of $~^{0}\boldsymbol{\hat{r}}_{}$; Fig. \ref{Affinity_composite2}c shows that this rotation is important. Indeed, this is a basic
difference between both approaches: whether the rotations of the chains follow
the continuum line rotations or if these rotations (which do not produce work according to this framework)
are such that the distribution remains isotropic (i.e. chains fluctuate freely).
Of course, intermediate cases are also possible, depending on the relative importance of the system network entropy respect to the entropy of the chains.

Then, if we assume that $\Lambda_{ch}$ (or $\lambda_{ch}^C$) is the state variable for the
deformation in the chains, then we can define ---cf. Eq. \eqref{Pch}
\begin{equation}
\boldsymbol{S}%
_{ch}:=2\frac
{d\psi_{ch}}{d\Lambda_{ch}}~^{0}\boldsymbol{\hat{r}}\otimes~^{0}\boldsymbol{\hat{r}}%
\end{equation}
and further define an averaged second Piola-Kirchhoff (PK) stress by
integration in the unit sphere as ---cf. Eq. \eqref{Pchmean}%
\begin{align}
\boldsymbol{\bar{S}}_{ch} =\left\langle\boldsymbol{{S}}_{ch}\right\rangle &  =\int_{\Omega}\boldsymbol{S}_{ch}\frac{d\Omega}{\Omega}=\frac
{1}{\Omega}\int_{\Omega}2\frac{d\psi_{ch}}{d\Lambda_{ch}}~^{0}\boldsymbol{\hat{r}%
}\otimes~^{0}\boldsymbol{\hat{r}}d\Omega\nonumber\\
&  =\frac{1}{\Omega}\int_{\Omega}2\frac{d\psi_{ch}}{d\Lambda_{ch}}%
\frac{d\Lambda_{ch}}{d\boldsymbol{C}}d\Omega=\frac{1}{\Omega}\int_{\Omega}\frac
{1}{\lambda_{ch}^{C}}\frac{d\psi_{ch}}{d\lambda_{ch}^{C}}~^{0}\boldsymbol{\hat{r}%
}\otimes~^{0}\boldsymbol{\hat{r}}d\Omega\label{Sbarch}
\end{align}
This is the typical chain rule and approach followed in the affine approach; see e.g. Eq. (128) of \cite{Steinmann}. It is also the typical additive scheme in terms of 2nd Piola-Kirchhoff stresses employed in soft tissues as chained derivatives of energy terms taken respect to Rivlin-Spencer invariants of $\boldsymbol{C}$; see e.g.  \cite{Holzapfel}, Eq. (12) in  \cite{Chagnon}, Eq. (5.9) in \cite{Volokh} and therein referenced models. Note that this is equivalent to assigning a surface weight $d\Omega/\lambda
_{ch}^{C}$ to the chain, an expression which would also follow applying
Nanson's formula to the chain stretch considering the deformation gradient of
the chain. Then, we assume that the principal values of the continuum 2nd PK
stresses are given by ---cf. its push-forward $\sigma_i=\lambda_i^2S_i$ in Eq. (4) of \cite{Wu-Giessen-MRC} %
\begin{align}
\left.  S_{i}=\frac{1}{\lambda_{i}^{2}}p+2\frac{d\Psi}{d\Lambda_{i}}\right.
&  =\frac{1}{\lambda_{i}^{2}}p+\boldsymbol{\bar S}_{ch}:\boldsymbol{\hat{n}}_{i}%
\otimes\boldsymbol{\hat{n}}_{i}\nonumber\\
&  =\frac{1}{\lambda_{i}^{2}}p+\left(  \frac{1}{\Omega}\int_{\Omega}%
2\frac{d\psi_{ch}}{d\Lambda_{ch}}~^{0}\boldsymbol{\hat{r}}\otimes~^{0}%
\boldsymbol{\hat{r}}~d\Omega\right)  :\boldsymbol{\hat{n}}_{i}\otimes\boldsymbol{\hat{n}%
}_{i}\nonumber\\
&  =\frac{1}{\lambda_{i}^{2}}p+\frac{1}{\Omega}\int_{\Omega}\frac{1}%
{\lambda_{ch}^{C}}\frac{d\psi_{ch}}{d\lambda_{ch}^{C}}\left(  \boldsymbol{\hat{n}%
}_{i}\cdot~^{0}\boldsymbol{\hat{r}}\right)  ^{2}d\Omega\label{Si-0}%
\end{align}
It is obvious that the averaging methods of Eq. (\ref{Pi-0}) and Eq.
(\ref{Si-0}) are different, as they are also $\lambda_{ch}\left(
~^{0}\boldsymbol{\hat{r}}\right)  $ and $\lambda_{ch}^{C}\left(  ~^{0}%
\boldsymbol{\hat{r}}\right)  $; and $\lambda_{ch}$ cannot be written as a scalar
function of $\lambda_{ch}^{C}$ (and vice-versa) because all the principal
stretches (or invariants) are involved in the relation.

The stress power in this case is ---cf. Eq. \eqref{EQPwr}%
\begin{align}
\left.  \mathcal{P}\equiv\dot{\Psi}=%
{\textstyle\sum\limits_{i=1}^{3}}
\tfrac{1}{2}S_{i}\dot{\Lambda}_{i}=%
{\textstyle\sum\limits_{i=1}^{3}}
\frac{d\Psi}{d\Lambda_{i}}\dot{\Lambda}_{i}\right.   &  =%
{\textstyle\sum\limits_{i=1}^{3}}
\left(  \frac{1}{\Omega}\int_{\Omega}\frac{d\psi_{ch}}{d\Lambda_{ch}}%
\frac{d\Lambda_{ch}}{d\Lambda_{i}}~d\Omega\right)  ~\dot{\Lambda}%
_{i}\nonumber\\
&  =\frac{1}{\Omega}\int_{\Omega}\frac{d\psi_{ch}}{d\Lambda_{ch}}\left(
{\textstyle\sum\limits_{i=1}^{3}}
\frac{d\Lambda_{ch}}{d\Lambda_{i}}\dot{\Lambda}_{i}\right)  ~d\Omega\\
&  =\frac{1}{\Omega}\int_{\Omega}\frac{d\psi_{ch}}{d\Lambda_{ch}}\dot{\Lambda
}_{ch}~d\Omega=\frac{1}{\Omega}\int_{\Omega}\dot{\psi}_{ch}~d\Omega
=\mathcal{\bar{P}}\nonumber
\end{align}
Interestingly, we can push-forward the averaged stress\ in Eq. \eqref{Sbarch} to obtain the Piola
stress through the continuum mapping, which in the standard tests with
$\boldsymbol{R}=\boldsymbol{I}$ is, without the pressure term%
\begin{align}
\left.  \boldsymbol{F}\boldsymbol{\bar S}_{ch}=\boldsymbol{U}\boldsymbol{\bar S}_{ch}\right.   &
=\frac{1}{\Omega}\int_{\Omega}2\frac{d\psi_{ch}}{d\Lambda_{ch}}\boldsymbol{U}%
\frac{d\Lambda_{ch}}{d\boldsymbol{C}}d\Omega\nonumber\\
&  =\frac{1}{\Omega}\int_{\Omega}\frac{1}{\lambda_{ch}^{C}}\frac{d\psi_{ch}%
}{d\lambda_{ch}^{C}}%
{\textstyle\sum\limits_{i=1}^{3}}
\left[  \lambda_{i}\left(  \boldsymbol{\hat{n}}_{i}\cdot~^{0}\boldsymbol{\hat{r}%
}\right)  \boldsymbol{\hat{n}}_{i}\otimes~^{0}\boldsymbol{\hat{r}}\right]  d\Omega
\end{align}
If applied to a principal plane, principal direction in the standard
homogeneous test%
\begin{align}
P_i=\frac{p}{\lambda_i}+\boldsymbol{\hat{n}}_{i}\cdot\left(  \boldsymbol{F}\boldsymbol{\bar S}_{ch}\right)
\cdot\boldsymbol{\hat{n}}_{i} &  =\frac{p}{\lambda_i}+\frac{1}{\Omega}\int_{\Omega}\frac{\lambda_{i}%
}{\lambda_{ch}^{C}}\frac{d\psi_{ch}}{d\lambda_{ch}^{C}}\left(  \boldsymbol{\hat
{n}}_{i}\cdot~^{0}\boldsymbol{\hat{r}}\right)  ^{2}d\Omega\nonumber\\
&  =\frac{p}{\lambda_i}+\frac{\lambda_{i}}{\Omega}\int_{\Omega}\frac{d\psi_{ch}}{d\lambda_{ch}%
^{C}}\left(  \boldsymbol{\hat{n}}_{i}\cdot~^{0}\boldsymbol{\hat{r}}\right)  ^{2}%
\frac{d\Omega}{\lambda_{ch}^{C}}%
\end{align}
which is to be compared to Eq. (\ref{Pi-0}). Note that, (1) the surfaces $\Omega$ and $d\Omega$ are now weighted according to their stretches $\lambda_i$ and $\lambda_{ch}^C$, see Fig. \ref{Affinity_composite2}b;  and (2) the
chain stretches $\lambda_{ch}$ and $\lambda_{ch}^{C}$ are very different: there
is no one-to-one relation between them for a general chain direction, see Fig. \ref{Affinity_composite2}a.
 These two differences, respectively, could possibly be behind the needed non-affine micro-macro correction of the \emph{``macro-area-stretch''} (Eqs. (58)-(60) in \cite{Microsphere})
and the non-affine micro-macro correction of the \emph{``macro-stretch''} (Eqs. (45)-(47) in \cite{Microsphere}) to accommodate macroscopic experimental results via maximization of the entropy respect to an assumed fluctuation field (Sec. 4.3.2 in \cite{Microsphere}).

Of course, an intermediate option regarding orientational affinity is possible, although not explored. For example using $\lambda_{ch}^*=\boldsymbol{(U}^p:\boldsymbol{\hat{n}}\otimes \boldsymbol{\hat{n}})^{1/p}$, with $p=2$ for the affine case and $p=1$ for the unconstrained case. In general $p$ may be a function of the network characteristics and of the deformation.
However, note that solutions of this type are different from the typical non-affine stretches in the literature.

\section{The relevance of the choice of the chain stretch in general
deformations\label{toy example}}
\subsection{Comparison of two examples of micro-macro connection\label{SEC:MMconnection}}
In a phenomenological model, $\Psi\left(  \lambda_{1},\lambda_{2},\lambda
_{3}\right)  $ and $\Psi\left(  \Lambda_{1},\Lambda_{2},\Lambda_{3}\right)  $
are mathematically different functions but may be made physically equivalent, because
arguments are related in a one-to-one pattern, namely $\Lambda_{i}=\left(
\lambda_{i}^{C}\right)  ^{2}=\lambda_{i}^{2}$. However, in chain models,
different chain stretch variables give physically different models, see Eq.
(\ref{Lambdasnotequal})
and Fig. \ref{Affinity_composite2}c\begin{equation}
\Psi^{(1)}\left(  \boldsymbol{C}\right) = \dfrac{1}{\Omega}\int_{\Omega}\psi^{(1)}_{ch}\left(
\Lambda_{ch}\right)  d\Omega \neq \dfrac{1}{\Omega}\int_{\Omega}\psi^{(2)}_{ch}\left(
\lambda_{ch}\right)  d\Omega = \Psi^{(2)}\left(  \boldsymbol{U}\right)
\label{Eq:1y2}\end{equation}
Since in principal directions of deformation $\lambda_{ch}^2=(\lambda_{ch}^{C})^2=\Lambda_{ch}$, and by Fig. \ref{Affinity_composite2}a differences between both stretches are not very large, one
may think that the practical relevance of this observation in the predictions
of a model should be small. To investigate this, consider two simple models in which the strain energy is considered only due to the sum of the entropy of the individual chains %
\begin{align}
\psi_{ch}^{\left(  M1\right)  }\left(  \Lambda_{ch}\right)  &=\dfrac{3\mu}%
{2}\left(  \Lambda_{ch}-1\right)  \quad\text{with}\quad\Lambda_{ch}=\boldsymbol{C}:\left(
\boldsymbol{\hat r}\otimes\boldsymbol{\hat r}\right)=(\lambda_{ch}^C)^2 \label{M1} \\
\psi_{ch}^{\left(  M2\right)  }\left(  \lambda_{ch}\right)  &=\dfrac{3\mu}%
{2}\left(  \lambda_{ch}^{2}-1\right)  \quad\text{with}\quad \lambda_{ch}=\boldsymbol{U}%
:\left(  \boldsymbol{\hat r}\otimes\boldsymbol{\hat r}\right)
\label{M2}\end{align}
from which the macroscopic energies\ $\alpha=1,2$ follow by
$\Psi^{\left(  \alpha\right)  }\left( \bullet\right)  =\tfrac{1}{\Omega
}\int_{\Omega}\psi_{ch}^{\left(  \alpha\right)  }\left(  \bullet\right)
d\Omega$, and the derivatives respect to the principal continuum stretches are%
\begin{equation}
\dfrac{\partial\Psi^{\left(  M1\right)  }\left(  \lambda_{1},\lambda
_{2},\lambda_{3}\right)  }{\partial\lambda_{i}}=\dfrac{1}{\Omega}\int_{\Omega
}\dfrac{d\psi^{\left(  M1\right)  }\left(  \Lambda_{ch}\right)  }{d\Lambda
_{ch}}\dfrac{\partial\Lambda_{ch}}{\partial\lambda_{i}}d\Omega
\label{eq:deriv1}%
\end{equation}%
\begin{equation}
\dfrac{\partial\Psi^{\left(  M2\right)  }\left(  \lambda_{1},\lambda
_{2},\lambda_{3}\right)  }{\partial\lambda_{i}}=\dfrac{1}{\Omega}\int_{\Omega
}\dfrac{\psi^{\left(  M2\right)  }\left(  \lambda_{ch}\right)  }{d\lambda_{ch}%
}\dfrac{\partial\lambda_{ch}}{\partial\lambda_{i}}d\Omega\label{eq:deriv2}%
\end{equation}
Both models would be identical if $\lambda_{ch}=\lambda_{ch}^C$. The macroscopic predictions may be obtained through a numerical angular integration, see \cite{MMM}. However, it is well known that
\begin{equation}\frac{1}{\Omega}%
\int_{\Omega}\Lambda_{ch}d\Omega=\frac{1}{3}I_{1}
\end{equation}
where $I_{1}$ is the
first invariant of $\boldsymbol{C}$. Therefore $\Psi^{\left(  M1\right)  }\left(
\lambda_{1},\lambda_{2},\lambda_{3}\right)  $ may also be integrated analytically. The result
 is the well-known Neo-Hookean model, which is the model obtained from the classical Gaussian network theory, where $\mu=nk_BT$ ($n$ being the chain density)---see Sec. 4.2, Eq. (4.9a) of \cite{Treloar_book}\begin{equation}\Psi^{\left(  M1\right)  }\left(  \lambda_{1},\lambda_{2},\lambda_{3}\right)
\equiv
\Psi^{\left(  \text{N-H}\right)  }\left(  \lambda_{1},\lambda_{2},\lambda_{3}\right)
=\frac{\mu}{2}\left(  I_{1}-3\right)
\end{equation}
Model M2 would be also a Gaussian model but considering an unconstrained orientational distribution of chains. For the incompressible materials at hand,  $P_{3}=0$ will be enforced
for all the deformation states considered to  eliminate the  pressure-like penalization term from the equations, leading to the following set of
equations for computing the nominal stresses in principal directions of deformation%

\begin{equation}
P_{i}=\dfrac{\partial\Psi\left(  \lambda_{1},\lambda_{2},\lambda_{3}\right)
}{\partial\lambda_{i}}-\dfrac{\lambda_{3}}{\lambda_{i}}\dfrac{\partial
\Psi\left(  \lambda_{1},\lambda_{2},\lambda_{3}\right)  }{\partial\lambda_{3}%
};\quad i=1,2 \label{eq:Nominal}%
\end{equation}

To evaluate the aforementioned influence of the independent chain variable used to
describe the problem, two states of deformation are computed for the constant $\mu=1\operatorname{MPa}$. For these comparisons we consider two tests defining two families of states of deformation, both with  $P_{3}=0$ and complying with the incompressibility constraint
\begin{itemize}
\item
s1:  biaxial test in which $\lambda_{2}=3.1$ and $\lambda_{1}$ varies in the range from $0.2$ to $3.1$, and  $\lambda_{3}=1/\left(  \lambda_{1}\lambda_{2}\right) $
\item
s2: uniaxial test in which $\lambda_{1}$ will be in the range from $0.8$ to $2.2$ and $\lambda_{2}=\lambda_{3}=1/\sqrt{\lambda_{1}}$
\end{itemize}

The biaxial test (s1) is interesting because keeping $\lambda_2=3.1$ for all the range of $\lambda_1$, the angular distribution of the chains under the affinity assumption is not isotropic in the testing plane $1-2$ (except for the specific point $\lambda_1=3.1$). In the uniaxial test (s2), the distribution is isotropic at the beginning of the loading, for $\lambda_1=1$. Hence, to emphasize the differences between models, in the sections below we will use the $P_1-\lambda_1$ curve of the biaxial (s1) test to characterize the materials.

\begin{figure}[htbp!]
        \begin{minipage}{\textwidth}
        \centering
        \includegraphics[width=.4\textwidth]{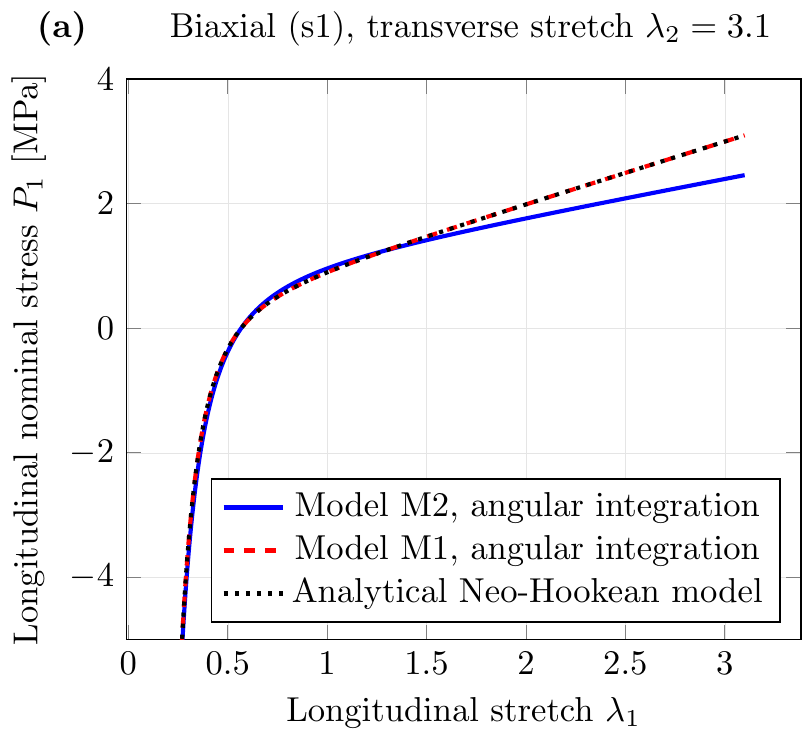}\quad
        \includegraphics[width=.4\textwidth]{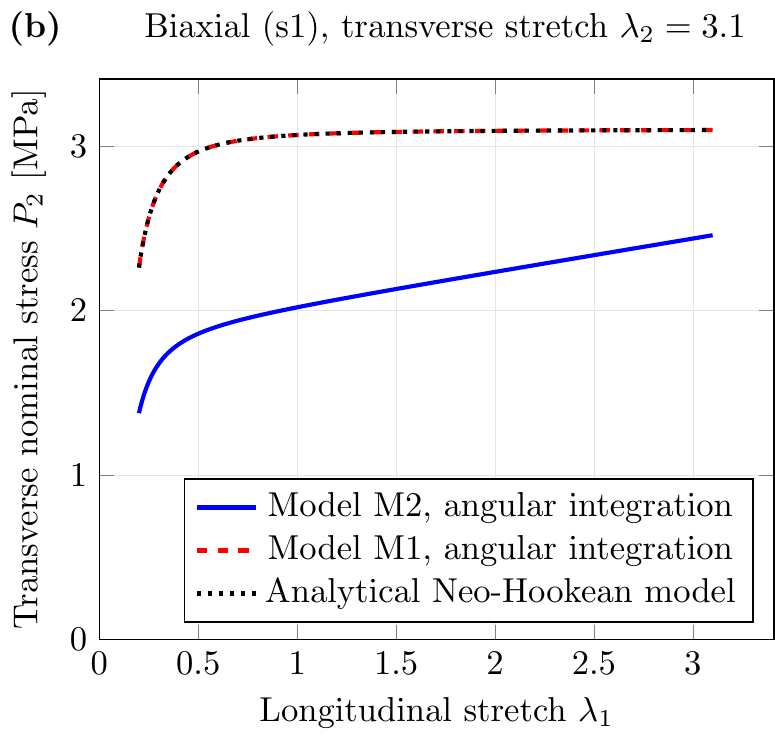}\\
        \includegraphics[width=.4\textwidth]{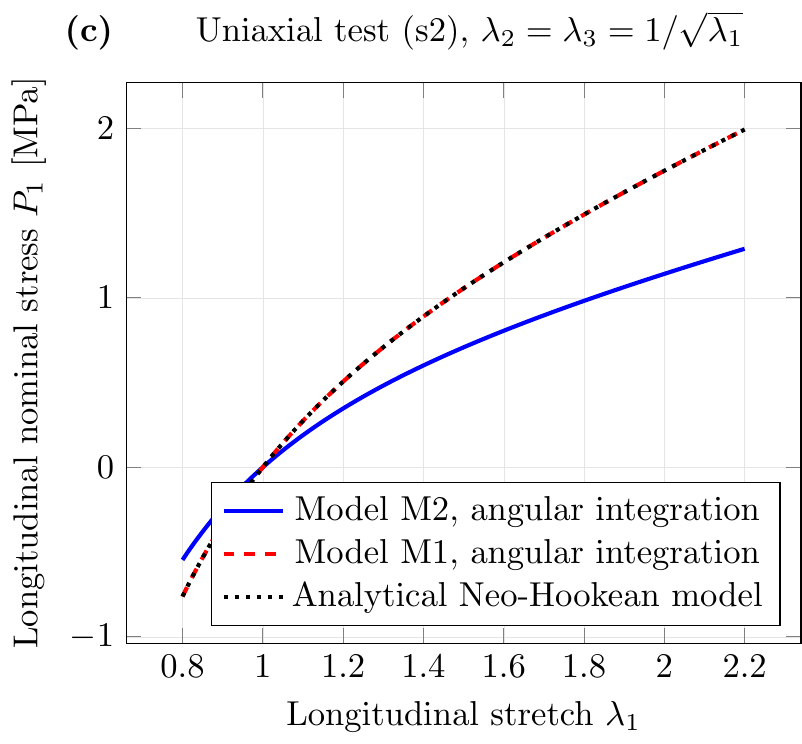}\quad
        \includegraphics[width=.4\textwidth]{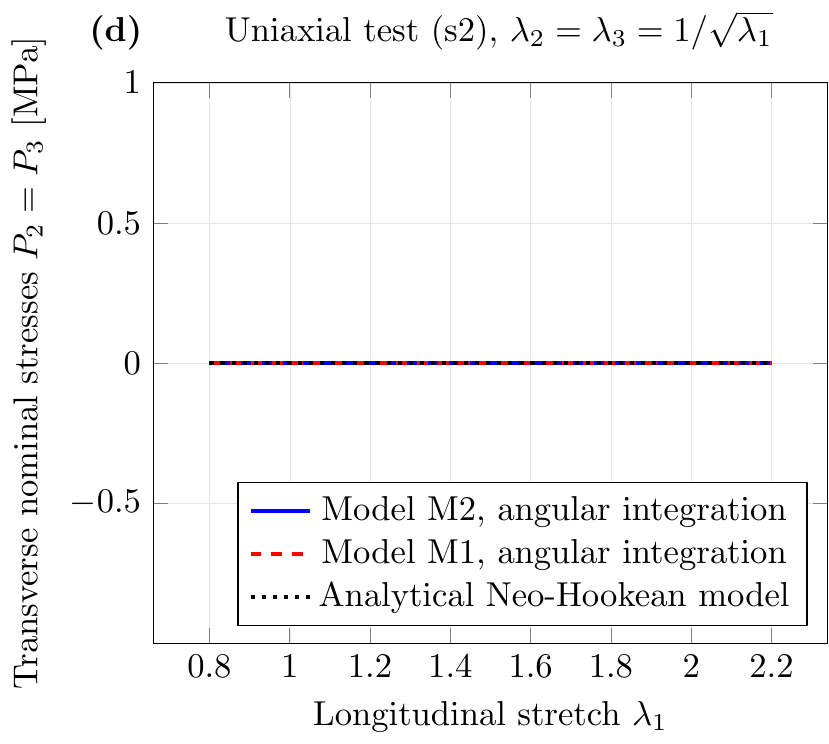}
        \caption{Relevance of different micro-macro connections through different chain stretches. Comparison of material behavior when using model M1 and model M2 through angular integrations, and the analytical Neo-Hookean model (equivalent to model M1). a) Biaxial test (s1) in which $\lambda_2=3.1$ is fixed and $\lambda_1$ varies; $P_1$ vs. $\lambda_1$. b) Cross-effect shown in $P_2$ vs. $\lambda_1$. c) Uniaxial test (s2) in axis 1; $P_1$ vs. $\lambda_1$. d) Transverse stresses $P_2=P_3$ in the uniaxial test (s2) for $\lambda_2=\lambda_3=1/\sqrt{\lambda_1}$, showing fulfillment of the incompressibility condition. Note that red dashed curves and black dotted curves are superimposed because the M1 model is the same as the classical Neo-Hookean model from the Gaussian network theory, see Sec. 4.2 of \cite{Treloar_book}}
        \label{exps1y3}
        \end{minipage}
\end{figure}

For angular integration in the sphere, we used the $ 21$ points from \cite{Bazant}. We plot the two non-vanishing nominal stresses, namely the longitudinal one $P_1$ (in the axis of main varying stretch) and the transverse one $P_2$. Using both M1 and M2 models, the nominal stresses $P_1$ and $P_2$ obtained for both (s1) and (s2) tests are presented in Fig. \ref{exps1y3}.


%
%


 As commented, expected, and observed in Fig. \ref{exps1y3}, model M1 integrated angularly is completely equivalent to the well-known Neo-Hookean model. Note that Fig. \ref{exps1y3} also serves as a check of the accuracy of the numerical angular integration scheme employed.  However, in that figure it can also be observed that the behaviors of models M1 and M2 are substantially different, meaning that despite the resemblance of Eqs. \eqref{M1} and \eqref{M2} and that $\lambda_{ch}=\lambda_{ch}^C$ in principal directions of macroscopic deformation, the chain stretch variable used to link the chain behavior to that of the continuum has a substantial impact.

An interesting aspect of the Neo-Hookean model can be noted. For large stretches during a biaxial test with constant $\lambda_2$, we have that the transverse and the longitudinal moduli are
\begin{equation}
\frac{\partial P_2}{\partial \lambda_1}=\frac{2\mu}{\lambda_1^3\lambda_2^3}\quad \text{and}\quad \frac{\partial P_1}{\partial \lambda_1}=\mu\left(1+ \frac{3}{\lambda_1^4\lambda_2^2}\right)
\label{Eq NH issues}
\end{equation}

The first one is interesting, because it shows that for moderate stretches and a constant modulus $\mu$, the affine models show no appreciable increase in transverse stresses $P_2$ when the material is strained in the longitudinal direction. For example for stretches of, say, $\lambda_1=\lambda_2=2$, we have $\partial P_2/\partial\lambda_1=\mu/32$. This is the horizontal behavior observed in Fig. \ref{exps1y3}b for the Neo-Hookean model. A relevant positive slope could only be achieved for large stretches if the modulus $\mu$ has a strong increase, locking-type behavior. Note however that model $M2$ has a hardening in the transverse direction when the longitudinal one is stretched. The second one in Eq. \eqref{Eq NH issues} is that a constant $\mu$ means also
a constant slope $\mu$ for large stretches. This can also be observed in Fig. \ref{exps1y3}b. These observations are interesting as a reference for analysing predictions of affine models during biaxial tests in real materials; see Sec. \ref{SEC:5}.

\begin{figure}[htbp!]
        \begin{minipage}{\textwidth}
        \centering
        \includegraphics[width=.4\textwidth]{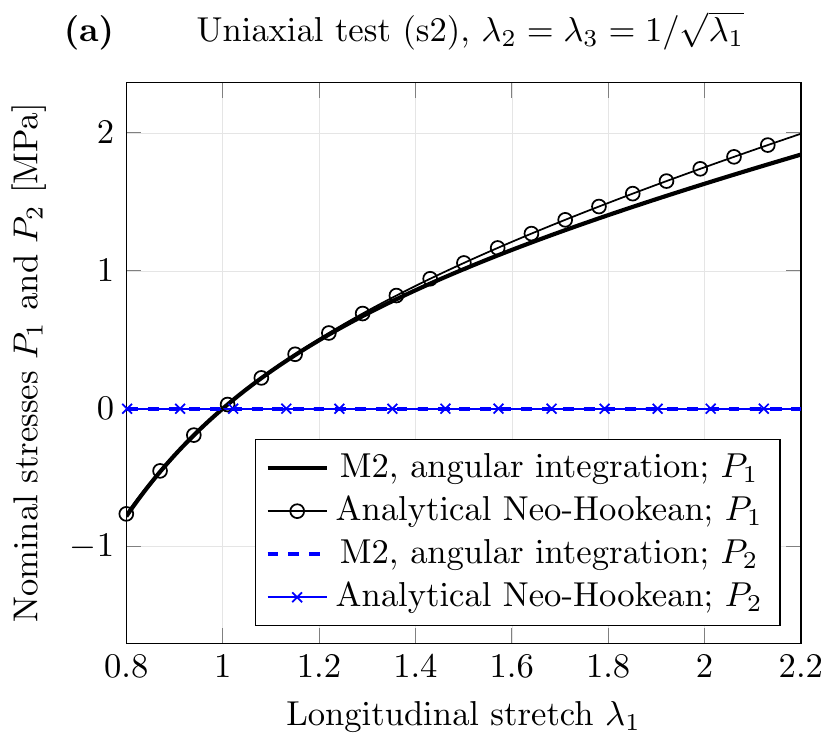}\quad
        \includegraphics[width=.4\textwidth]{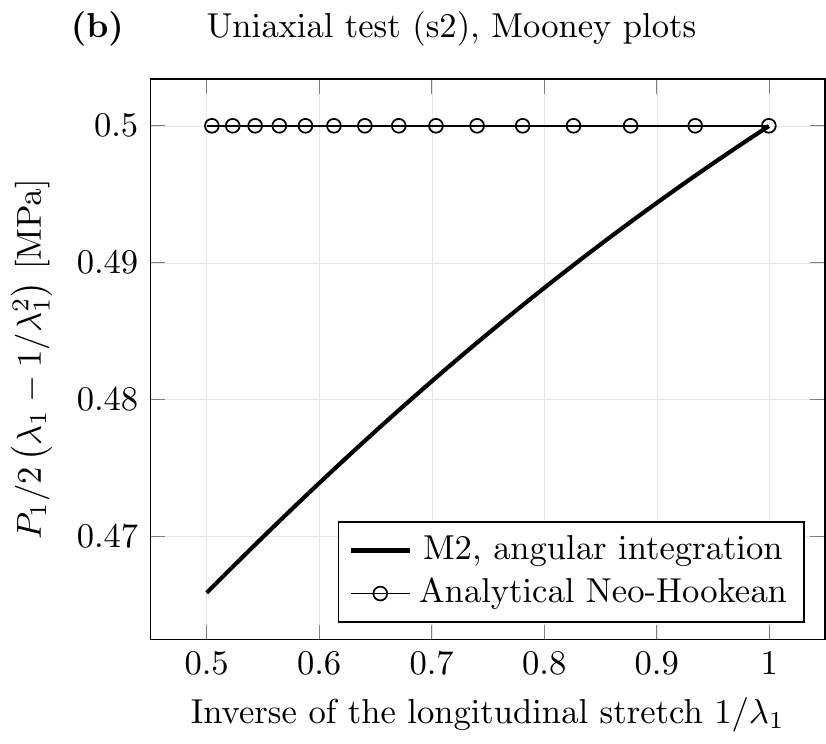}
        \includegraphics[width=.4\textwidth]{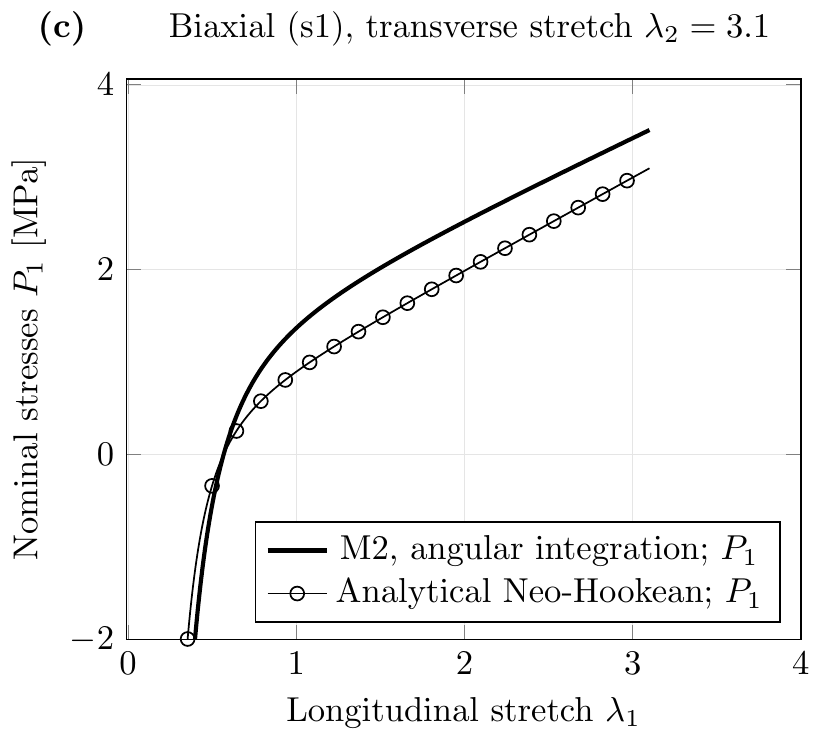}\quad
        \includegraphics[width=.4\textwidth]{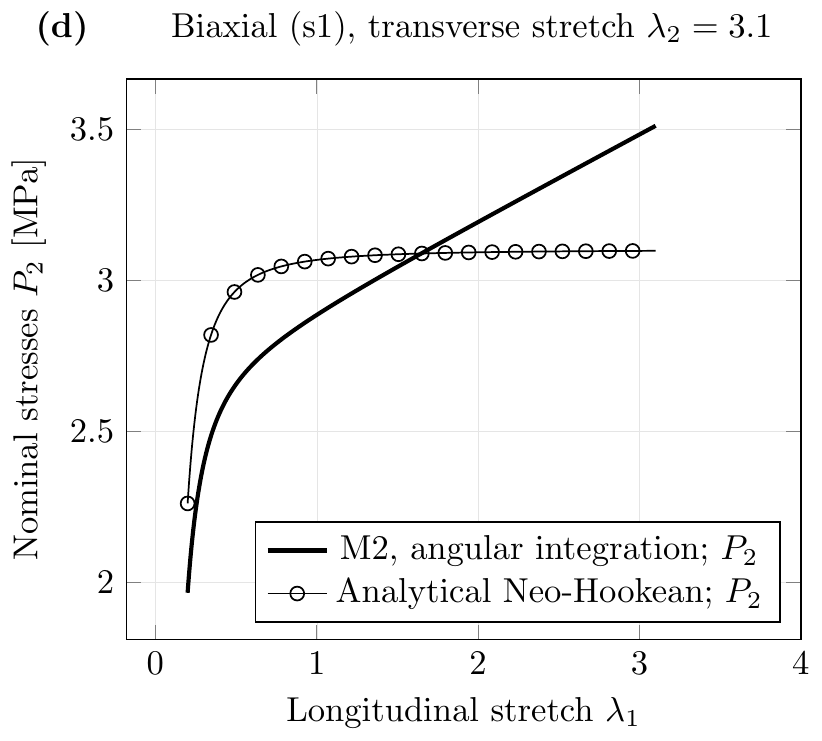}
        \caption{Comparison of material behavior when using model M2 through angular integrations with scaled modulus $\mu^{*}=\mu/0.7$ and the analytical Neo-Hookean model (equivalent to model M1). a) Uniaxial test (s2) in axis 1; $P_1$ vs. $\lambda_1$ and $P_2$ vs. $\lambda_{1}$. b) Uniaxial test (s2) Mooney plots. c) biaxial test (s1) in which $\lambda_2=3.1$ is fixed and $\lambda_1$ varies;  $P_1$ vs. $\lambda_1$. d) biaxial test (s1) in which $\lambda_2=3.1$ is fixed and $\lambda_1$ varies; $P_2$ vs. $\lambda_{1}$. }
        \label{expsmodslope}
        \end{minipage}
\end{figure}

Note also that during a tensile test, the apparent Young modulus is different for both models, despite that resemblance. Indeed, a perturbation analysis shows that both moduli are related by a factor of $0.7$, being $3\mu$ the Neo-Hookean case. However, even when both uniaxial moduli are made equal (scaling $\mu$ of M2 by $1/0.7$), the behaviour from both models is clearly different, see Fig. \ref{expsmodslope},  because the physics behind both models is different. An interesting observation to which we will refer in Sec. \ref{SEC:5} can also be made. In Fig. \ref{expsmodslope}a both models have the same initial slopes (from the constant $\mu$)  for the uniaxial test, and they keep a large similitude for the  quite large range shown. This results in also the same slopes ($\approx \mu $, see Eq. \eqref{Eq NH issues}) for the longitudinal curve $P_1-\lambda_1$ for $\lambda_1\gtrsim1$,  Fig. \ref{expsmodslope}c; but note that whereas model M2 has a positive $P_2-\lambda_1$ slope for the transverse behavior ($\lambda_1\gtrsim1$), Fig. \ref{expsmodslope}d, the Neo-Hookean model gets an almost vanishing one, see Eq. \eqref{Eq NH issues}. Remarkably, this means that the transverse tension in biaxial tests with a fixed transverse stretch ($\lambda_2\gtrsim1$)~ remains insensitive to longitudinal stretch increments (with $\lambda_1\gtrsim1$) already in the moderate tension ranges. This contradicts experimental observations (e.g. Table I in \cite{RivlinSaunders}; see also experiments of  \cite{Kawabata,Kawamura} and Sec. \ref{SEC:5}), which show an increase of tension from transverse stretching in the order of $P_2/P_1\approx 1/4$ (depending on the stretches), similar to that
shown by model M2.

Moreover, another important observation is given in the Mooney plot for the tensile test shown in Fig. \ref{expsmodslope}b. Mooney motivated his model $\Psi(I_1,I_2)=C_1(I_1-3)+C_2(I_2-3)$ in this type of plot, which for the uniaxial test results in
\begin{equation}
\frac{P_1}{2(\lambda_1-1/\lambda_1^2)} = C_1+\frac{C_2}{\lambda_1}\label{Mooneyplot}
\end{equation}
In Mooney's model, the constants are $C_1=\partial\Psi/\partial I_1$ and $C_2=\partial\Psi/\partial I_2$. As mentioned in the Introduction Section (claim C0) the classical statistical theory does not predict any term in $I_2$, so in Mooney's plot it shows an horizontal line at Mooney's $C_1$ value of $C_1=\mu/2$; see Fig. \ref{expsmodslope}b. However, experiments in polymers consistently show a positive slope, which in view of Eq. \eqref{Mooneyplot}, Mooney related to $C_2$. Using several types of tests, \cite{RivlinSaunders} did a detailed experimental analysis  of
the ratios $(\partial\Psi/\partial I_2) / (\partial\Psi/\partial I_1)$, which is $C_2/C_1$ in Mooney's model. \cite{RivlinSaunders} found a relation  $(\partial\Psi/\partial I_2) / (\partial\Psi/\partial I_1)=1/8$ for small stretches to $1/30$ for large stretches. Whereas from experiments it is seen that $\partial\Psi/\partial I_2$ has a strong dependence on  both $I_1$ and $I_2$ (it decreases for increasing values of the invariants), the $1/6$ to $1/30$ are values in the order of the slope of Fig. \ref{expsmodslope}b. However, \cite{RivlinSaunders} and \cite{Obata} warned about Mooney's simplistic interpretation of the almost straight slope as a validation of Mooney's model, because at small stretches, experiments show also a relevant dependence of $\partial\Psi/\partial I_1$ with deformation; see detailed discussion in \cite[Ch.10]{Treloar_book}. In this line, we must remark that the Mooney-Rivlin invariants where developed from the apparition of $I_1$ in the Gaussian-Neo-Hookean (affine) theory ($I2$ has been introduced to handle the ``small deviations''), but model M2 is not based on such invariants, so we should not expect a simple expression in terms of $I_1$ and $I_2$.

As a conclusion of this subsection, the Neo-Hookean model (hence model M1) is widely accepted to represent the behaviour of polymers at moderate stretches, those under which the Gaussian treatment is a good approximation. As most polymer network models, it uses two \textit{fundamental assumptions} (cf. \cite{Treloar_book} Sec. 4.2, pp. 60-61, assumptions \#4 and \#5): (a1) affine deformations in the network
(chain junction points move embedded in the continuum, so the change of length in the chains is that of the line segment in the continuum); (a2)
the entropy of the network is just the sum of the entropies of the individual chains (so the network configurational entropy  due to the non-isotropic distribution of the orientations of the chains from the affinity constraint, Secs. \ref{SEC:2-level entropy} and \ref{Sec:networkentropy}, is neglected). In contrast, model M2 brakes these assumptions, replacing them by: (Ra1) the network is free to fluctuate, the same way as the links in the polymer chain; (Ra2) there is no constraint to establish the  orientation of the chains,  meaning that chains will remain isotropically oriented.
The latter assumptions seem to include a proper dependence in $I_2$ in the statistical theory (Claim C0 in the Introduction), and comply better with experimental observations.

\subsection{The macro-micro-macro approach using different micro-macro connections\label{SEC:MMM}}

In \cite{MMM} we presented a Macro-Micro-Macro (MMM) approach to model polymers. The approach is an extension of WYPiWYG hyperelasticity (\cite{Sussman-Bathe,WYPIWYG,LatorreWYIPWYG}) to travel scales to reverse-engineer the chain behavior in a polymer from a single experimental test (e.g. a tensile test). Once that average chain behavior is obtained, it may be employed to simulate the behavior of the polymer under any (homogeneous or non-homogeneous) loading condition, either through angular integration or, more efficiently, through pre-computed constitutive manifolds, see \cite{MMM}.

\begin{figure}[htbp!]
        \begin{minipage}{\textwidth}
        \centering
        \includegraphics[width=.43\textwidth]{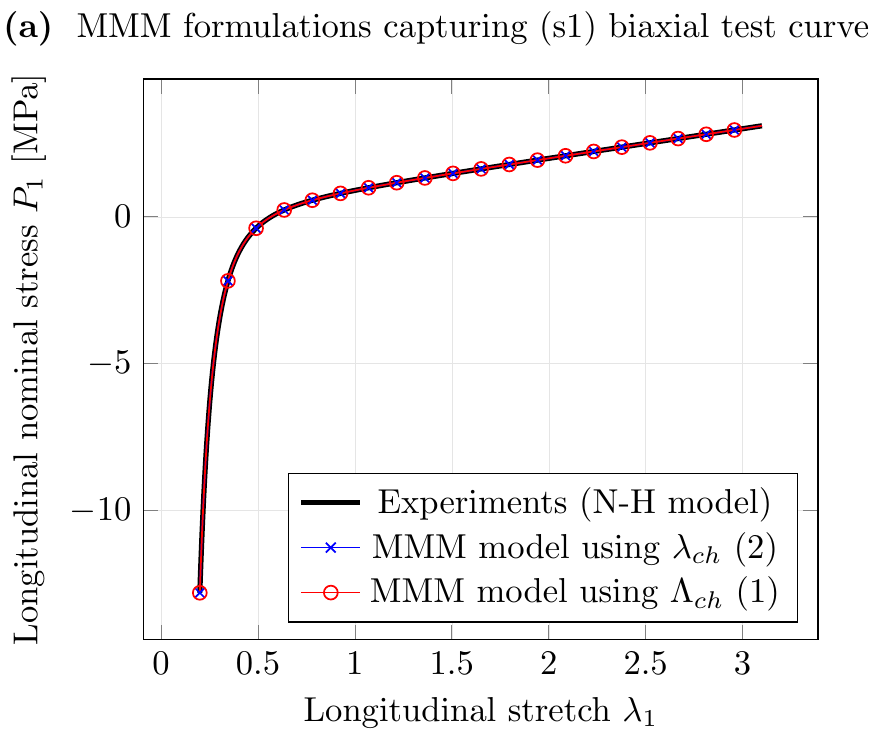}\quad
        \includegraphics[width=.4\textwidth]{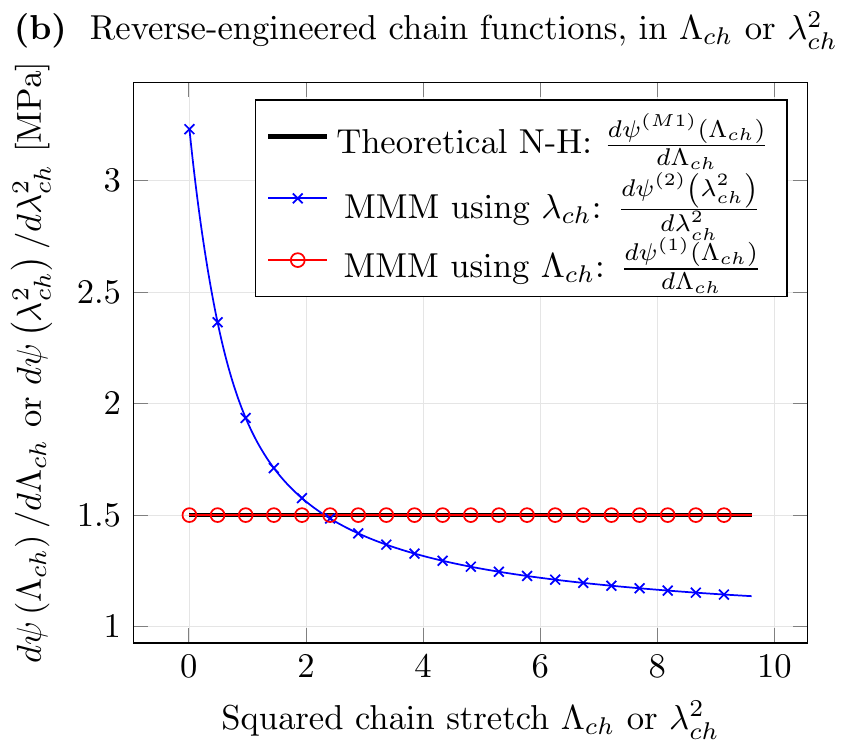}\\
        \includegraphics[width=.4\textwidth]{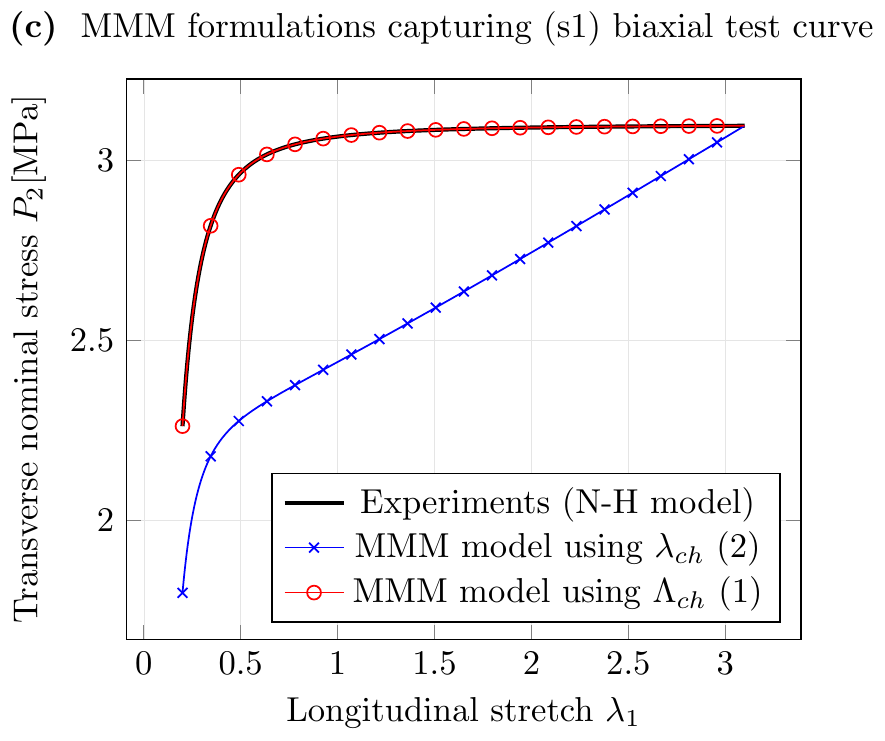}\quad
        \includegraphics[width=.4\textwidth]{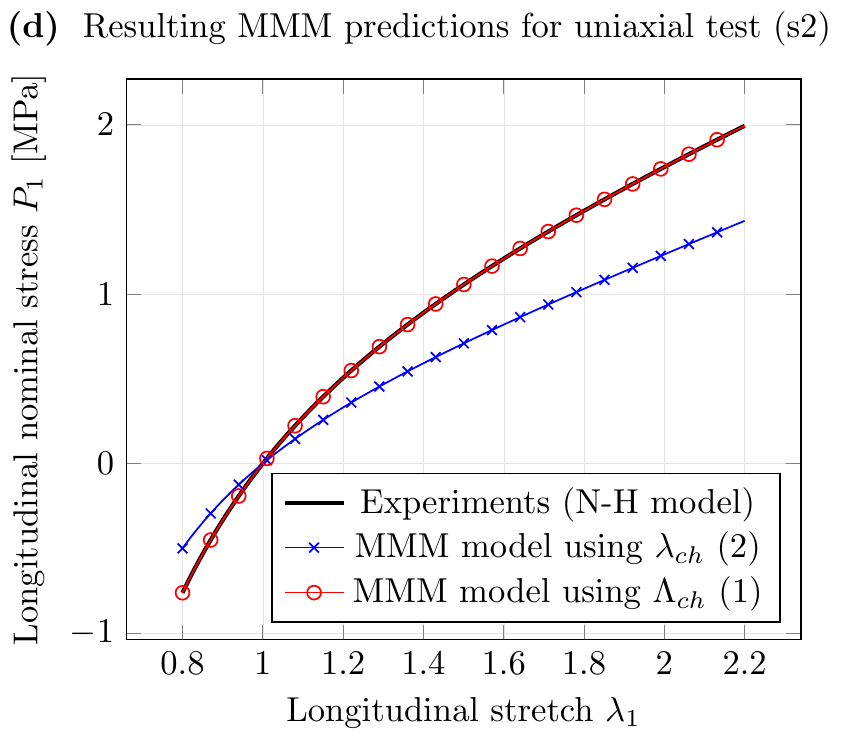}
        \caption{Macro-Micro-Macro reverse engineering of the Neo-Hookean model using only ``experimental''  data from a biaxial test with fixed $\lambda_2=3.1$  and varying $\lambda_1$. The two microstructural assumptions are employed in the MMM approach: (1) affine network deformations using $\Lambda_{ch}=\boldsymbol{C}:\boldsymbol{\hat{r}}\otimes\boldsymbol{\hat{r}}$; and (2) free network using $\lambda_{ch}=\boldsymbol{U}:\boldsymbol{\hat{r}}\otimes\boldsymbol{\hat{r}}$. a) Resulting predictions for the test employing to characterize the model (determine the chain function); note that both approaches capture exactly (numerically speaking) the prescribed test data. b) Reverse-engineered chain functions, compared to the theoretical Neo-Hookean one from which ``experimental'' data has been obtained; note that they are plotted in terms of squared stretches (modulus alike), namely $d\psi/d\lambda_{ch}=2\lambda_{ch} d\psi/d\lambda^2_{ch}$ is an increasing function. c) Theoretical and predicted transverse nominal stresses for the (s1) test; note that, as expected, the MMM model (1) also captures accurately the behavior, but the MMM model (2) fails to do so despite capturing exactly the behavior in the other axis. d) Predictions of both models with the captured chain functions for a different test: the uniaxial test; a similar conclusion is obtained as for case c).}
        \label{FIG:NHrecovery}
        \end{minipage}
\end{figure}

\begin{figure}[htbp!]
        \begin{minipage}{\textwidth}
        \centering
        \includegraphics[width=.4\textwidth]{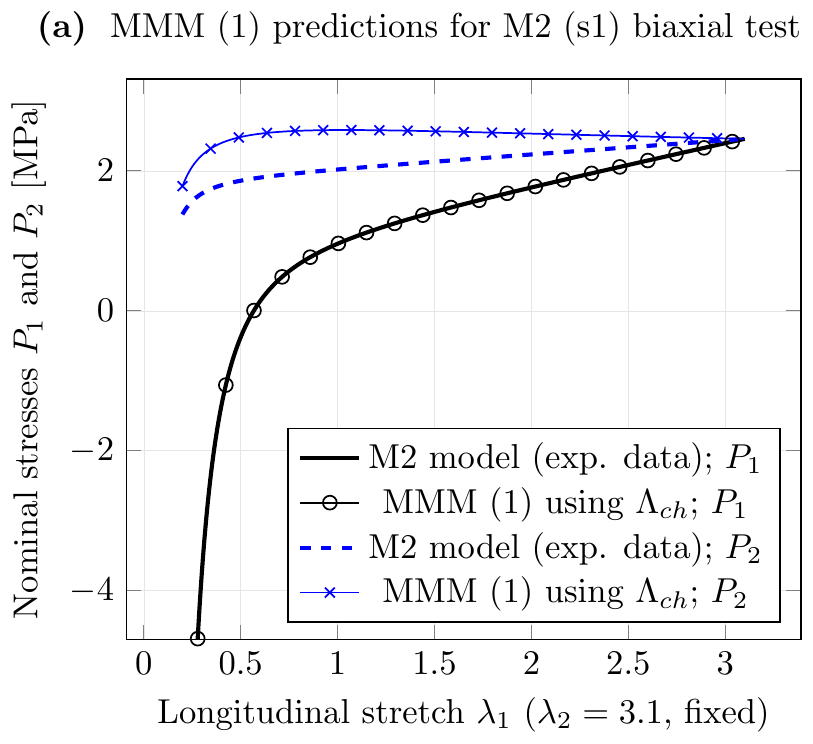}\quad
        \includegraphics[width=.4\textwidth]{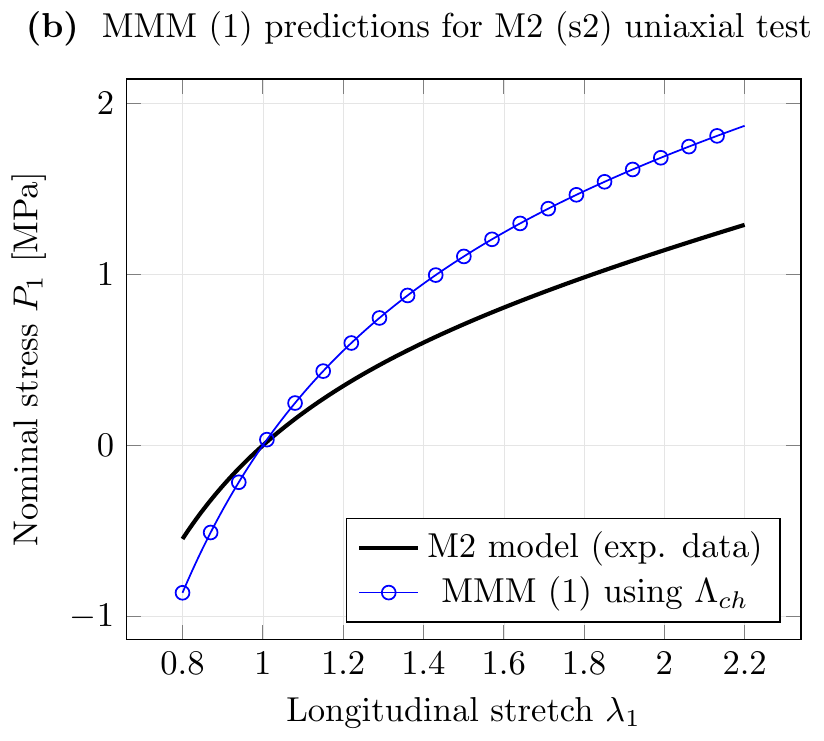}
        \caption{Reverse engineering of the M1 model of Eq. \eqref{M2} using the MMM approach with $\Lambda_{ch}$ (affine assumption). For all the plots in this figure, the MMM model is also calibrated from the $P_1-\lambda_1$ curve of the biaxial (s1) test (with $\lambda_2=3.1$ fixed). a) Predictions for the biaxial test (s1), both longitudinal $P_1$ and transverse $P_2$ nominal stresses. b) Predictions for the uniaxial (s2) test.}
        \label{FIG:M2recovery}
        \end{minipage}
\end{figure}

The MMM approach is quite general and simply hinges in a proper description of the microstructure of the material and its micro-macro connection. In \cite{MMM} we employed the same network layout and micro-macro connection as in model M2, namely
\begin{equation}
\Psi^{(2)}(\lambda_1,\lambda_2,\lambda_3) = \frac{1}{\Omega}\int_{\Omega}\psi_{ch}^{(2)}(\lambda_{ch}) d\Omega \label{M2AI}
\end{equation}
However, the same procedure may be applied employing the assumptions in model M1, namely
\begin{equation}
\Psi^{(1)}(\lambda_1,\lambda_2,\lambda_3) = \frac{1}{\Omega}\int_{\Omega}\psi_{ch}^{(1)}(\Lambda_{ch}) d\Omega \label{M1AI}
\end{equation}
We emphasize that with either assumption Eq. \eqref{M2AI} or Eq. \eqref{M1AI}, the respective chain energy function (the unknown function) may be obtained using the MMM procedure \emph{from a single test} (the known function) and that the chain behavior will be such that the presented test will be captured exactly if desired (numerically speaking) as happens with all WYPiWYG procedures. Importantly,  of course it does not mean that we will be able to capture the behavior of the material under any loading condition. The accuracy of such simulations will be parallel to the accuracy of the micro-macro description of the model, i.e. whether the assumptions (a1), (a2), (Ra1), (Ra2) are accurate. In Fig. \ref{FIG:NHrecovery} we show this observation. In that figure we use the ``experimental'' data from the analytical Neo-Hookean model for a biaxial test with a fixed $\lambda_2=3.1$ in axis 2, and a varying $\lambda_1$ in axis 1. It is seen in Fig. \ref{FIG:NHrecovery}a that both MMM models capture accurately such behavior, but the corresponding chain functions shown in Fig. \ref{FIG:NHrecovery}b (plotted in comparable squared stretches, modulus alike) are clearly different. Furthermore, having captured the test in Fig. \ref{FIG:NHrecovery}a means little about having characterized the actual behavior of the material under general deformations, as the transverse stresses in Fig. \ref{FIG:NHrecovery}c and the stresses for a uniaxial test in Fig. \ref{FIG:NHrecovery}d show.

Of course a minimum squares approach with several tests may be employed as typically done in determining material parameters of polymer models; overall better predictions will be obtained for the tests, but model MMM (2) will never be capable of capturing the Neo-Hookean model because the intrinsic structure of the model (i.e. assumptions (a1) and (a2)) are different.

At this point it should be evident to the reader that the opposite also holds. If we try to capture model M2, Eq. \eqref{M2}, with the MMM approach using $\Lambda_{ch}$, Eq. \eqref{M1AI}, we will be uncapable of obtaining satisfactory results. This observation is summarized in Fig. \ref{FIG:M2recovery}.

As a conclusion of this subsection to keep in mind for the next section, the MMM approach is capable of accurately capturing the chain behavior of a material if that material behaves according to the assumptions employed in the MMM formulation, but fails to give reasonably accurate results if these assumptions do not hold.

\section{Comparison of assumptions against experimental data in polymers\label{SEC:5}}

With the observations from the previous section, to capture experimental data of actual polymer materials we now employ the MMM method using both assumptions: unconstrained isotropic network configuration (use of $\lambda_{ch}$) and constrained affinely oriented chains (use of $\Lambda_{ch}$). The experimental data used for this study are the true biaxial family of tests from  \cite{Kawabata}. This material is almost identical to that from  \cite{Treloar_exp} (same composition of unfilled cross-linked rubber with 8ph sulfur, and very close experimental results for uniaxial, equibiaxial and pure shear tests), whose experiments are used in most works to test the predictive power of material models. Remarkably, \cite{Kawabata} experiments are complete for an incompressible material, because they are true biaxial tests in which both $\lambda_1$ and $\lambda_2$ are prescribed independently ($\lambda_3=1/(\lambda_1\lambda_2$), so in those experimental sets all possible states of deformation are tested (obviously within the experimental range of deformations). Hence, if a model is capable of accurately predicting the behavior of all \cite{Kawabata} sets of experiments we can claim that the model represents accurately the material behaviour under any loading condition.

\begin{figure}[htbp!]
        \begin{minipage}{\textwidth}
        \centering
        \includegraphics[width=.4\textwidth]{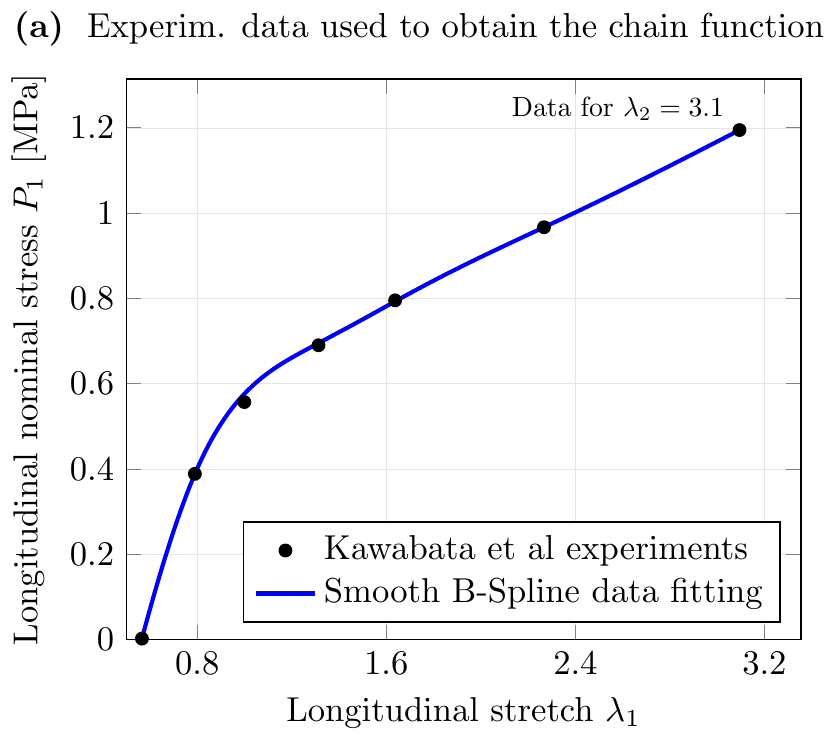}\quad
        \includegraphics[width=.4\textwidth]{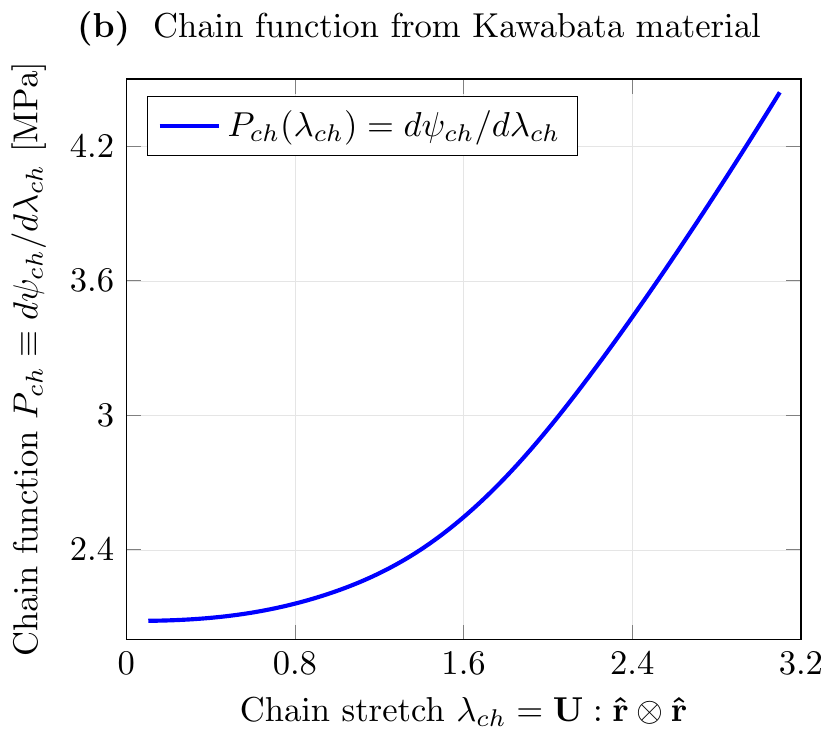}
        \caption{Reverse-engineering of the chain function from Kawabata et al material. (a) Experimental data from Kawabata et al for $\lambda_2=3.1$ employed in the determination of the chain function, and smooth spline to interpolate experimental data in the MMM procedure; only these data from a single test curve have been employed to obtain the chain function. (b) Resulting chain function for the Kawabata material under the assumption of isotropic chain distribution during deformation (non-affine, unconstrained orientation of network chains)}
        \label{FIG:Pchkawabata}
        \end{minipage}
\end{figure}

\begin{figure}[htbp!]
        \begin{minipage}{\textwidth}
        \centering
        \includegraphics[width=.43\textwidth]{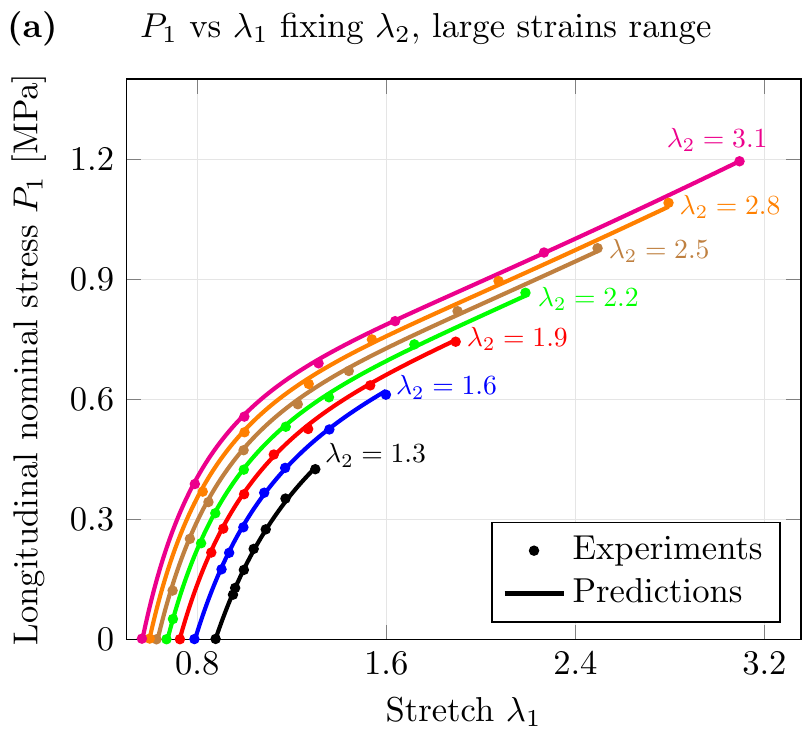}\quad
        \includegraphics[width=.43\textwidth]{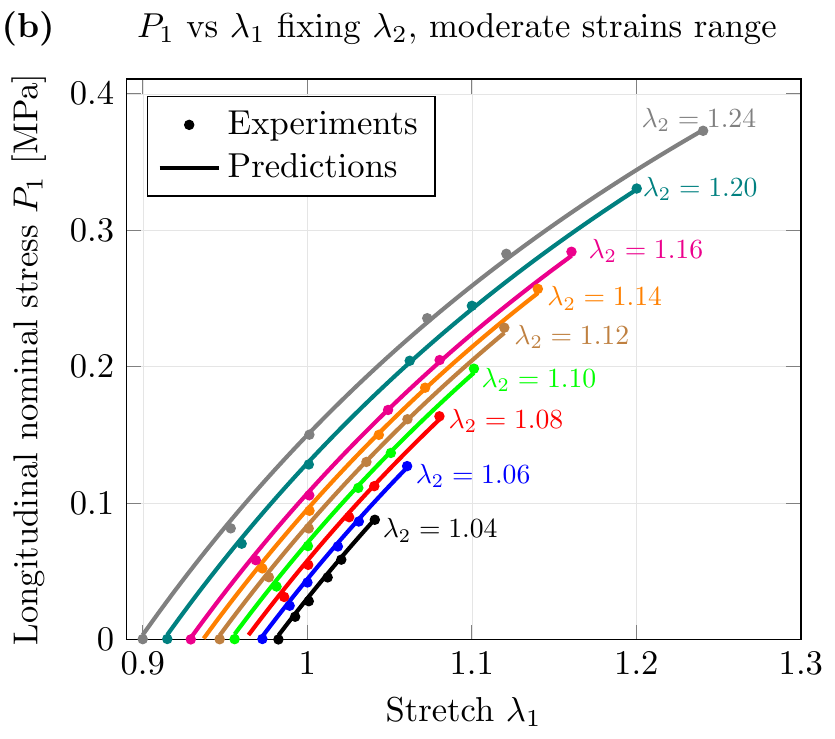} \\
        \includegraphics[width=.43\textwidth]{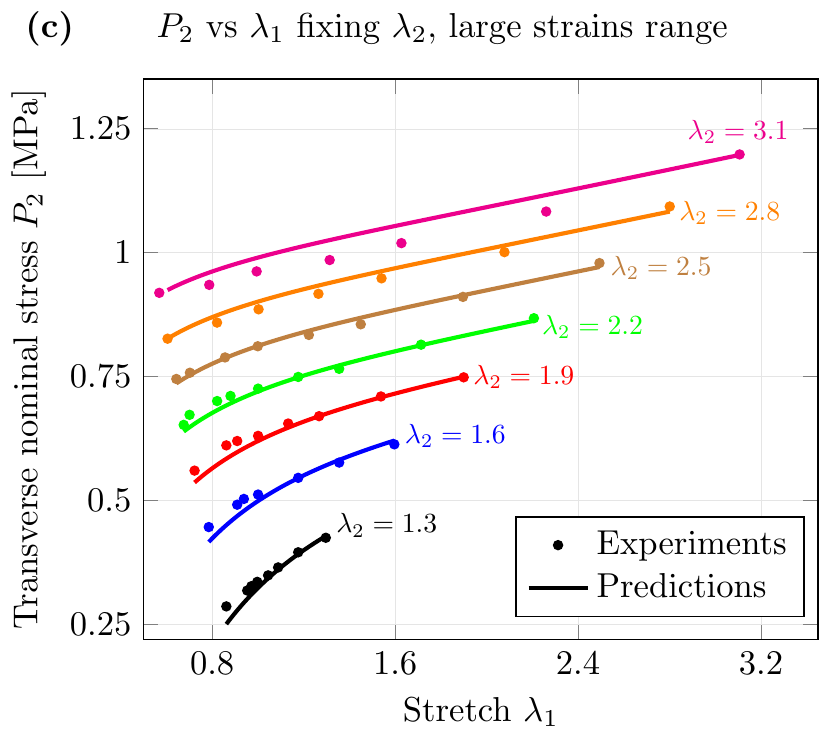} \quad
        \includegraphics[width=.43\textwidth]{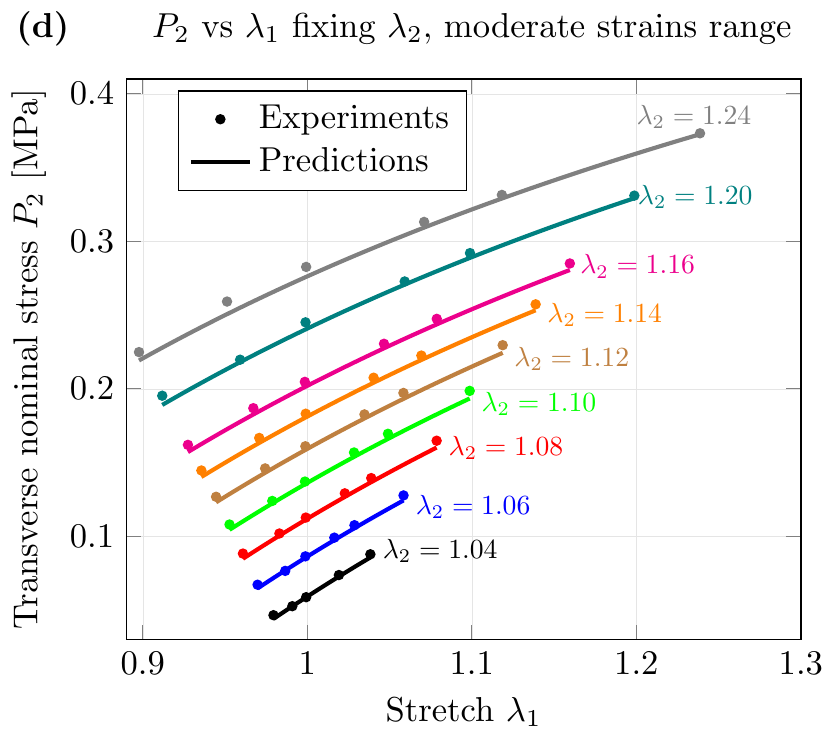} 
        \caption{Predictions of the Kawabata et al biaxial experiments \cite{Kawabata} using the present macro-micro-macro proposal \emph{under the non-affine orientational assumption}, where $P_{ch}(\lambda_{ch})$ has been characterized from the $P_1(\lambda_1)$ curve for $\lambda_2=3.1$. Dots represent experimental data, whereas continuous curves represent the predictions. (a) Longitudinal nominal stresses $P_1$ as a function of the longitudinal stretch $\lambda_1$ for different fixed values of the transverse stretch $\lambda_2$; large stretches range. (b) Longitudinal nominal stresses $P_1$ as a function of the longitudinal stretch $\lambda_1$ for different fixed values of the transverse stretch $\lambda_2$; moderate stretches range, shown in a different plot for legibility. (c) Transverse nominal stresses $P_2$ as a function of the longitudinal stretch $\lambda_1$ for different fixed values of the transverse stretch $\lambda_2$; large stretches range. (d) Transverse nominal stresses $P_2$ as a function of the longitudinal stretch $\lambda_1$ for different fixed values of the transverse stretch $\lambda_2$; moderate stretches range, shown in a different plot for legibility.}
        \label{P2kawabatafirstrank.tikz}
        \end{minipage}
\end{figure}

%

\begin{figure}[htbp!]
        \begin{minipage}{\textwidth}
        \centering
        \includegraphics[width=.43\textwidth]{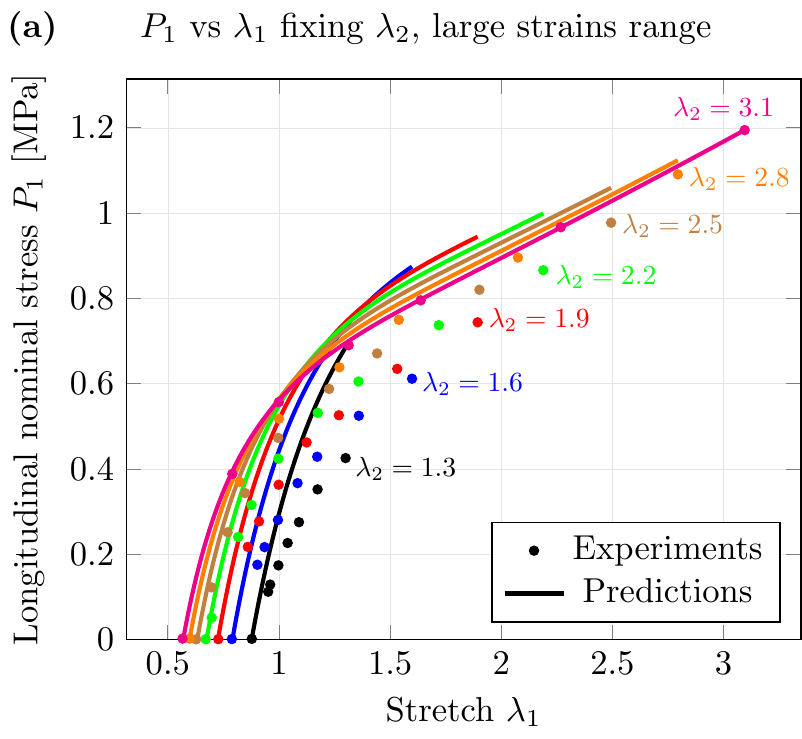}\quad
        \includegraphics[width=.43\textwidth]{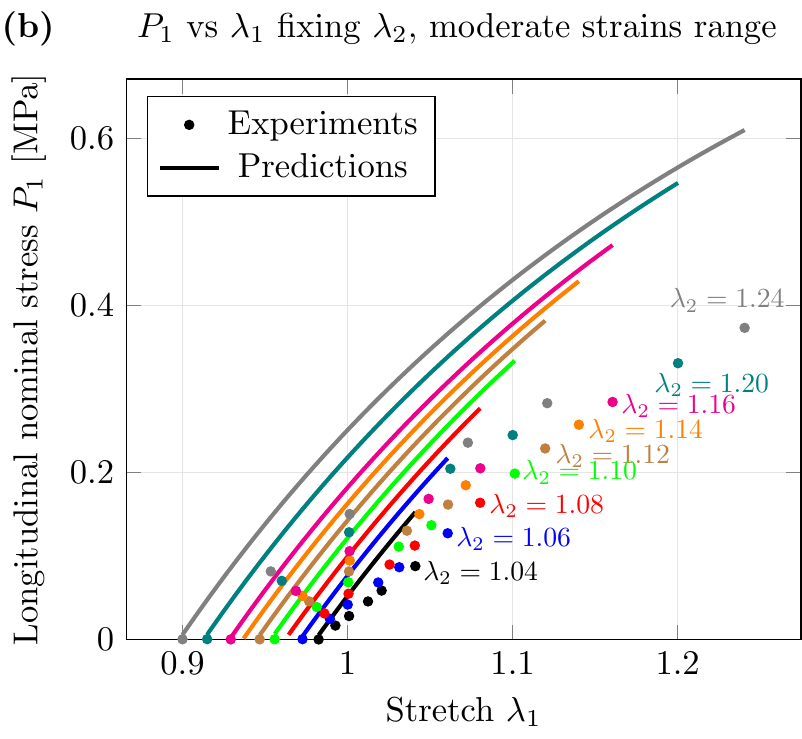} \\
        \includegraphics[width=.43\textwidth]{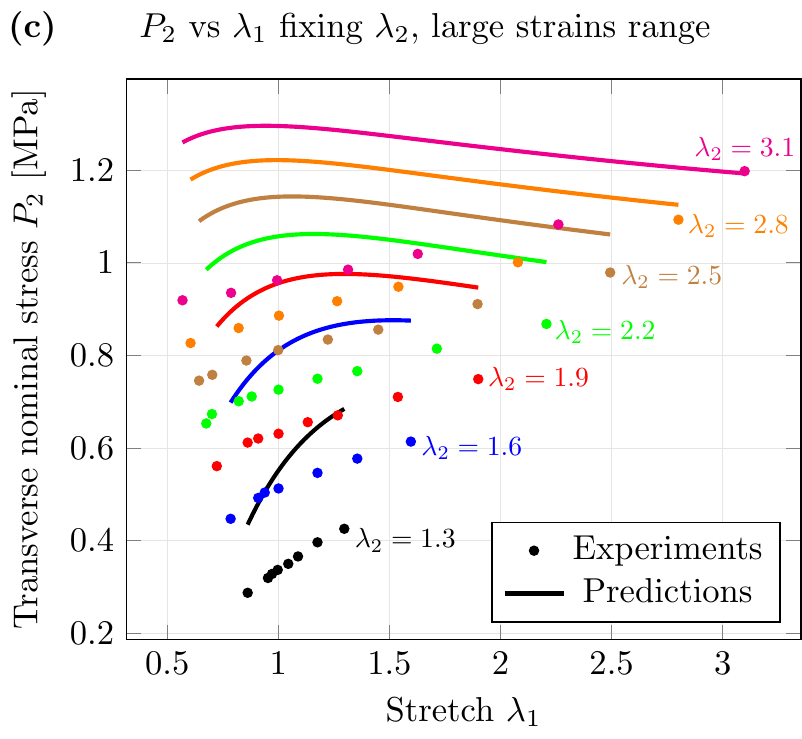} \quad
        \includegraphics[width=.43\textwidth]{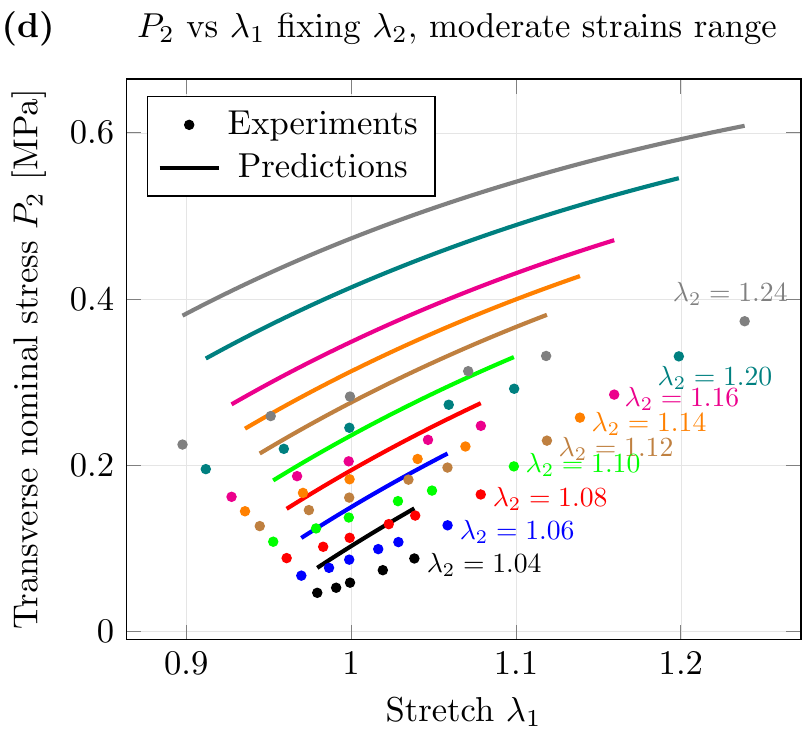} 
        \caption{Predictions of the Kawabata et al biaxial experiments \cite{Kawabata} using the present macro-micro-macro proposal \emph{under the affine orientational assumption}, where the chain function $P_{ch}(\lambda_{ch}^C)$ has been characterized from the $P_1(\lambda_1)$ curve for $\lambda_2=3.1$. Dots represent experimental data, whereas continuous curves represent the predictions. (a) Longitudinal nominal stresses $P_2$ as a function of the longitudinal stretch $\lambda_1$ for different fixed values of the transverse stretch $\lambda_2$; large stretches range. (b) Longitudinal nominal stresses $P_1$ as a function of the longitudinal stretch $\lambda_1$ for different fixed values of the transverse stretch $\lambda_2$; moderate stretches range, shown in a different plot for legibility. (c) Transverse nominal stresses $P_2$ as a function of the longitudinal stretch $\lambda_1$ for different fixed values of the transverse stretch $\lambda_2$; large stretches range. (d) Transverse nominal stresses $P_2$ as a function of the longitudinal stretch $\lambda_1$ for different fixed values of the transverse stretch $\lambda_2$; moderate stretches range, shown in a different plot for legibility.}
        \label{P2kawabatafirstrankusingLambdaC}
        \end{minipage}
\end{figure}


We first address the characterization of the chain function \emph{using a single test curve}, namely the curve for fixed $\lambda_2=3.1$ and varying $\lambda_1$. This curve consists of \emph{just} the \emph{seven experimental points} shown in Fig. \ref{FIG:Pchkawabata}a. With these seven experimental points (interpolated through the B-spline also shown in Fig. \ref{FIG:Pchkawabata}a), using the MMM method \citep{MMM}, we reverse-engineer the chain function $d\psi_{ch}/d\lambda_{ch}$, a result shown in Fig. \ref{FIG:Pchkawabata}b. With this chain function, via angular integration, we predict all the tests from \cite{Kawabata} both the longitudinal stress $P_1$ and the transverse stress $P_2$, and for both the small stretch range and the large stretch range. Figure \ref{P2kawabatafirstrank.tikz} shows the comparison of the predictions against experimental data. Remarkably, from just the mentioned seven experimental points used to characterize the material, we are capable of accurately predicting the behavior of the material under any loading condition. Indeed, this accuracy does not only hold for the \cite{Kawabata} material: similar accuracy has been obtained for different types of silicones tested also through true biaxial tests by \cite{Kawamura}; see predictions in \cite{MMM}. We have used $ 21$ quadrature points as in \cite{Bazant}, but we note that in our case no relevant difference is appreciated when using other more intensive quadrature rules, see discussion in \cite{MMM}.

From Fig. \ref{P2kawabatafirstrank.tikz} we can conclude that the assumption of unconstrained network chain orientations, or equivalently the use of $\lambda_{ch}=\boldsymbol{U}:\boldsymbol{\hat{r}}\otimes\boldsymbol{\hat{r}}$ as state variable to determine the chain stretch, seems adequate to describe the polymer behavior under general state of deformations. However, one may question that the predictive capability may de attributed to the MMM approach  and its suitability to reverse-engineer the chain behavior, and not to the mentioned assumption. To verify this, we have also employed the MMM approach using the affinity assumption for the orientation of the network chains; equivalently $\Lambda_{ch}$. Following the same procedure, with the same seven mentioned experimental points and with the same interpolating B-spline, we reverse-engineered the chain function $d\psi_{ch}(\lambda_{ch}^C)/d\lambda_{ch}^C$. With that chain function, via angular integration, we performed the predictions for the \cite{Kawabata} experiments. This comparison is shown in Fig. \ref{P2kawabatafirstrankusingLambdaC}. It is observed that the MMM model using the affinity assumption gives large errors, except for the $P_1-\lambda_1$ curve with $\lambda_2=3.1$ which is indeed the curve employed to characterize the material. As in Fig. \ref{FIG:NHrecovery}, the prediction for the equibiaxial test at $\lambda_1=\lambda_2=3.1$ is accurate for both $P_1$ and $P_2$ (in plane chain orientations in this case are isotropic for both models, and the materials behave isotropically). From the conclusions of Sec. \ref{SEC:MMM}, the MMM model can only capture the behavior of a polymer under any loading conditions if the micro-macro assumptions represent properly the actual behavior, so the comparison of Figs. \ref{P2kawabatafirstrank.tikz} and \ref{P2kawabatafirstrankusingLambdaC} seem to favor the unconstrained (non-affine, isotropic) chain orientations in the network. Moreover, it is interesting to compare the shape of Fig. \ref{FIG:Pchkawabata}a (reproduced now by both MMM approaches accurately) to that of the $P_1-\lambda_1$ curves in Fig. \ref{expsmodslope}b for $\lambda_1\gtrsim1$: they have an almost constant slope. Recalling the comments regarding the slopes of transverse curves $P_2-\lambda_1$, the affine model resulted in a vanishing slope, whereas the non-affine one maintained a possitive slope. Note that the transverse \emph{experimental} data $P_2-\lambda_1$ in Fig. \ref{P2kawabatafirstrank.tikz}c has a positive (almost constant) slope, easily accomodated by the non-affine model. In contrast,  the affine model results in a negative slope for that curve.

However, even though sometimes questioned, the affine deformations approach has been employed by many researches in a large amount of material models for almost a century. Albeit not  accurate,  somewhat reasonable predictions using this assumption have been obtained for many materials, including the Kawabata et al material, so the results in Fig. \ref{P2kawabatafirstrankusingLambdaC} seem to contradict the vast experience using the affinity assumption. The poor results in Fig. \ref{P2kawabatafirstrankusingLambdaC} could be attributed, for example, to inadequacy of the MMM approach or to the lack of sophistication in the model. Such type of sophistication includes, for example,  constraining tubes for chains (changing the assigned surface) or the non-affinity in the chain stretch (free fluctuating field as in the microsphere model), among others. The qualitative observations regarding slopes could also be attributed to the limit in the validity of the Gaussian theory.

\begin{figure}[htbp!]
        \begin{minipage}{\textwidth}
        \centering
        \includegraphics[width=.43\textwidth]{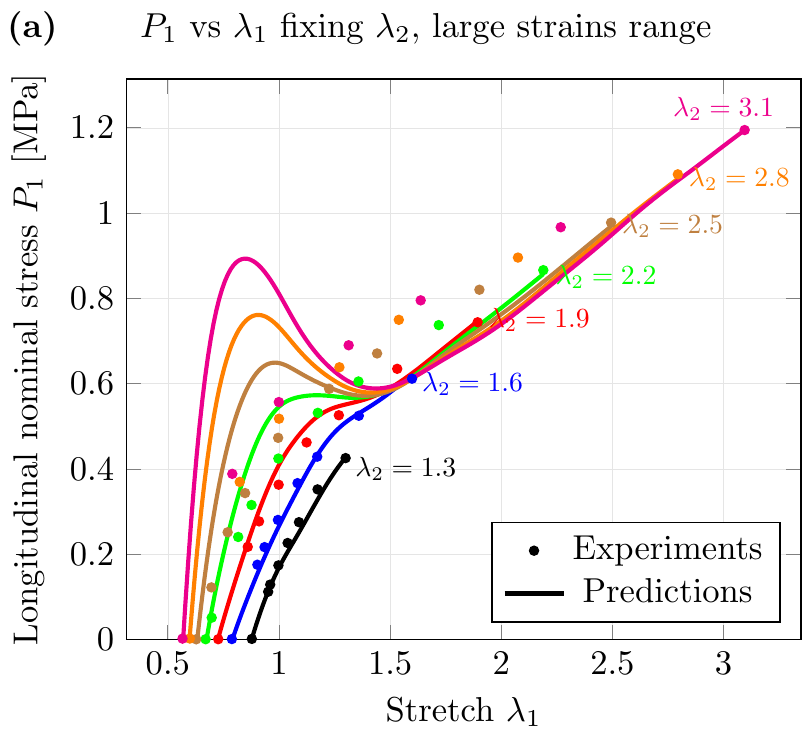}\quad
        \includegraphics[width=.43\textwidth]{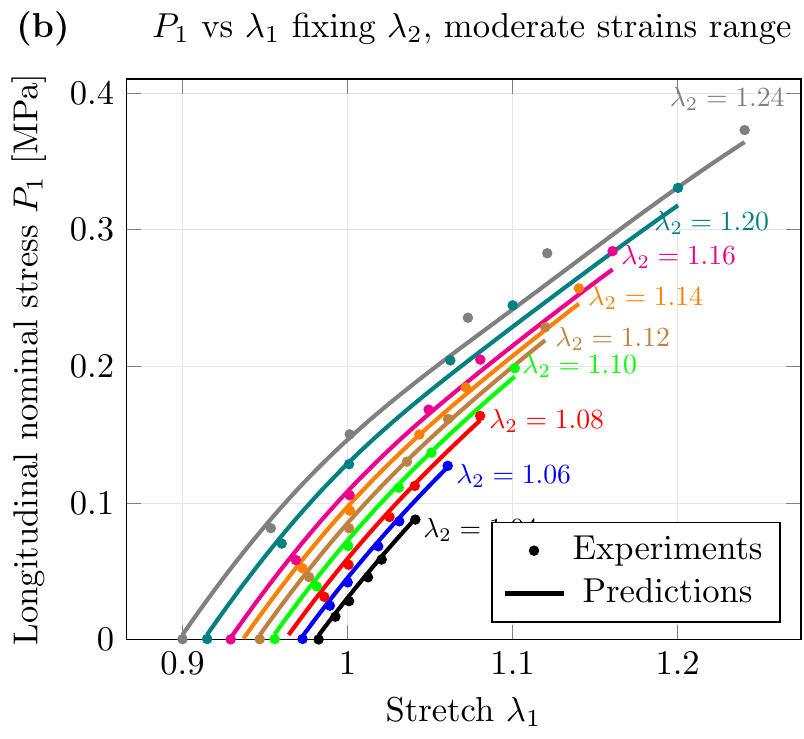} \\
        \includegraphics[width=.43\textwidth]{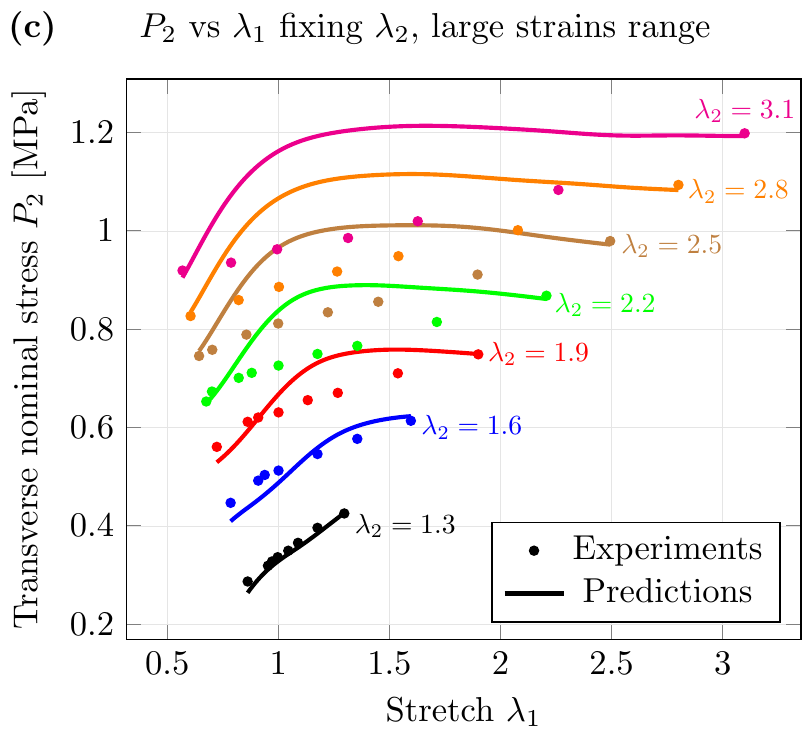} \quad
        \includegraphics[width=.43\textwidth]{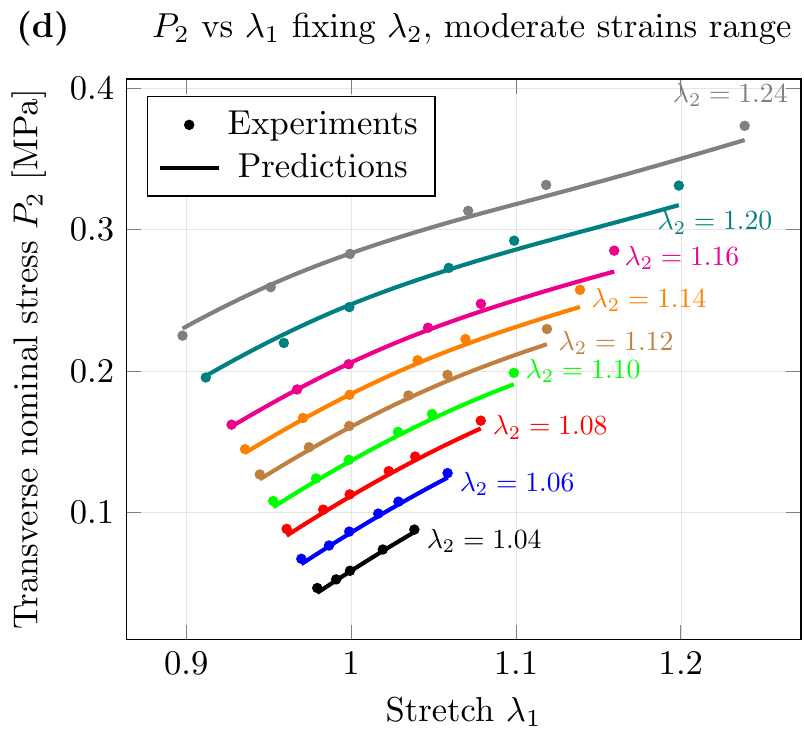} 
        \caption{Predictions of the Kawabata et al biaxial experiments \cite{Kawabata} using the present macro-micro-macro proposal  \emph{under the affine orientational assumption}, where the chain function $P_{ch}(\lambda_{ch}^C)$, where the chain function
 has been characterized from the uniaxial test. Dots represent experimental data, whereas continuous curves represent the predictions. (a) Longitudinal nominal stresses $P_1$ as a function of the longitudinal stretch $\lambda_1$ for different fixed values of the transverse stretch $\lambda_2$; large stretches range. (b) Longitudinal nominal stresses $P_1$ as a function of the longitudinal stretch $\lambda_1$ for different fixed values of the transverse stretch $\lambda_2$; moderate stretches range, shown in a different plot for legibility. (c) Transverse nominal stresses $P_2$ as a function of the longitudinal stretch $\lambda_1$ for different fixed values of the transverse stretch $\lambda_2$; large stretches range. (d) Transverse nominal stresses $P_2$ as a function of the longitudinal stretch $\lambda_1$ for different fixed values of the transverse stretch $\lambda_2$; moderate stretches range, shown in a different plot for legibility.}
        \label{P2kawabatafirstrankusingLambdaCuniaxial}
        \end{minipage}
\end{figure}

To further investigate the reason behind  the poor predictions in Fig. \ref{P2kawabatafirstrankusingLambdaC}, we have repeated the process employing in this case the uniaxial tension-compression test data from \cite{Kawabata} to reverse-engineer the chain function of the affine model. The data corresponding to the Kawabata uniaxial test, is extracted from the same set experiments given in Fig. \ref{P2kawabatafirstrankusingLambdaC}. The predictions of the \cite{Kawabata} experiments using the new chain function determined using the uniaxial data, are given in Fig. \ref{P2kawabatafirstrankusingLambdaCuniaxial}.

%

These plots show better results than those in Fig. \ref{P2kawabatafirstrankusingLambdaC}. However, it is observed that the good accuracy is obtained only for small-to-moderate stretches, being predictions for large stretches disappointing. Furthermore, the observation regarding the slopes still holds. Note that now both ends of the curves are prescribed in the process (they correspond to the prescribed B-spline from uniaxial tension-compression or tensile-equibiaxial tests), so the model tries to comply with that data, whatever the resulting chain function is. Our explanation to the improved results of Fig. \ref{P2kawabatafirstrankusingLambdaCuniaxial} over Fig.  \ref{P2kawabatafirstrankusingLambdaC} comes from the configuration from which the chain function has been reverse-engineered. The uniaxial test departs from an unloading configuration. In that configuration the affinity assumption has an isotropic network distribution of chains, so no conclusion may be extracted regarding the accuracy of either assumption. Only at large stretches the distribution becomes strongly anisotropic (specially during the tensile test). Then, when the chain function is reverse-engineered from a tensile test, moderate strain results will be very similar under both orientational assumptions, but at very large strains, differences in orientational distributions are significative, see Fig. \ref{chains_in_spherev0}e, so are the differences between both models. These comments are indeed behind the reason of using the biaxial test curve with a fixed $\lambda_2=3.1$ and variable $\lambda_1$ to reverse-engineer the chain function in the paper. For this test curve the orientational distribution of the chains for the affine assumption is strongly anisotropic, so if this curve is used, differences between both approaches are emphasized.
This observation may be in connection with the \cite{Microsphere} claim quoted as (C2)  in Sec. \ref{SEC:1} regarding the need for the non-affine correction to $\lambda_{ch}^C$ especially in the range of large deformations.

We recall that, to date, no micromechanical model under the assumption of affine orientation of chains has been capable of reproducing accurately all the tests of the \cite{Kawabata} material (or similar polymer) using only one test curve to characterize the model. When using several test curves to characterize the chain behavior (material parameters), as done in  \cite{Verron}, the different orientational distributions from the tests are somehow implicitly averaged, so the resulting chain function performs better over all ranges of deformations than when using a single one. However, this does not validate any assumption on the microstructural arrangement of the chains, because for example, Ogden's phenomenological model \citep{Ogden_model} gives similar or better performance using just two tests \citep{Understanding,OSS}. For this reason, it is widely used to characterize polymeric materials and available in most finite element codes. Indeed, as mentioned in the Introduction Section, Ogden's and Shariff's \citep{Shariff} models, both phenomenological models based in the  \cite{VL} decomposition, where best performers in the comparison made by \cite{Verron} for predicting the \cite{Kawabata} experiments. For example, the results shown in Fig. 3 of \cite{Verron} look marginally better than those using the \cite{Microsphere} microsphere model (Fig. 8 in \cite{Verron}); compare also Figs. 4, 11-13 of \cite{Itskov}.

The excellent performance of Ogden's and Shariff's models deserves some study here. These models are  based on the \cite{VL} decomposition:
\begin{equation}
\Psi(\lambda_1,\lambda_2,\lambda_3)=\omega(\lambda_1)+\omega(\lambda_2)+\omega(\lambda_3)
\end{equation}
where $\omega(\lambda)=\sum_{k=1}^{\infty}(\mu_k/\alpha_k)(\lambda^{\alpha_k}-1)$ is the particular fitting function chosen by Ogden; Shariff's model just changes the fitting function. Valanis and Landel already made an assessment of the decomposition using several types of tests concluding its suitability for a moderately large range of stretches (see additional validations in \cite{Obata,Treloar_book}). The same \cite{VL} decomposition has been used in the WYPiWYG models \citep{Sussman-Bathe,WYPIWYG,Understanding,OSS}, where the function $\omega(\lambda)$ is obtained in spline form, with a numerically exact fit just employing the \cite{KZ} series or solving a system of equations.

\begin{figure}[htbp!]
        \begin{minipage}{\textwidth}
        \centering
        \includegraphics[width=.43\textwidth]{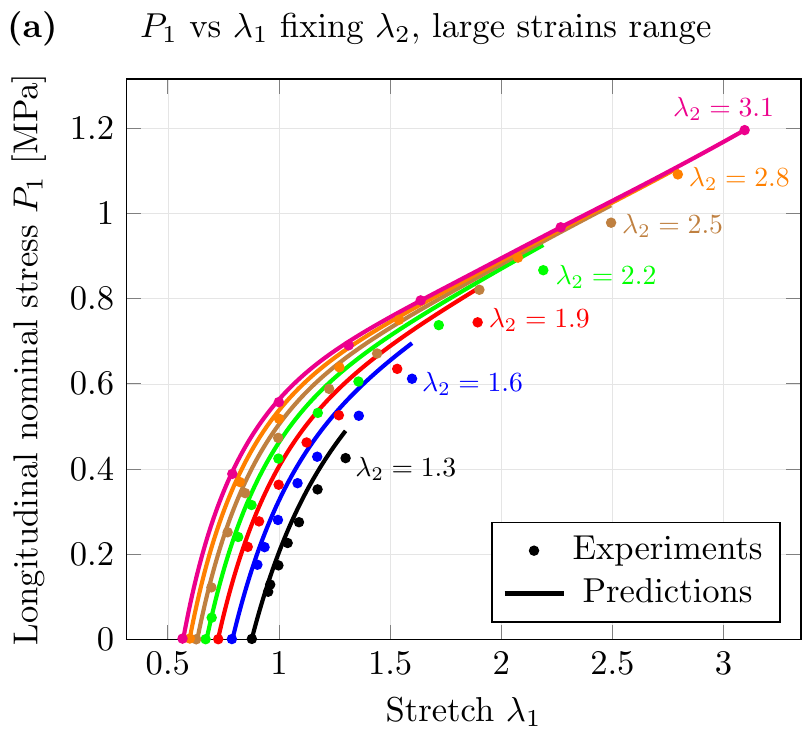}\quad
        \includegraphics[width=.43\textwidth]{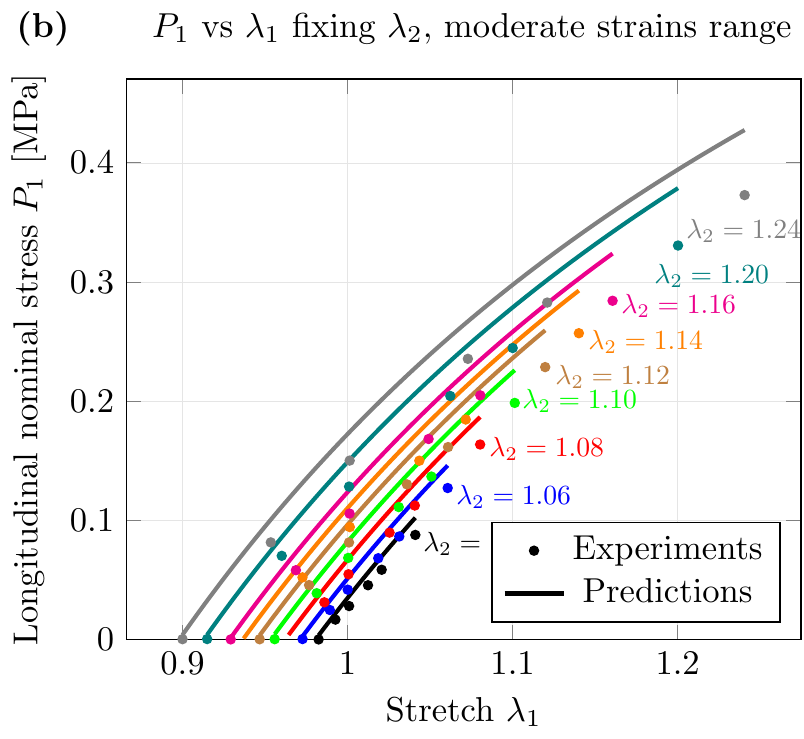} \\
        \includegraphics[width=.43\textwidth]{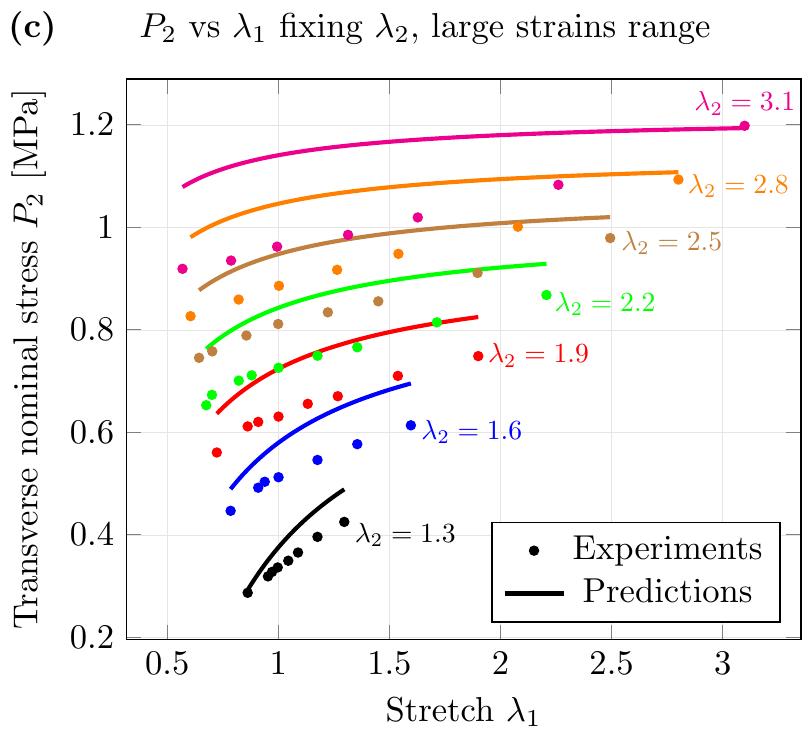} \quad
        \includegraphics[width=.43\textwidth]{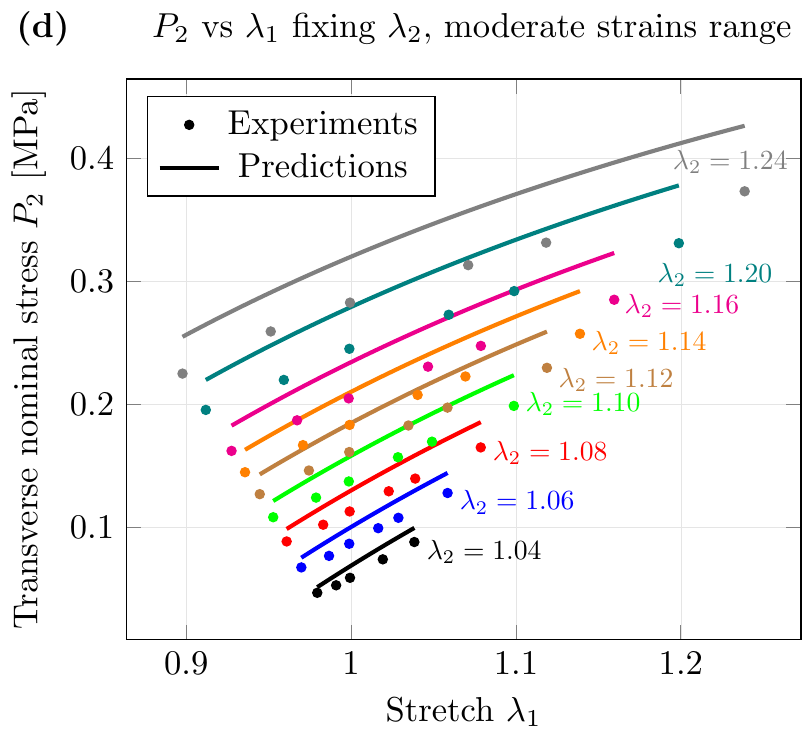} 
        \caption{Predictions of the Kawabata et al biaxial experiments \cite{Kawabata} using the present macro-micro-macro proposal \emph{with only three integration points oriented at the principal continuum deformation directions}, where $P_{ch}(\lambda_{ch})$ has been characterized from the $P_1(\lambda_1)$ curve for $\lambda_2=3.1$. Dots represent experimental data, whereas continuous curves represent the predictions. (a) Longitudinal nominal stresses $P_1$ as a function of the longitudinal stretch $\lambda_1$ for different fixed values of the transverse stretch $\lambda_2$; large stretches range. (b) Longitudinal nominal stresses $P_2$ as a function of the longitudinal stretch $\lambda_1$ for different fixed values of the transverse stretch $\lambda_2$; moderate stretches range, shown in a different plot for legibility. (c) Transverse nominal stresses $P_2$ as a function of the longitudinal stretch $\lambda_1$ for different fixed values of the transverse stretch $\lambda_2$; large stretches range. (d) Transverse nominal stresses $P_2$ as a function of the longitudinal stretch $\lambda_1$ for different fixed values of the transverse stretch $\lambda_2$; moderate stretches range, shown in a different plot for legibility.}
        \label{P23intpoint}
        \end{minipage}
\end{figure}

Emphasis in using as many tests as possible in fitting parameters of models is continuously placed in the literature e.g. \cite{Pancheri_Dorfmann,Verron,OSS,Criscione,Itskov,Urayama};\ claim (C3) in Sec. \ref{SEC:1}.  However, we can consider Valanis-Landel models as three-chain models, whose chains are aligned with the principal directions of deformation. The weight for each chain is $4\pi/3$. Indeed, the model is also the same as an angularly integrated model in which only three integration points are used. The direction vectors of those integration points are $\boldsymbol{\hat{n}}_i$, the principal directions of deformation. Then, we can apply our MMM method, with just 3 integration points, to obtain the chain function for this model in the same way as we did before. We use once again the same 7 experimental points from the $P_1-\lambda_1$ curve with $\lambda_2=3.1$ to reverse-engineer the chain function $P_{ch}(\lambda_{ch})=3/(4 \pi)\omega'(\lambda)$. With this chain function, we repeat the predictions for the \cite{Kawabata} experiments using this ``three-chain'' model. The results are shown in Fig. \ref{P23intpoint}.

%

%
Interestingly, despite using only one test curve, the same that resulted in poor predictions for the affine model, the predictions for this three chain model are rather good in both $P_1$ and $P_2$, and in both the low stretch and large stretch ranges,  with maximum errors of about $20\%$ and keeping the tendencies of the experimental curves. Indeed, Fig. \ref{P23intpoint}a may be compared with the results of the 8-chain model in Fig. 6 of \cite{Verron}. Whereas the 8-chain model gives good predictions for the low stretch region, it fails to give reasonable predictions for the true biaxial tests in the large stretch region, despite capturing simultaneously a tensile test  and an equibiaxial test rather well, as Fig. 6 of \cite{Verron} shows. Hence, this may be an explanation of the success of both the Ogden and the Shariff models.

However, the predictions in Fig. \ref{P23intpoint} are much more interesting if they are compared to those given in Figs. \ref{P2kawabatafirstrank.tikz} and \ref{P2kawabatafirstrankusingLambdaC}. For computing the 3-chain model in Fig. \ref{P23intpoint} we have not specified whether we used the affinity assumption or not, simply because in this 3-chain model both assumptions are essentially the same because the three chains are aligned with the principal directions of deformation (for these three chains in the model $\lambda^2_{ch}=\Lambda_{ch}$). Then, Fig. \ref{P2kawabatafirstrank.tikz} may be seen as the improvement over Fig. \ref{P23intpoint} when more chains (or integration points) are included in the model following the unconstrained network assumption. In contrast, Fig. \ref{P2kawabatafirstrankusingLambdaC} may be seen as the improvement over Fig. \ref{P23intpoint} when more chains (or integration points) are included in the model following the affine assumption. From the comparisons, it is clear that in the first case the results improve as expected, but in the second case, including more chains worsens the accuracy.
This is in line with the observations respect to the affine full network model (claim C1 in the Introduction).

\section{Conclusions}

In this paper we questioned the network affinity assumption in polymers respect to the orientational distribution of the chains during deformations. This assumption is tied to the use of the stretch from the right Cauchy-Green deformation tensor as the basic state variable (the micro-stretch) in the chain. These ingredients are employed by most polymer network and soft tissue models.

We have addressed the network configurational entropy to be added to the entropy of the individual chains. We observed that  the affinity assumption introduces an additional stress if the changes of entropy due to such assumption are considered. This additional stress is typically neglected.

We have shown that the use of an unconstrained configuration assumption, in line with a negligible network configuration stress, results in an isotropic distribution of the orientation of chains during deformation. This average orientation of the chains is equivalent to employing the chain-projected stretch from the stretch tensor as the state variable for the stretch function. In contrast, the affinity assumption employs the chain-projected stretch from the Cauchy-Green deformation tensor.

We also have shown that both assumptions result, in general, in very different predictions for moderate-to-large stretches, and discussed long-standing issues regarding the absence of $I_2$ terms in the classical statistical theory which contradicts experimental observations. These terms are implicit in the proposed unconstrained formulation.

We have employed our recent data-driven macro-micro-macro approach to reverse-engineer the chain behavior from a single test curve using both assumptions, and have compared the predictions for the true biaxial test series from \cite{Kawabata}. When using the unconstrained assumption and any single test curve to calibrate the material, we are capable of accurately predicting the behavior of the polymer for all tests. We have not been capable of doing so using the affinity assumption. Moreover, in our investigations, when the orientation of chains play a relevant role in the predictions, the network affinity assumption deviates more from experimental results. We are not aware of any model using this assumption capable of accurately capturing Kawabata's experiments (or any similar set) using only one test curve for obtaining the material parameters.
However, we did so without including any additional ingredients like chain tube constraints or minimizing free-fluctuating chain stretch fields.
 Indeed, we have made connections with the different observations in the literature
which highlight difficulties
in predicting experimental results and which justified needed improvements.

Probably, the free orientational network assumption has also some range of validity which needs to be studied. The chain rotations and their stretches are coupled and both entropy contributions may have a different balance in different cases, as it is apparently the case in collagenous tissues. Whether the changes in the network orientational entropy may be always neglected in isotropic polymers is an open question, but it seems probable that strongly crosslinked and swollen polymers may behave in an orientational more affine manner. That may also hold near chain locking stretches.

In summary, in our opinion, the assumptions behind the chain orientations in polymer networks during finite deformations should not be overlooked, because they have an important impact in the predictions and, hence, also in conclusions regarding the microstructure if these are just supported on the accuracy of predictions.

\section*{Acknowledgements}
Partial financial support for the present work has been facilitated by the Agencia Estatal de Investigaci\'{o}n, Ministerio de Ciencia, Innovaci\'{o}n y Universidades of Spain, under grant PGC-2018-097257-B-C32.

\bibliographystyle{elsarticle-harv}
\bibliography{AffineBib}

\begin{thebibliography}{82}
\expandafter\ifx\csname natexlab\endcsname\relax\def\natexlab#1{#1}\fi
\providecommand{\url}[1]{\texttt{#1}}
\providecommand{\href}[2]{#2}
\providecommand{\path}[1]{#1}
\providecommand{\DOIprefix}{doi:}
\providecommand{\ArXivprefix}{arXiv:}
\providecommand{\URLprefix}{URL: }
\providecommand{\Pubmedprefix}{pmid:}
\providecommand{\doi}[1]{\href{http://dx.doi.org/#1}{\path{#1}}}
\providecommand{\Pubmed}[1]{\href{pmid:#1}{\path{#1}}}
\providecommand{\bibinfo}[2]{#2}
\ifx\xfnm\relax \def\xfnm[#1]{\unskip,\space#1}\fi
\bibitem[{Alastru\'{e} et~al.(2009)Alastru\'{e}, Mart\'{i}nez, Doblar\'{e} and
  Menzel}]{Alastrue}
\bibinfo{author}{Alastru\'{e}, V.}, \bibinfo{author}{Mart\'{i}nez, M.},
  \bibinfo{author}{Doblar\'{e}, M.}, \bibinfo{author}{Menzel, A.},
  \bibinfo{year}{2009}.
\newblock \bibinfo{title}{Anisotropic micro-sphere-based finite elasticity
  applied to blood vessel modelling}.
\newblock \bibinfo{journal}{Journal of the Mechanics and Physics of Solids}
  \bibinfo{volume}{57}, \bibinfo{pages}{178--203}.
\bibitem[{Amores et~al.(2020)Amores, Ben\'{i}tez and Mont\'{a}ns}]{MMM}
\bibinfo{author}{Amores, V.}, \bibinfo{author}{Ben\'{i}tez, J.},
  \bibinfo{author}{Mont\'{a}ns, F.}, \bibinfo{year}{2020}.
\newblock \bibinfo{title}{Data-driven, structure-based hyperelastic manifolds:
  A macro-micro-macro approach to reverse-engineer the chain behavior and
  perform efficient simulations of polymers}.
\newblock \bibinfo{journal}{Computers and Structures} \bibinfo{volume}{231},
  \bibinfo{pages}{106209}.
\bibitem[{Anthony et~al.(1942)Anthony, Caston and Guth}]{Anthony}
\bibinfo{author}{Anthony, R.L.}, \bibinfo{author}{Caston, R.H.},
  \bibinfo{author}{Guth, E.}, \bibinfo{year}{1942}.
\newblock \bibinfo{title}{Equations of state for natural and synthetic
  rubber-like materials. {I. U}naccelerated natural soft rubber}.
\newblock \bibinfo{journal}{The Journal of Physical Chemistry}
  \bibinfo{volume}{46}, \bibinfo{pages}{826--840}.
\bibitem[{Argon(2013)}]{Argon}
\bibinfo{author}{Argon, A.}, \bibinfo{year}{2013}.
\newblock \bibinfo{title}{The physics of deformation and fracture of polymers}.
\newblock \bibinfo{publisher}{Cambridge University Presss}.
\bibitem[{Arruda and Boyce(1993)}]{Arruda-Boyce-8chain}
\bibinfo{author}{Arruda, E.}, \bibinfo{author}{Boyce, M.},
  \bibinfo{year}{1993}.
\newblock \bibinfo{title}{A three-dimensional constitutive model for the large
  stretch behavior of rubber elastic materials}.
\newblock \bibinfo{journal}{Journal of the Mechanics and Physics of Solids}
  \bibinfo{volume}{41}, \bibinfo{pages}{389--412}.
\bibitem[{Arruda and Boyce(2000)}]{Arruda-Boyce-review}
\bibinfo{author}{Arruda, E.}, \bibinfo{author}{Boyce, M.},
  \bibinfo{year}{2000}.
\newblock \bibinfo{title}{Constitutive models of rubber elasticity: A review}.
\newblock \bibinfo{journal}{Rubber Chemistry and Technology}
  \bibinfo{volume}{73}, \bibinfo{pages}{504--522}.
\bibitem[{Ball et~al.(1981)Ball, Doi, Edwards and Warner}]{Ball}
\bibinfo{author}{Ball, R.}, \bibinfo{author}{Doi, M.},
  \bibinfo{author}{Edwards, S.}, \bibinfo{author}{Warner, M.},
  \bibinfo{year}{1981}.
\newblock \bibinfo{title}{Elasticity of entangled networks}.
\newblock \bibinfo{journal}{Polymer} \bibinfo{volume}{22},
  \bibinfo{pages}{1010--1018}.
\bibitem[{Bazant and Oh(1986)}]{Bazant}
\bibinfo{author}{Bazant, Z.}, \bibinfo{author}{Oh, B.}, \bibinfo{year}{1986}.
\newblock \bibinfo{title}{Efficient numerical integration on the surface of a
  sphere}.
\newblock \bibinfo{journal}{ZAMM-Journal of Applied Mathematics and Mechanics /
  Zeitschrift f\"ur Angewande Mathematik unk Mechanik} \bibinfo{volume}{66},
  \bibinfo{pages}{37--49}.
\bibitem[{Bergstr\"om(2015)}]{Bergstrom}
\bibinfo{author}{Bergstr\"om, J.}, \bibinfo{year}{2015}.
\newblock \bibinfo{title}{Mechanics of Solid Polymers}.
\newblock \bibinfo{publisher}{Elsevier, Amsterdam}.
\bibitem[{Billiar and Sacks(1997)}]{Billiar}
\bibinfo{author}{Billiar, K.}, \bibinfo{author}{Sacks, M.},
  \bibinfo{year}{1997}.
\newblock \bibinfo{title}{A method to quantify the fiber kinematics of planar
  tissues under biaxial stretch}.
\newblock \bibinfo{journal}{Journal of Biomechanics} \bibinfo{volume}{30},
  \bibinfo{pages}{753--756}.
\bibitem[{Carroll(2011)}]{Carroll}
\bibinfo{author}{Carroll, M.}, \bibinfo{year}{2011}.
\newblock \bibinfo{title}{A strain energy function for vulcanized rubbers}.
\newblock \bibinfo{journal}{Journal of Elasticity} \bibinfo{volume}{103},
  \bibinfo{pages}{173--187}.
\bibitem[{Chagnon et~al.(2015)Chagnon, Rebouah and Favier}]{Chagnon}
\bibinfo{author}{Chagnon, G.}, \bibinfo{author}{Rebouah, M.},
  \bibinfo{author}{Favier, D.}, \bibinfo{year}{2015}.
\newblock \bibinfo{title}{Hyperelastic energy densities for soft biological
  tissues: A review}.
\newblock \bibinfo{journal}{Journal of Elasticity} \bibinfo{volume}{120},
  \bibinfo{pages}{129--160}.
\bibitem[{Crespo et~al.(2017)Crespo, Latorre and Mon\'{a}ns}]{WYPIWYG}
\bibinfo{author}{Crespo, J.}, \bibinfo{author}{Latorre, M.},
  \bibinfo{author}{Mon\'{a}ns, F.}, \bibinfo{year}{2017}.
\newblock \bibinfo{title}{{WYPIWYG} hyperelasticity for isotropic, compressible
  materials}.
\newblock \bibinfo{journal}{Computational Mechanics} \bibinfo{volume}{59},
  \bibinfo{pages}{73--92}.
\bibitem[{Crespo and Mont\'ans(2019)}]{CrespoIJES}
\bibinfo{author}{Crespo, J.}, \bibinfo{author}{Mont\'ans, F.},
  \bibinfo{year}{2019}.
\newblock \bibinfo{title}{General solution procedures to compute the stored
  energy density of conservative solids directly from experimental data}.
\newblock \bibinfo{journal}{International Journal of Engineering Science}
  \bibinfo{volume}{141}.
\bibitem[{Criscione(2003)}]{Criscione}
\bibinfo{author}{Criscione, J.}, \bibinfo{year}{2003}.
\newblock \bibinfo{title}{Rivlin's representation formula is ill-conceived for
  the determination of response functions via biaxial testing}.
\newblock \bibinfo{journal}{Journal of Elasticity} \bibinfo{volume}{70},
  \bibinfo{pages}{129--147}.
\bibitem[{Doi and Edwards(1988)}]{Doi-Edwards}
\bibinfo{author}{Doi, M.}, \bibinfo{author}{Edwards, S.}, \bibinfo{year}{1988}.
\newblock \bibinfo{title}{The theory of polymer dynamics}.
\newblock \bibinfo{publisher}{Clarendon Press, Oxford}.
\bibitem[{Edwards(1967)}]{Edwards67}
\bibinfo{author}{Edwards, S.}, \bibinfo{year}{1967}.
\newblock \bibinfo{title}{Statistical mechanics with topological constraints:
  {I}}.
\newblock \bibinfo{journal}{Proceedings of the Physical Society}
  \bibinfo{volume}{91}, \bibinfo{pages}{513--519}.
\bibitem[{Edwards and Vilgis(1988)}]{Edwards-Vilgis}
\bibinfo{author}{Edwards, S.}, \bibinfo{author}{Vilgis, T.},
  \bibinfo{year}{1988}.
\newblock \bibinfo{title}{The tube model theory of rubber elasticity}.
\newblock \bibinfo{journal}{Reports on Progress in Physics}
  \bibinfo{volume}{51}, \bibinfo{pages}{243--297}.
\bibitem[{Flory(1969)}]{Flory}
\bibinfo{author}{Flory, P.}, \bibinfo{year}{1969}.
\newblock \bibinfo{title}{Statistical Mechanics of Chain Molecules}.
\newblock \bibinfo{publisher}{Interscience, New York}.
\bibitem[{Flory(1976)}]{Flory76}
\bibinfo{author}{Flory, P.}, \bibinfo{year}{1976}.
\newblock \bibinfo{title}{Statistical thermodynamics of random entworks}.
\newblock \bibinfo{journal}{Proceedings of the Royal Society of London, A:
  Mathematical, Physical and Engineering Sciences} \bibinfo{volume}{351},
  \bibinfo{pages}{351--380}.
\bibitem[{Flory and Erman(1982)}]{Flory-Erman}
\bibinfo{author}{Flory, P.}, \bibinfo{author}{Erman, B.}, \bibinfo{year}{1982}.
\newblock \bibinfo{title}{Theory of elasticity polymer networks. 3.}
\newblock \bibinfo{journal}{Macromolecules} \bibinfo{volume}{15},
  \bibinfo{pages}{800--806}.
\bibitem[{Flory and Rehner(1943)}]{Flory-Rehner}
\bibinfo{author}{Flory, P.}, \bibinfo{author}{Rehner, J.J.},
  \bibinfo{year}{1943}.
\newblock \bibinfo{title}{Statistical mechanics of cross-linked polymer
  networks}.
\newblock \bibinfo{journal}{The Journal of Chemical Physics}
  \bibinfo{volume}{11}, \bibinfo{pages}{512--520}.
\bibitem[{Gent(1996)}]{Gent}
\bibinfo{author}{Gent, A.}, \bibinfo{year}{1996}.
\newblock \bibinfo{title}{A new constitutive relation for rubber}.
\newblock \bibinfo{journal}{Rubber Chemistry and Technology}
  \bibinfo{volume}{69}, \bibinfo{pages}{59--61}.
\bibitem[{Gilbert et~al.(2006)Gilbert, Sacks, Grashow, Woo, Badylak and
  Chancellor}]{Gilbert}
\bibinfo{author}{Gilbert, T.}, \bibinfo{author}{Sacks, M.},
  \bibinfo{author}{Grashow, J.}, \bibinfo{author}{Woo, S.Y.},
  \bibinfo{author}{Badylak, S.}, \bibinfo{author}{Chancellor, M.},
  \bibinfo{year}{2006}.
\newblock \bibinfo{title}{Fiber kinematics of small intestinal submucosa under
  biaxial and uniaxial stretch}.
\newblock \bibinfo{journal}{Journal of Biomechanical Engineering}
  \bibinfo{volume}{128}, \bibinfo{pages}{890--898}.
\bibitem[{Gusev and Schwarz(2019)}]{Gusev}
\bibinfo{author}{Gusev, A.}, \bibinfo{author}{Schwarz, F.},
  \bibinfo{year}{2019}.
\newblock \bibinfo{title}{Molecular dynamics validation and applications of the
  maximum entropy homogenization procedure for predicting the elastic
  properties of gaussian polymer networks}.
\newblock \bibinfo{journal}{Macromolecules} \bibinfo{volume}{52},
  \bibinfo{pages}{9445--9455}.
\bibitem[{Hart-Smith(1966)}]{Hart-Smith}
\bibinfo{author}{Hart-Smith, L.}, \bibinfo{year}{1966}.
\newblock \bibinfo{title}{Elasticity parameters for finite deformations of
  rubber-like materials}.
\newblock \bibinfo{journal}{Zeitschrift F\"ur Angewandte Mathematik Und Physik
  ZAMP} \bibinfo{volume}{17}, \bibinfo{pages}{608--626}.
\bibitem[{Heinrich et~al.(2003)Heinrich, Kaliske, K\"uppel, Mark, Straube and
  Vilgis}]{Heinrich-Kaliske-critica}
\bibinfo{author}{Heinrich, G.}, \bibinfo{author}{Kaliske, M.},
  \bibinfo{author}{K\"uppel, M.}, \bibinfo{author}{Mark, J.},
  \bibinfo{author}{Straube, E.}, \bibinfo{author}{Vilgis, T.},
  \bibinfo{year}{2003}.
\newblock \bibinfo{title}{The thermoelasticity of rubberlike materials and
  related constitutive laws}.
\newblock \bibinfo{journal}{Journal of Macromolecular Science}
  \bibinfo{volume}{40}, \bibinfo{pages}{87--93}.
\bibitem[{Holzapfel(2000)}]{Holzapfel}
\bibinfo{author}{Holzapfel, G.}, \bibinfo{year}{2000}.
\newblock \bibinfo{title}{Nonlinar Solid Mechanics. {A} Continuum Approach for
  Engineering}.
\newblock \bibinfo{publisher}{John Wiley \& Sons, Chichester}.
\bibitem[{Horgan and Saccomandi(2002)}]{Horgan-Sacco}
\bibinfo{author}{Horgan, C.}, \bibinfo{author}{Saccomandi, G.},
  \bibinfo{year}{2002}.
\newblock \bibinfo{title}{Constitutive modeling of rubber-like and biological
  materials with limiting chain extensibility}.
\newblock \bibinfo{journal}{Mathematics and Mechanics of Solids}
  \bibinfo{volume}{7}, \bibinfo{pages}{353--371}.
\bibitem[{Horgan and Schwartz(2005)}]{Horgan-Sch}
\bibinfo{author}{Horgan, C.}, \bibinfo{author}{Schwartz, J.},
  \bibinfo{year}{2005}.
\newblock \bibinfo{title}{Constitutive modeling and the trousers test for
  fracture of rubber-like materials}.
\newblock \bibinfo{journal}{Journal of the Mechanics and Physics of Solids}
  \bibinfo{volume}{53}, \bibinfo{pages}{545--564}.
\bibitem[{James and Guth(1943)}]{James-Guth}
\bibinfo{author}{James, H.}, \bibinfo{author}{Guth, E.}, \bibinfo{year}{1943}.
\newblock \bibinfo{title}{Theory of the elastic properties of rubber}.
\newblock \bibinfo{journal}{The Journal of Chemical Physics}
  \bibinfo{volume}{11}, \bibinfo{pages}{455--481}.
\bibitem[{Kaliske and Heinrich(1999)}]{Kaliske-Heinrich}
\bibinfo{author}{Kaliske, M.}, \bibinfo{author}{Heinrich, G.},
  \bibinfo{year}{1999}.
\newblock \bibinfo{title}{An extended tube-model for rubber elasticity:
  Statistical-mechanical theory and finite element implementation}.
\newblock \bibinfo{journal}{Rubber Chemistry and Technology}
  \bibinfo{volume}{72}, \bibinfo{pages}{602--632}.
\bibitem[{Kawabata et~al.(1981)Kawabata, Matsuda, Tei and Kawai}]{Kawabata}
\bibinfo{author}{Kawabata, S.}, \bibinfo{author}{Matsuda, M.},
  \bibinfo{author}{Tei, K.}, \bibinfo{author}{Kawai, H.}, \bibinfo{year}{1981}.
\newblock \bibinfo{title}{Experimental survey of the strain energy density of
  isoprene rubber}.
\newblock \bibinfo{journal}{Macromolecules} \bibinfo{volume}{14},
  \bibinfo{pages}{154--162}.
\bibitem[{Kawamura et~al.(2001)Kawamura, Urayama and Kohjiya}]{Kawamura}
\bibinfo{author}{Kawamura, T.}, \bibinfo{author}{Urayama, K.},
  \bibinfo{author}{Kohjiya, S.}, \bibinfo{year}{2001}.
\newblock \bibinfo{title}{Multiaxial deformations of end-linked
  poly(dimethylsiloxane) networks. 1. phenomenological approach tro strain
  energy density function}.
\newblock \bibinfo{journal}{Macromolecules} \bibinfo{volume}{34},
  \bibinfo{pages}{8252--8260}.
\bibitem[{Kearsley and Zapas(1980)}]{KZ}
\bibinfo{author}{Kearsley, E.}, \bibinfo{author}{Zapas, L.},
  \bibinfo{year}{1980}.
\newblock \bibinfo{title}{Some methods of measurement of an elastic strain
  energy function of the {V}alanis-{L}andel type}.
\newblock \bibinfo{journal}{Journal of Rheology} \bibinfo{volume}{24},
  \bibinfo{pages}{483--500}.
\bibitem[{Ki\^em and Itskov(2016)}]{Itskov}
\bibinfo{author}{Ki\^em, V.}, \bibinfo{author}{Itskov, M.},
  \bibinfo{year}{2016}.
\newblock \bibinfo{title}{Analytical network-averaging of the tube model:
  Rubber elasticity.}
\newblock \bibinfo{journal}{Journal of the Mechanics and Physics of Solids}
  \bibinfo{volume}{95}, \bibinfo{pages}{254--269}.
\bibitem[{Kroon(2011)}]{Kroon}
\bibinfo{author}{Kroon, M.}, \bibinfo{year}{2011}.
\newblock \bibinfo{title}{An 8-chain model for rubber-like materials accounting
  for non-affine chain deformations and topological constraints}.
\newblock \bibinfo{journal}{Journal of Elasticity} \bibinfo{volume}{102},
  \bibinfo{pages}{99--116}.
\bibitem[{Kuhn(1934)}]{Kuhn1}
\bibinfo{author}{Kuhn, W.}, \bibinfo{year}{1934}.
\newblock \bibinfo{title}{\"uber die gestalt fadenf\"ormiger molek\"ule in
  l\"osungen}.
\newblock \bibinfo{journal}{Kolloid-Zeitschrift} \bibinfo{volume}{59},
  \bibinfo{pages}{208--216}.
\bibitem[{Kuhn(1936)}]{Kuhn2}
\bibinfo{author}{Kuhn, W.}, \bibinfo{year}{1936}.
\newblock \bibinfo{title}{Beziehungen zwischen molek\"ulgr\"osse, statistischer
  molek\"ulgestalt und elastischen eigenschaften hochpolymerer stoffe}.
\newblock \bibinfo{journal}{Kolloid-Zeitschrift} \bibinfo{volume}{76}.
\bibitem[{Kuhn and Gr\"un(1942)}]{Kuhn-Grun}
\bibinfo{author}{Kuhn, W.}, \bibinfo{author}{Gr\"un, F.}, \bibinfo{year}{1942}.
\newblock \bibinfo{title}{Beziehungen zwischen elastischen konstanten und
  dehnungsdoppelbrechung hochelastischer stoffe}.
\newblock \bibinfo{journal}{Kolloid-Zeitschrift} \bibinfo{volume}{101},
  \bibinfo{pages}{248--271}.
\bibitem[{Lang(2017)}]{Lang}
\bibinfo{author}{Lang, M.}, \bibinfo{year}{2017}.
\newblock \bibinfo{title}{Relation between cross-link fluctuations and
  elasticity in entangled polymer networks}.
\newblock \bibinfo{journal}{Macromolecules} \bibinfo{volume}{50},
  \bibinfo{pages}{2547--2555}.
\bibitem[{Latorre et~al.(2017a)Latorre, De~Rosa and
  Mont\'{a}ns}]{Understanding}
\bibinfo{author}{Latorre, M.}, \bibinfo{author}{De~Rosa, E.},
  \bibinfo{author}{Mont\'{a}ns, F.}, \bibinfo{year}{2017}a.
\newblock \bibinfo{title}{Understanding the need of the compression branch to
  characterize hyperelastic materials}.
\newblock \bibinfo{journal}{International Journal of Non-Linear Mechanics}
  \bibinfo{volume}{89}, \bibinfo{pages}{14--24}.
\bibitem[{Latorre and Mont\'{a}ns(2014)}]{LatorreWYIPWYG}
\bibinfo{author}{Latorre, M.}, \bibinfo{author}{Mont\'{a}ns, F.},
  \bibinfo{year}{2014}.
\newblock \bibinfo{title}{{W}hat-{Y}ou-{P}rescribe-{I}s-{W}hat-{Y}ou-{G}et
  orthotropic hyperelasticity}.
\newblock \bibinfo{journal}{Computational Mechanics} \bibinfo{volume}{53},
  \bibinfo{pages}{1279--1298}.
\bibitem[{Latorre et~al.(2017b)Latorre, Pe\~na and Mont\'ans}]{Pena}
\bibinfo{author}{Latorre, M.}, \bibinfo{author}{Pe\~na, E.},
  \bibinfo{author}{Mont\'ans, F.}, \bibinfo{year}{2017}b.
\newblock \bibinfo{title}{Determination and finite element validation of the
  {WYPIWYG} strain energy of superficial fascia from experimental data}.
\newblock \bibinfo{journal}{Annals of Biomedical Engineering}
  \bibinfo{volume}{45}, \bibinfo{pages}{799--810}.
\bibitem[{Lopez-Pamies(2010)}]{Lopez-Pamies}
\bibinfo{author}{Lopez-Pamies, O.}, \bibinfo{year}{2010}.
\newblock \bibinfo{title}{A new $i_1$-based hyperelastic model for rubber
  elastic materials}.
\newblock \bibinfo{journal}{Comptes Rendus Mecanique} \bibinfo{volume}{338},
  \bibinfo{pages}{3--11}.
\bibitem[{Mark and Erman(1988)}]{Mark_Eman_book}
\bibinfo{author}{Mark, J.}, \bibinfo{author}{Erman, B.}, \bibinfo{year}{1988}.
\newblock \bibinfo{title}{Rubberlike elasticity: A molecular primer}.
\newblock \bibinfo{publisher}{Wiley, New York}.
\bibitem[{Markmann and Verron(2006)}]{Verron}
\bibinfo{author}{Markmann, G.}, \bibinfo{author}{Verron, E.},
  \bibinfo{year}{2006}.
\newblock \bibinfo{title}{Comparison of hyperelastic models for rubber-like
  materials}.
\newblock \bibinfo{journal}{Rubber Chemistry and Technology}
  \bibinfo{volume}{79}, \bibinfo{pages}{835--858}.
\bibitem[{Menzel and Waffenschmidt(2009)}]{Menzel}
\bibinfo{author}{Menzel, A.}, \bibinfo{author}{Waffenschmidt, T.},
  \bibinfo{year}{2009}.
\newblock \bibinfo{title}{A microsphere-based remodelling formulation for
  anisotropic biological tissues}.
\newblock \bibinfo{journal}{Philosophical Transactions of the Royal Society A:
  Mathematical, Physical and Engineering Sciences} \bibinfo{volume}{367},
  \bibinfo{pages}{3499--3523}.
\bibitem[{Meyer et~al.(1932)Meyer, Susich and Valk\'o}]{Meyer}
\bibinfo{author}{Meyer, K.}, \bibinfo{author}{Susich, G.},
  \bibinfo{author}{Valk\'o, E.}, \bibinfo{year}{1932}.
\newblock \bibinfo{title}{Die elastischen eigenschaften der organischen
  hochpolymeren und ihre kinetische deutung}.
\newblock \bibinfo{journal}{Kolloid-Zeitschrift} \bibinfo{volume}{59},
  \bibinfo{pages}{208--216}.
\bibitem[{Miehe et~al.(2004)Miehe, G\"oktepe and Lulei}]{Microsphere}
\bibinfo{author}{Miehe, C.}, \bibinfo{author}{G\"oktepe, S.},
  \bibinfo{author}{Lulei, F.}, \bibinfo{year}{2004}.
\newblock \bibinfo{title}{A micro-macro approach to rubber-like
  materials---{P}art {I}: the non-affine micro-sphere model of rubber
  elasticity}.
\newblock \bibinfo{journal}{Journal of the Mechanics and Physics of Solids}
  \bibinfo{volume}{52}, \bibinfo{pages}{2617--2660}.
\bibitem[{Mooney(1940)}]{Mooney}
\bibinfo{author}{Mooney, M.}, \bibinfo{year}{1940}.
\newblock \bibinfo{title}{A theory of large elastic deformations}.
\newblock \bibinfo{journal}{Journal of Applied Physics} \bibinfo{volume}{11},
  \bibinfo{pages}{582}.
\bibitem[{M\"uller and Weiss(2005)}]{MullerWeiss_book}
\bibinfo{author}{M\"uller, I.}, \bibinfo{author}{Weiss, W.},
  \bibinfo{year}{2005}.
\newblock \bibinfo{title}{Entropy and Energy}.
\newblock \bibinfo{publisher}{Springer, Berlin}.
\bibitem[{Obata et~al.(1970)Obata, Kawabata and Kawai}]{Obata}
\bibinfo{author}{Obata, Y.}, \bibinfo{author}{Kawabata, S.},
  \bibinfo{author}{Kawai, H.}, \bibinfo{year}{1970}.
\newblock \bibinfo{title}{Mechanical properties of natural rubber vulcanizates
  in finite deformation}.
\newblock \bibinfo{journal}{Journal of Polymer Science Part A-2: Polymer
  Physics} \bibinfo{volume}{8}, \bibinfo{pages}{903--919}.
\bibitem[{Ogden(1972)}]{Ogden_model}
\bibinfo{author}{Ogden, R.}, \bibinfo{year}{1972}.
\newblock \bibinfo{title}{Large deformation isotropic elasticity: on the
  correlation of theory and experiment for incompressible rubberlike solids}.
\newblock \bibinfo{journal}{Proceedings of the Royal Society of London A}
  \bibinfo{volume}{326}.
\bibitem[{Ogden et~al.(2004)Ogden, Saccomandi and Sgura}]{OSS}
\bibinfo{author}{Ogden, R.}, \bibinfo{author}{Saccomandi, G.},
  \bibinfo{author}{Sgura, I.}, \bibinfo{year}{2004}.
\newblock \bibinfo{title}{Fitting hyperelastic models to experimental data}.
\newblock \bibinfo{journal}{Computational Mechanics} \bibinfo{volume}{34},
  \bibinfo{pages}{484--502}.
\bibitem[{Pancheri and Dorfmann(2014)}]{Pancheri_Dorfmann}
\bibinfo{author}{Pancheri, F.}, \bibinfo{author}{Dorfmann, L.},
  \bibinfo{year}{2014}.
\newblock \bibinfo{title}{Strain-controlled biaxial tension of natural rubber:
  new experimental data}.
\newblock \bibinfo{journal}{Rubber Chemistry and Technology}
  \bibinfo{volume}{87}, \bibinfo{pages}{120--138}.
\bibitem[{Qin et~al.(2012)Qin, So and Milner}]{Qin}
\bibinfo{author}{Qin, J.}, \bibinfo{author}{So, J.}, \bibinfo{author}{Milner,
  S.}, \bibinfo{year}{2012}.
\newblock \bibinfo{title}{ube diameter of stretched and compressed permanently
  entangled polymers}.
\newblock \bibinfo{journal}{Macromolecules} \bibinfo{volume}{45},
  \bibinfo{pages}{9816--9822}.
\bibitem[{Rivlin(1948)}]{Rivlin}
\bibinfo{author}{Rivlin, R.}, \bibinfo{year}{1948}.
\newblock \bibinfo{title}{Large elastic deformations of isotropic materials:
  {IV}. {F}urther developments of the general theory}.
\newblock \bibinfo{journal}{Philosophical Transactions of the Royal Society A}
  \bibinfo{volume}{241}, \bibinfo{pages}{379--397}.
\bibitem[{Rivlin and Saunders(1951)}]{RivlinSaunders}
\bibinfo{author}{Rivlin, R.}, \bibinfo{author}{Saunders, D.},
  \bibinfo{year}{1951}.
\newblock \bibinfo{title}{Large elastic deformations of isotropic materials
  {VII. E}xperiments on the deformation of rubber}.
\newblock \bibinfo{journal}{Philosophical Transactions of the Royal Society A:
  Mathematical, Physical and Engineering Sciences} \bibinfo{volume}{243},
  \bibinfo{pages}{251--288}.
\bibitem[{Romero et~al.(2017)Romero, Latorre and Mont\'ans}]{Romero}
\bibinfo{author}{Romero, X.}, \bibinfo{author}{Latorre, M.},
  \bibinfo{author}{Mont\'ans, F.}, \bibinfo{year}{2017}.
\newblock \bibinfo{title}{Determination of the wypiwyg strain energy density of
  skin through finite element analysis of the experiments on circular
  specimens}.
\newblock \bibinfo{journal}{Finite Elements in Analysis and Design}
  \bibinfo{volume}{134}, \bibinfo{pages}{1--15}.
\bibitem[{Rubinstein and Panyukov(1997)}]{Rubinstein}
\bibinfo{author}{Rubinstein, M.}, \bibinfo{author}{Panyukov, S.},
  \bibinfo{year}{1997}.
\newblock \bibinfo{title}{Nonaffine deformation and elasticity of polymer
  networks}.
\newblock \bibinfo{journal}{Macromolecules} \bibinfo{volume}{30},
  \bibinfo{pages}{8036--8044}.
\bibitem[{S\'{a}ez et~al.(2012)S\'{a}ez, V., Pe\~{n}a, Doblar\'{e} and
  Mart\'{i}nez}]{Saez}
\bibinfo{author}{S\'{a}ez, P.}, \bibinfo{author}{V., A.},
  \bibinfo{author}{Pe\~{n}a, E.}, \bibinfo{author}{Doblar\'{e}, M.},
  \bibinfo{author}{Mart\'{i}nez, M.}, \bibinfo{year}{2012}.
\newblock \bibinfo{title}{Anisotropic microsphere-based approach to damage in
  soft fibered tissue}.
\newblock \bibinfo{journal}{Biomechanics and Modelling in Mechanobiology}
  \bibinfo{volume}{11}, \bibinfo{pages}{595--608}.
\bibitem[{Shariff(2000)}]{Shariff}
\bibinfo{author}{Shariff, M.}, \bibinfo{year}{2000}.
\newblock \bibinfo{title}{Strain energy function for filled and unfilled
  rubberlike material}.
\newblock \bibinfo{journal}{Rubber Chemistry and Technology}
  \bibinfo{volume}{73}, \bibinfo{pages}{1--18}.
\bibitem[{Shariff(2017)}]{Shariff_inv}
\bibinfo{author}{Shariff, M.}, \bibinfo{year}{2017}.
\newblock \bibinfo{title}{On the spectral constitutive modelling of
  transversely isotropic soft tissue: Physical invariants}.
\newblock \bibinfo{journal}{International Journal of Engineering Science}
  \bibinfo{volume}{120}, \bibinfo{pages}{199--219}.
\bibitem[{Staudinger(1920)}]{Staudinger}
\bibinfo{author}{Staudinger, H.}, \bibinfo{year}{1920}.
\newblock \bibinfo{title}{\"uber polymerisation}.
\newblock \bibinfo{journal}{Berichte der deutschen chemischen Gesellschaft}
  \bibinfo{volume}{53}, \bibinfo{pages}{1073--1985}.
\bibitem[{Steinmann et~al.(2012)Steinmann, Hossain and Possart}]{Steinmann}
\bibinfo{author}{Steinmann, P.}, \bibinfo{author}{Hossain, M.},
  \bibinfo{author}{Possart, G.}, \bibinfo{year}{2012}.
\newblock \bibinfo{title}{Hyperelastic models for rubber-like materials:
  consistent tangent operators and suitability for treloar's data}.
\newblock \bibinfo{journal}{Archive of Applied Mechanics} \bibinfo{volume}{82},
  \bibinfo{pages}{1183--1217}.
\bibitem[{Sussman and Bathe(2008)}]{Sussman-Bathe}
\bibinfo{author}{Sussman, T.}, \bibinfo{author}{Bathe, K.},
  \bibinfo{year}{2008}.
\newblock \bibinfo{title}{A model of incompressible isotropic hyperelastic
  material behavior using spline interpolations of tension-compression test
  data}.
\newblock \bibinfo{journal}{Communications in Numerical Methods in Engineering}
  \bibinfo{volume}{25}, \bibinfo{pages}{53--63}.
\bibitem[{Treloar(1944)}]{Treloar_exp}
\bibinfo{author}{Treloar, L.}, \bibinfo{year}{1944}.
\newblock \bibinfo{title}{Stress-strain data for vulcanized rubber under
  various types of deformation}.
\newblock \bibinfo{journal}{Rubber Chemistry and Technology}
  \bibinfo{volume}{17}, \bibinfo{pages}{813--825}.
\bibitem[{Treloar(1975)}]{Treloar_book}
\bibinfo{author}{Treloar, L.}, \bibinfo{year}{1975}.
\newblock \bibinfo{title}{The physics of Rubber Elasticity}.
\newblock \bibinfo{publisher}{Oxford University Press}.
\bibitem[{Urayama(2006)}]{Urayama}
\bibinfo{author}{Urayama, K.}, \bibinfo{year}{2006}.
\newblock \bibinfo{title}{An experimentalist's view of the physics of rubber
  elasticity}.
\newblock \bibinfo{journal}{Journal of Polymer Science} \bibinfo{volume}{44},
  \bibinfo{pages}{3440--3444}.
\bibitem[{Valanis and Landel(1967)}]{VL}
\bibinfo{author}{Valanis, K.}, \bibinfo{author}{Landel, R.},
  \bibinfo{year}{1967}.
\newblock \bibinfo{title}{The strain energy function of a hyperelastic material
  in terms of the extension ratios}.
\newblock \bibinfo{journal}{Journal of Applied Physics} \bibinfo{volume}{38},
  \bibinfo{pages}{2997}.
\bibitem[{Verron and Gros(2017)}]{Verron-Gros}
\bibinfo{author}{Verron, E.}, \bibinfo{author}{Gros, A.}, \bibinfo{year}{2017}.
\newblock \bibinfo{title}{An equal force theory for network models of soft
  materials with arbitrary molecular weight distribution}.
\newblock \bibinfo{journal}{Journal of the Mechanics and Physics of Solids}
  \bibinfo{volume}{106}, \bibinfo{pages}{176--190}.
\bibitem[{Volokh(2016)}]{Volokh}
\bibinfo{author}{Volokh, K.}, \bibinfo{year}{2016}.
\newblock \bibinfo{title}{Mechanics of Soft Materials}.
\newblock \bibinfo{publisher}{Springer, Singapore}.
\bibitem[{Waffenschmidt et~al.(2012)Waffenschmidt, Menzel and Kuhl}]{EllenKuhl}
\bibinfo{author}{Waffenschmidt, T.}, \bibinfo{author}{Menzel, A.},
  \bibinfo{author}{Kuhl, E.}, \bibinfo{year}{2012}.
\newblock \bibinfo{title}{Anisotropic density growth of bone. a computational
  micro-sphere approach}.
\newblock \bibinfo{journal}{International Journal of Solids and Structures}
  \bibinfo{volume}{49}, \bibinfo{pages}{1928--1946}.
\bibitem[{Wall(1942)}]{Wall}
\bibinfo{author}{Wall, F.}, \bibinfo{year}{1942}.
\newblock \bibinfo{title}{Statistical thermodynamics of rubber (and part
  {II})}.
\newblock \bibinfo{journal}{The Journal of Chemical Physics}
  \bibinfo{volume}{10 (2 and 7)}, \bibinfo{pages}{132--134 and 485--488}.
\bibitem[{Ward and Hadley(1993)}]{Ward_book}
\bibinfo{author}{Ward, I.}, \bibinfo{author}{Hadley, D.}, \bibinfo{year}{1993}.
\newblock \bibinfo{title}{An introduction to the mechanical properties of solid
  polymers}.
\newblock \bibinfo{publisher}{John Wiley \& Sons, Chichester}.
\bibitem[{Wen et~al.(2012)Wen, Basu, Janmey and Yodh}]{Wen}
\bibinfo{author}{Wen, Q.}, \bibinfo{author}{Basu, A.}, \bibinfo{author}{Janmey,
  P.}, \bibinfo{author}{Yodh, A.}, \bibinfo{year}{2012}.
\newblock \bibinfo{title}{Non-affine deformations in polymer hydrogels}.
\newblock \bibinfo{journal}{Soft Matter} \bibinfo{volume}{8}.
\bibitem[{Wu and van~der Giessen(1992)}]{Wu-Giessen-MRC}
\bibinfo{author}{Wu, P.}, \bibinfo{author}{van~der Giessen, E.},
  \bibinfo{year}{1992}.
\newblock \bibinfo{title}{On improved {3-D} non-{G}aussian network models for
  rubber elasticity}.
\newblock \bibinfo{journal}{Mechanics Research Communications}
  \bibinfo{volume}{19}.
\bibitem[{Wu and van~der Giessen(1993)}]{Wu-Giessen}
\bibinfo{author}{Wu, P.}, \bibinfo{author}{van~der Giessen, E.},
  \bibinfo{year}{1993}.
\newblock \bibinfo{title}{On improved network models for rubber elasticity and
  their applications to orientation hardening in glassy polymers}.
\newblock \bibinfo{journal}{Journal of the Mechanics and Physics of Solids}
  \bibinfo{volume}{41}, \bibinfo{pages}{427--456}.
\bibitem[{Xiang et~al.(2018)Xiang, Zhong, Wang, Mao, Yu and Qu}]{Xiang}
\bibinfo{author}{Xiang, Y.}, \bibinfo{author}{Zhong, D.},
  \bibinfo{author}{Wang, P.}, \bibinfo{author}{Mao, G.}, \bibinfo{author}{Yu,
  H.}, \bibinfo{author}{Qu, S.}, \bibinfo{year}{2018}.
\newblock \bibinfo{title}{A general constitutive model of soft elastomers}.
\newblock \bibinfo{journal}{Journal of the Mechanics and Physics of Solids}
  \bibinfo{volume}{117}, \bibinfo{pages}{110--122}.
\bibitem[{Yeoh(1993)}]{Yeoh}
\bibinfo{author}{Yeoh, O.}, \bibinfo{year}{1993}.
\newblock \bibinfo{title}{Some forms of the strain energy function for rubber}.
\newblock \bibinfo{journal}{Rubber Chemistry and Technology}
  \bibinfo{volume}{66}, \bibinfo{pages}{754--771}.
\bibitem[{Yeoh and Fleming(1997)}]{Yeah-Fleming}
\bibinfo{author}{Yeoh, O.}, \bibinfo{author}{Fleming, P.},
  \bibinfo{year}{1997}.
\newblock \bibinfo{title}{A new attempt to reconcile the statistical and
  phenomenological theories of rubber elasticity}.
\newblock \bibinfo{journal}{Journal of Polymer Science, Part B: Polymer
  Physics} \bibinfo{volume}{35}, \bibinfo{pages}{1919--1931}.

\end{thebibliography}

\end{document}